\documentclass[useAMS,usenatbib]{mn2e}

\usepackage{graphicx}
\usepackage{psfig}

\newcommand\apj{ApJ}
\newcommand\apjl{ApJ}
\newcommand\apjs{ApJS}
\newcommand\aap{A$\&$A}
\newcommand\mnras{MNRAS}
\newcommand\prd{Phys.~Rev.~D}
\newcommand\nat{Nature}
\newcommand\physrep{Phys.~Rep.}

\title[Excursion sets and non-Gaussianity]{Statistics of the excursion
  sets in models with local primordial non-Gaussianity}
\author[Graziano Rossi et al.] {Graziano Rossi$^{1}$\thanks{Email: graziano@kias.re.kr}, 
 Pravabati Chingangbam$^{1,2}$\thanks{Email: prava@kias.re.kr} and Changbom
  Park$^{1}$\thanks{Email: cbp@kias.re.kr}\\
\\
$^{1}$ Korea Institute for Advanced Study, Hoegiro 87, Dongdaemun-Gu, Seoul $130-722$, South Korea\\
$^{2}$ Astrophysical Research Center for the Structure and Evolution
  of the Cosmos, Sejong University,\\
  ~~98 Gunja Dong, Gwangjin gu, Seoul $143747$, South Korea}
\date{Accepted ?? Received ?? ; in original form ??}
\date{\today}
\pagerange{\pageref{firstpage}--\pageref{lastpage}}
\pubyear{2010}

\begin{document}
\maketitle
\label{firstpage}



\def\gnl{g_{\rm NL}}
\def\fnl{f_{\rm NL}}
\def\be{\begin{equation}}
\def\ee{\end{equation}}
\def\bea{\begin{eqnarray}}
\def\eea{\end{eqnarray}}
\def\bn{\begin{enumerate}}
\def\en{\end{enumerate}}
\def\nn{\nonumber}
\def\pa{{\partial}}
\def\l{\left}
\def\r{\right}



\begin{abstract}

We use the statistics of regions above or below a temperature threshold
(excursion sets) to study the cosmic microwave background (CMB)
anisotropy in models with primordial non-Gaussianity of the local type. 
By computing the full-sky spatial distribution and clustering of pixels above/below
threshold from a large set of simulated maps with different levels  
of non-Gaussianity, we find that a positive value of 
the dimensionless non-linearity parameter $f_{\rm NL}$
enhances the number density of the cold CMB excursion sets along with their
clustering strength, and reduces that of the hot ones. 
We quantify the robustness of this effect, which may be important to discriminate between the simpler
Gaussian hypothesis and non-Gaussian scenarios,
arising either from non-standard inflation or alternative early-universe models. 
The clustering of hot and cold pixels
exhibits distinct non-Gaussian signatures,
particularly at angular scales of about 75 arcmin (i.e. around the Doppler peak),
which increase linearly with $f_{\rm NL}$. Moreover, the clustering
changes strongly as a function of the smoothing angle.  
We propose several statistical tests to maximize the 
detection of a local primordial non-Gaussian signal, and provide some
theoretical insights within this framework, 
including an optimal selection of the threshold level. 
We also describe a procedure which aims at minimizing the  
cosmic variance effect, the main limit within this statistical
framework.

\end{abstract}






\section{Introduction}

Since some level of non-Gaussianity is generically expected in all inflation
models, due to interactions of the inflaton with gravity and/or
from inflaton self-interactions, seeking for deviations from the
Gaussian paradigm has recently become a major effort -- and a minor
industry -- in cosmology.
Properties of the primordial perturbations are uniquely imprinted in
the cosmic microwave background (CMB) anisotropy distribution; hence, 
its analysis is a powerful way of
looking at the specifics of the inflationary models
(or alternatives to inflation).
At the present time, the main challenge is either to detect or to constrain mild or weak
departures from primordial Gaussian initial
conditions, as the level of non-Gaussianity predicted in the simplest
single-field slow-roll inflation is slightly below the minimum value
detectable by the Planck satellite, and not within reach of future galaxy surveys. 
This is essentially why primordial non-Gaussianity is regarded as one of the most promising
probes of the inflationary universe (Komatsu et al. 2009b), and it has 
received a recent boost, both theoretically and observationally, 
mainly because of the \textit{W}ilkinson \textit{M}icrowave \textit{A}nisotropy
\textit{P}robe (WMAP) data which seems to favor a slightly positive value of the
dimensionless non-linearity parameter $f_{\rm NL}$ (Yadav \& Wandelt
2008; Komatsu et al. 2009a, 2010; Smith et al. 2009).

From the theoretical side, much effort has been directed towards   
the development of competing scenarios for perturbation generation
which go beyond the single-field slow-roll
paradigm, for instance by the inclusion in the Lagrangian of non-trivial kinetic
terms, the presence of more than one
light field during inflation, the temporary violation of slow-roll, or
a non-adiabatic initial vacuum state for the inflaton.
Examples are the curvaton model, the modulated reheating, DBI or ghost
inflation, or multi-field scenarios, some of which imply
large departures from Gaussianity 
(see, for instance, among the plethora of papers on this
subject, Linde \& Mukhanov 1997; Lyth \& Wands 2002; Acquaviva et al. 2003;
Lyth, Ungarelli \& Wands 2003; Maldacena 2003; Alishahiha et al. 2004;
Arkani-Hamed et al. 2004; Bartolo et al. 2004; Dvali, Gruzinov \& Zaldarriaga 2004; 
Chen 2005; Seery \& Lidsey 2005;
Bartolo, Matarrese \& Riotto 2006; Lyth \& Riotto 2006; Sasaki et al. 2006;
Creminelli et al. 2007; Creminelli \& Senatore 2007; Koyama et al. 2007; 
Buchbinder et al. 2008; Chen et al. 2008, 2009;
Lehners \& Steinhardt 2008;
Matarrese \& Verde 2008; Sasaki 2008; 
Bartolo \& Riotto 2009;
Brandenberger 2009;
Naruko \& Sasaki 2009; 
Senatore, Tassev \& Zaldarriaga 2009;
Silvestri \& Trodden 2009; Bartolo, Matarrese \& Riotto 2010). 

From the observational point of view, the main goal is to constrain
the level of primordial non-Gaussianity directly from a real data set, and this
is usually achieved by constructing and applying a variety of non-Gaussian
estimators such as
the 3-point function (Hinshaw et al. 1994; Gangui et al 1994),
the genus statistics or the topological genus density (Coles 1988;
Gott et al. 1990; Smoot et al. 1994; Colley \& Gott 2003; Park 2004; Gott et al. 2007),
the other Minkowski functionals (Schmalzing \& Gorski 1998; Winitzki \& Kosowsky 1998; 
Banday, Zaroubi \& Gorski 2000; Hikage et al. 2006, 2008b; Matsubara 2010), the
bispectrum and trispectrum (Spergel et al. 2007; Komatsu et al. 2009;
Rudjord et al. 2009; Liguori et al. 2010), tensor
modes (Coulson, Crittenden \& Turok 1994), wavelets (Cabella et
al. 2005; Curto et al. 2009; Vielva \& Sanz 2009), pixel and peak statistics (Adler 1981;
Bond \& Efstathiou 1987; Coles and Barrow 1987; Kogut et
al. 1995, 1996; Barreiro et al. 1997, 1998; Heavens 1998; Heavens \&
Sheth 1999; Heavens \& Gupta
2001; Hern{\'a}ndez-Monteagudo et al. 2004; Rossi et al. 2009; Hou et al. 2010), phase
correlations, multifractals, and so forth (see also Komatsu, Spergel \&
Wandelt 2005; Chen \& Szapudi
2006; Munshi \& Heavens 2010).
In this process, many observational challenges and experimental artifacts
come into play; therefore, it is perhaps not surprising that controversial
results and a long  
list of anomalies have been reported so far, ranging from a low value of the quadrupole
till North-South or parity asymmetries, strange alignments in the data, and much more
(see for example Chiang et al. 2003, 2007; Tegmark et al. 2003; de Oliveira-Costa et al. 2004; Eriksen et al. 2004, 2007; 
Schwarz et al. 2004; 
Cruz et al. 2005, 2006, 2007, 2008;
Land \& Magueijo 2005, 2007; 
Naselsky et al. 2005; 
Copi et al. 2006, 2007; Vielva et al. 2007;
Gurzadyan et al. 2008; Pietrobon et al. 2009; R{\"a}th et al 2009; Kim \& Naselsky 2010).

Deviations from Gaussian initial conditions (if any) also carry important
consequences on many aspects of the large-scale structure (LSS) of the
Universe, and galaxy
surveys can provide constraints on non-Gaussianity competitive with
those from the CMB alone. 
There are in fact modifications in the statistics of voids
(Kamionkowski, Verde \& Jimenez 2009), in the
distribution of neutral hydrogen and in the intergalactic medium
(Viel et. al 2009),  
in the high-mass tail of the halo distribution (Chiu et al. 1998;
Matarrese, Verde \& Jimenez
2000; Sefusatti \& Komatsu 2007; LoVerde et al. 2008), 
in the large-scale skewness of the galaxy distribution (Chodorowski \& Bouchet 1996),
in the number counts of clusters and of density peaks (Desjacques
et al. 2009; Jeong \& Komatsu 2009),
in the measurement of the scale dependence of
the bias of LSS tracers (Carbone et al. 2008; Dalal et al. 2008; Verde \& Matarrese 2009; Desjacques \& Seljak 2010),
in the reionization history (Crociani et al. 2009), 
in the galaxy power spectrum and bispectrum
(Scoccimarro 2000; Scoccimarro et al. 2004;
Mangilli \& Verde 2009), 
in the topology (Park et al. 1998, 2005; Gott et al. 2008; Hikage et al. 2008a), and in the abundance and
clustering of galaxies and dark matter halos (Verde et
al. 2001; Afshordi \& Tolley 2008;
Grossi et al. 2008; LoVerde et al. 2008; 
Matarrese \& Verde 2008; McDonald 2008; 
Slosar et al. 2008; 
Taruya et al. 2008; Pillepich et al. 2010).

Despite all these remarkable theoretical and
observational efforts, till date the experimental detection
of a significant deviation from the Gaussian paradigm remains still
challenging and not convincing. In this respect, 
we need to explore alternative statistics more
sensitive to deviations from Gaussianity, and
to search for unique features which may 
allow one to distinguish among the myriad of inflation models 
available in the literature.
It is important to adopt different and complementary 
statistical approaches, and  
not just a single view, because
non-Gaussianity can take innumerable forms.
In fact, while Gaussian random processes are theoretically desirable since they are
the only ones for which the knowledge of all spectral parameters completely determines all
the statistical properties, as soon as
we introduce departures from Gaussianity
a more complicated scenario emerges,
and there is no such statistics which describes fully and
uniquely the non-Gaussian nature of a sample. 
In particular, moving away from
standard estimators like the bispectrum,
trispectrum, three and four-point functions, skewness, etc, we are 
interested here in rare events,
which can often maximize deviations
from what is predicted by a Gaussian distribution.

The main goal of the present work is
to extend and apply the statistics of the excursion sets, 
regions above or below a temperature threshold, 
to models with primordial non-Gaussianity.
Specifically, we focus on the \textit{local} parametrization of 
non-Gaussianity (Salopek \&
Bond 1990), by including quadratic corrections to the curvature
perturbation.
We simulate a large set of full-sky maps with different $f_{\rm NL}$ values,
and compute the number density and the spatial clustering of 
the CMB excursion set regions. We also provide the theoretical formalism to
interpret our results. 
The excursion set statistics is fully characterized in the
context of Gaussian random fields (Kaiser 1984; Bardeen et al. 1986),
and it has been used in a variety of studies (see for example Jensen
\& Szalay 1986; Bond \&
Efstathiou 1987; Barreiro et al. 2001; Kashlinsky et al. 2001 and references therein). There are also some
extensions to non-Gaussian
conditions in the literature (i.e. Coles \& Barrow 1987; Coles 1988; Barreiro et al. 1998).
Our analysis differs from those of the previous authors primarily because we use a more realistic model
for non-Gaussianity supported by $f_{\rm NL}$ type simulations, and because we also propose some new statistical
tools, tests, and theoretical insights within this framework.
In particular, while in precedent studies it has always been shown
that the Gaussian correlation function of the excursion sets and peaks
(a subset of the
excursion sets) is easily distinguishable
from a non-Gaussian one, even if the underlying bispectra are not
statistically different (i.e. Kogut et al. 1995; Barreiro et al. 1998; 
Heavens \& Gupta 2001), we suggest here that it may not be the case if the model of
non-Gaussianity is of the local type, and the resolution adopted is
not optimal. 

Our work is also motivated by another reason. 
In a previous analysis (Rossi et al. 2009), we compared the
pixel clustering statistics -- properly extended to
handle inhomogeneous noise -- against WMAP five-year
data, and we detected deviations from the Gaussian theoretical expectations. 
In particular, we found a remarkable difference in the
clustering of hot and cold pixels at relatively small angular scales. 
A similar trend has also been reported in the literature by Tojeiro et
al. (2006), and by Hou, Banday \& Gorski (2010), although at much
larger scales.
Whether or not this discrepancy may arise from primordial
non-Gaussianity of the local type is another key question of this analysis. 

The layout of the paper is as follows.
Section \ref{NG theory} contains the theoretical tools
developed and used in this study. In Section \ref{fnl_model}
we briefly describe the local $f_{\rm NL}$ model. In Section \ref{map_simulations}  
we explain how the simulated non-Gaussian maps are constructed.
In Section \ref{excursion_set_formalism} we provide the basic
formalism for the excursion sets statistics, in the context of $f_{\rm NL}$ scenarios.    
Expressions for the one- and two-dimensional probability distribution
functions (PDFs) are given, under the assumption of weak
non-Gaussianity; this is done via a perturbative approach by the multidimensional Edgeworth expansion around a Gaussian
distribution function. Those PDFs are then used to characterize the number density
and the clustering statistics above/below threshold as a function of
$f_{\rm NL}$ (some details are provided in Appendix
\ref{edgeworth_proxy}). 
In Section \ref{history} we relate the excursion sets
formalism to other commonly used topological estimators.
In Section \ref{NG maps analysis}, computations of the number density
and the clustering statistics above/below threshold from non-Gaussian maps are presented
and interpreted according
to our theory predictions.  
Specifically, Section \ref{nd_subsection} shows the abundance of
the excursion set regions in a variety of ways, while
in Section \ref{nd_stat_subsection} we highlight some statistical tests
developed using the number density. We also 
argue that there are optimal thresholds which can maximize the
non-Gaussian contribution, as well as levels which do not allow to
distinguish a Gaussian signal from a non-Gaussian one. 
In Section \ref{clustering_subsection} we present the clustering of
hot and cold pixels for one of the optimal
temperature thresholds as a function of the smoothing scale, and in  
Section \ref{clustering_stat_subsection} we
propose a new statistical test derived from the clustering
statistics. This procedure aims at
minimizing the cosmic variance effect, and involves the
computation of the power spectrum for any given CMB map.
A final part (Section \ref{NG_conclusions})
summarizes our findings, and highlights ongoing and future work.
We leave in Appendix \ref{noise_analytic}, \ref{errors_analytic} and
\ref{spurious_ng} some technical details regarding
experimental artifacts such as inhomogeneous noise, incomplete sky
coverage, errorbar estimates and 
confusion effects caused by spurious non-Gaussianities;   
all these experimental complications will be examined in more detail in the
forthcoming publications.



\section{Theoretical framework} \label{NG theory}

In this paper we study the statistics of the excursion sets in CMB
temperature maps, to examine its sensitivity to primordial
non-Gaussianity. Even though the chosen statistics should be sensitive
to a wide class of non-Gaussian fields, in the present work we
consider the local $f_{\rm NL}$ model in detail.


\subsection{The local $f_{\rm NL}$ model} \label{fnl_model}

\begin{figure*}
\begin{center}
\includegraphics[angle=0,width=0.45\textwidth]{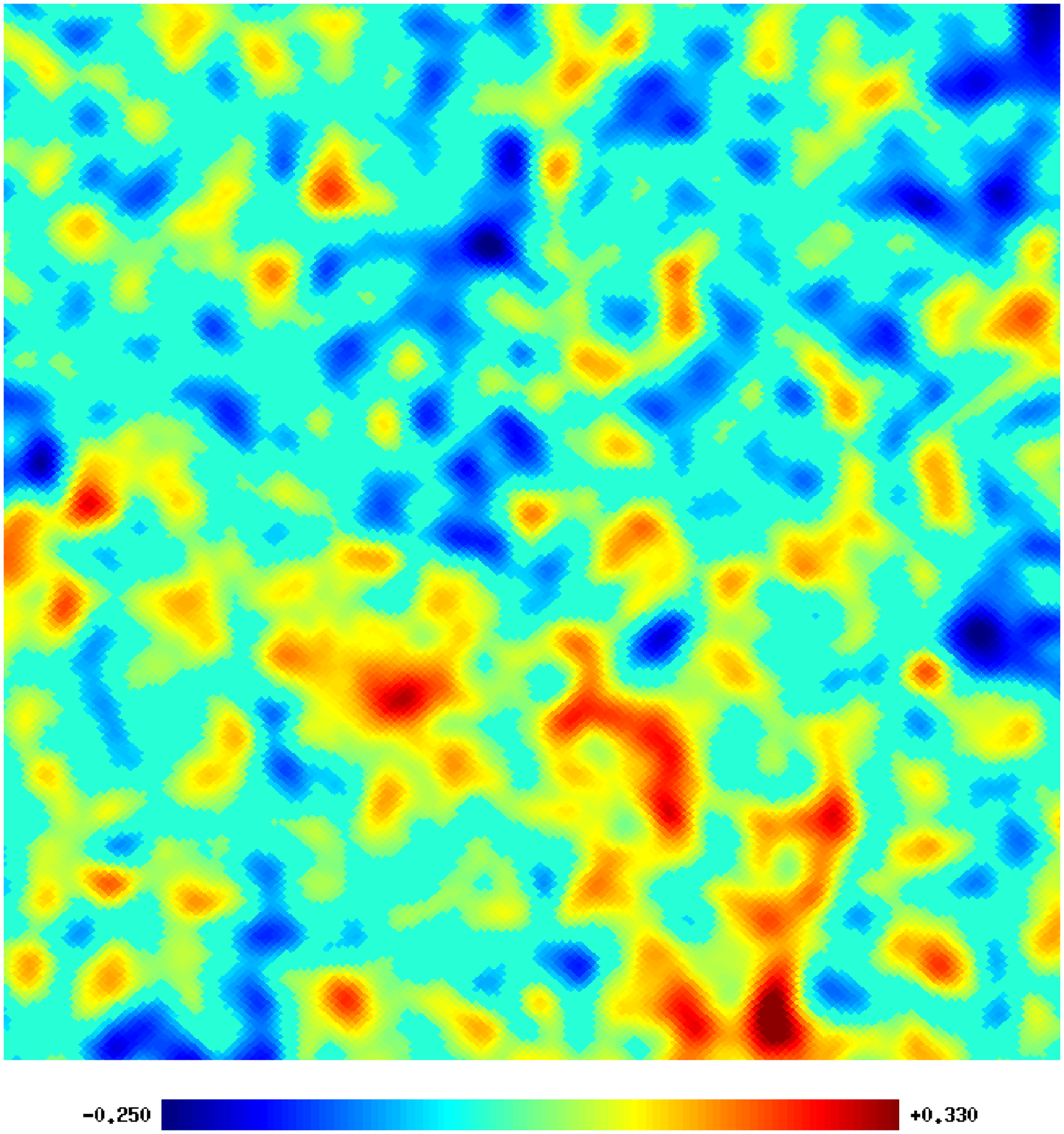}
\includegraphics[angle=0,width=0.45\textwidth]{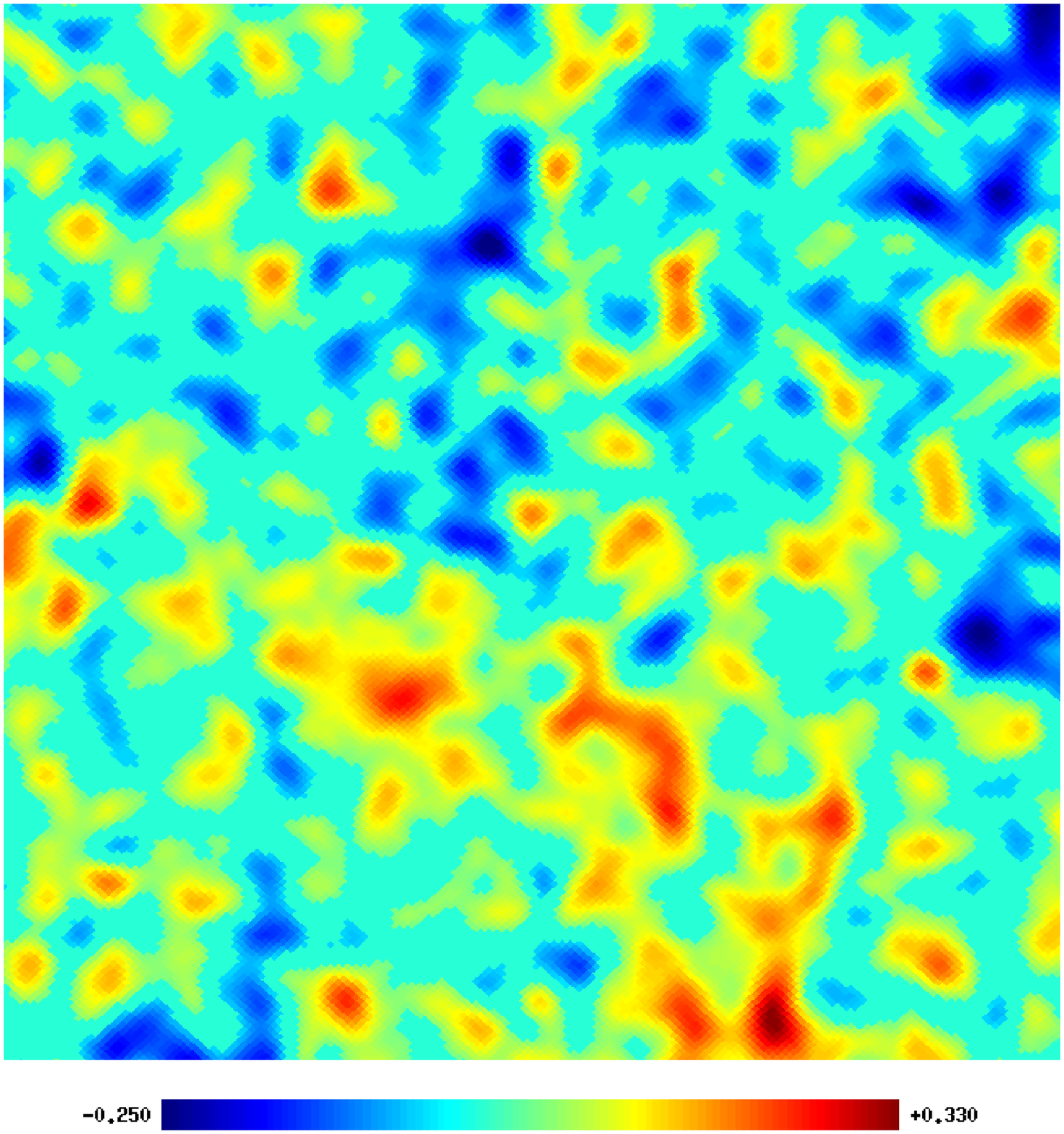}
\caption{A small patch ($\simeq 10^{\circ}\times10^{\circ}$) of the simulated CMB sky with primordial Gaussianity 
(left panel) and non-Gaussianity of the local type with $f_{\rm
  NL}=500$ (right panel), when
  smoothed with a Gaussian beam of FWHM=30 arcmin;
  regions below the threshold $\nu=0.50$ or above $\nu= -0.50$ are set to zero.
  The temperature scale is in mK, ranging from -0.250 to +0.330.}
\label{sim_patch_fig}
\end{center}
\end{figure*}

Considerable interest
has been recently focused on local type $f_{\rm NL}$, by which the non-Gaussianity of
Bardeen's curvature perturbations is \textit{locally} characterized in real
space, up to second order, by:
\begin{equation}
\Phi({\bf x})  = \phi({\bf x}) + f_{\rm NL} [\phi^2({\bf x}) - \langle
  \phi^2 \rangle]
\label{fnl_expansion_eq}
\end{equation}
and in Fourier space by
\begin{equation}
\Phi({\bf k}) = \phi({\bf k}) + \fnl \int \frac{{\rm d}^3 k'}{(2\pi)^3} 
\,\phi ({\bf k} + {\bf k'}) \,\phi({\bf k'}),
\label{phi_ng_eq}
\end{equation}
where $\phi$ is a Gaussian field (Salopek \& Bond 1990; Gangui et al. 1994; Verde et al. 2000; Komatsu
\& Spergel 2001). 
The local type non-Gaussianity
is sensitive to the bispectrum $B_{\rm \Phi} (k_1,k_2,k_3)$ with squeezed
configuration triangles (i.e. $k_1 \ll k_2 \sim k_3$; Babich et al. 2004), defined as
\begin{eqnarray}
\langle \Phi_{\rm \bf k_1} \Phi_{\rm \bf k_2} \Phi_{\rm \bf k_3}
\rangle &=& \delta_{\rm D}^3 ({\bf k}_{123}) B_{\rm \Phi}(k_1,k_2,k_3)
\nonumber \\ 
&=& \delta_{\rm D}^3 ({\bf k}_{123})
f_{\rm NL} F(k_1,k_2,k_3)
\label{bispectrum_eq}
\end{eqnarray}
where $\delta_{\rm D}$ is the Dirac delta, ${\bf k}_{\rm 123} = {\bf
  k}_{\rm 1} + {\bf k}_{\rm 2} + {\bf k}_{\rm 3}$ and
$f_{\rm NL}$ is a dimensionless parameter (or more generally
a non-linearity function), while the function F
describes the dependence on the shape of triangular configurations
defined by the three wave-numbers $k_1, k_2, k_3$.

This parametrization was originally motivated by the single-field inflation
scenarios, and it became quite popular shortly thereafter because 
it is possible to cast many inflationary models, including
the curvaton scenario (Lyth et al. 2003), in the form of equation (\ref{bispectrum_eq}); 
namely, one can express departures from non-Gaussianity in terms of a
generic function F, which may assume different model-dependent shapes
and it is broadly classified into three classes (local squeezed, non-local
equilateral, orthogonal), and the parameter or function $f_{\rm NL}$. 
Alternatives to inflation like New Ekpyrotic and cyclic models are
also expected to produce a large level of non-Gaussianity of this type
(Koyama et al. 2007; Buchbinder et al. 2008; Lehners \& Steinhardt 2008).
Therefore, the power of this formalism is that it allows one to rule out
a large class of models by putting constraints on $f_{\rm NL}$, 
and to reconstruct the inflationary
action starting from a measurement of a few observables like $f_{\rm NL}$ itself.

Note that in this paper we always use $f_{\rm NL}$ in its local meaning,
even if the usual superscript
\textit{local} is not present, and also that there  
are two distinct definitions of $f_{\rm
NL}$ in the literature, corresponding to a CMB and a LSS
convention. In the CMB convention adopted here, the local non-Gaussianity is
defined by equations (\ref{fnl_expansion_eq}-\ref{bispectrum_eq}) 
with the curvature perturbations
$\Phi$ evaluated at early times during the matter domination era, when their
value was constant. In the LSS convention, one usually assumes $\Phi$
to be the value linearly extrapolated at present time, and therefore
it includes the late-time effect of the accelerated expansion in a
cold dark matter cosmology with a cosmological constant (LCDM). 

Current limits on the primordial non-Gaussianity parameter $f_{\rm NL}$
at 95\% confidence level (CL)  
from the CMB alone are claimed to be $-4 < f_{\rm NL} < 80$ (Smith et al. 2009), 
$-18 < f_{\rm NL} < 80$ (Curto et al. 2009), 
$-36 < f_{\rm NL} < 58$ (Smidt et al. 2010), 
and $-10 < f_{\rm NL} < 74$ (Komatsu et al. 2010).
Those obtained from the LSS are similarly competitive; see for instance    
$-29 < f_{\rm NL} < 70$ by Slosar et al. (2008).


\subsection{Simulating non-Gaussian maps} \label{map_simulations}

The simulated non-Gaussian maps used in this analysis are constructed
following the method outlined in Liguori et al (2003).
The main point of their procedure
is to calculate the spherical harmonic coefficients $a_{\rm \ell m}$'s 
as an integral in real, rather than in Fourier space. 
Briefly, the CMB temperature fluctuations are expanded in terms of 
spherical harmonics as 
$\delta T(\hat{n}) = \sum_{\rm \ell m} a_{\rm \ell m} Y_{\rm \ell m}(\hat{n})$. The  $a_{\rm \ell
  m}$'s are then computed by convolving the primordial potential
fluctuations with the radiation transfer function $\Delta_{\rm \ell}$ 
(independently computed using CMBFAST developed by Seljak \& Zaldarriaga 1996), as
\begin{eqnarray}
a_{\rm \ell m}  &=& 4\pi(-i)^{\rm \ell} \int \frac{{\rm d}^3 k} {(2\pi)^3} \,\Phi({\bf k}) \,  
            \Delta_{\rm \ell}(k) \,Y^*_{\rm \ell m}(\hat{\bf k}) 
            \nonumber \\
 &=&\frac{(-i)^{\rm \ell}}{2\pi^2} \int \! {\rm d}k \,\, k^2 \,
            \Phi_{\rm \ell m}(k) \,
\Delta_{\rm \ell}(k) \nonumber \\ 
&=& \int \! {\rm d}r \, r^2 \Phi_{\rm \ell m}(r) \Delta_{\rm \ell}(r) \;
\label{alm_eq}
\end{eqnarray}
where 
\begin{eqnarray} 
\Phi_{\rm \ell m}(k) &=& \int \! {\rm d}\Omega_{\rm \hat{k}} \, 
\Phi(\mathbf{k}) \, Y_{\rm \ell m}(\hat{k}) \nonumber \\ 
&=& 4 \pi (i)^{\rm \ell} \int
\! {\rm d}r \, r^2 \, \Phi_{\rm \ell m}(r) \, j_{\rm \ell}(kr)
\label{phi_ell_eq} 
\end{eqnarray}
\begin{equation}
\Phi_{\rm \ell m}(r) =
\frac{(-i)^{\rm \ell}}{2 \pi^2} \int \! {\rm d}k \, k^2 \, \Phi_{\rm \ell m}(k) \,
j_{\rm \ell}(kr) 
\label{philmr_eq} 
\end{equation} 
\begin{equation}
\Delta_{\rm \ell}(r) =
\frac{2}{\pi} \int \! {\rm d}k \, k^2 \, \Delta_{\rm \ell}(k) j_{\rm \ell}(kr) \;.
\label{rtf_eq} 
\end{equation} 
$\Phi({\bf k})$ is the Fourier transform of the real space 
potential $\Phi({\bf x})$ defined in equations (\ref{fnl_expansion_eq}) and (\ref{phi_ng_eq}), 
$\Phi_{\rm \ell m}(r)$
is the real space harmonic potential, $\Phi_{\rm \ell m}(k)$
is its inverse and $j_{\rm \ell}$'s are spherical Bessel functions.

In presence of non-Gaussianity of the local type, from
equations (\ref{fnl_expansion_eq}-\ref{philmr_eq}) and for a constant $f_{\rm NL}$
it follows immediately that
\begin{equation}
\Phi_{\rm \ell m} = \Phi_{\rm \ell m}^{\rm G} + f_{\rm NL} 
\Phi_{\rm \rm \ell m}^{\rm f_{\rm NL}}
\label{phi_lm_gen_eq}
\end{equation} 
\begin{equation}
a_{\rm \ell m} = a_{\rm \ell m}^{\rm G} + f_{\rm NL}\,a_{\rm \ell m}^{\rm f_{\rm NL}},
\label{alm_fnl_eq}
\end{equation}
where in both equations the first right-hand side terms are the
Gaussian contributions, while the second ones account for the $f_{\rm NL}$ part. 
Note that those terms are integrals over the corresponding potentials
(i.e. $\Phi_{\rm \ell m}^{\rm G}$ involves $\phi$ only, while
$\Phi_{\rm \ell
  m}^{\rm f_{\rm NL}}$ accounts for $\phi^2$ -- see again equations
\ref{fnl_expansion_eq} and \ref{phi_ng_eq}).
The Gaussian part in (\ref{phi_lm_gen_eq}) is obtained in real space from
\begin{equation}
\Phi^{\rm G}_{\rm \ell m}(r) = \int \! {\rm d}r' \, r'^2 \, n_{\rm \ell m}(r') 
W_{\rm \ell}(r,r')
\label{real_conv_eq} 
\end{equation} 
where $n_{\rm \ell m}(r)$ are independent complex Gaussian variables, 
$W_{\rm \ell}(r,r')$ are filter functions defined by
\begin{equation}
W_{\rm \ell}(r,r') =
\frac{2}{\pi} \int \! {\rm d}k \, k^2 \, \sqrt{P_{\rm \Phi}(k)} \,
j_{\rm \ell}(kr)
j_{\rm \ell}(kr') \;
\label{filter_eq} 
\end{equation} 
and obtained as in Chingangbam \& Park (2009),
and $P_{\rm \Phi}(k)$ is the primordial power spectrum adopted.
After computing the Gaussian part of the potential 
$\phi ({\bf x}) = \sum_{\rm \ell m} \Phi^{\rm G}_{\rm \ell
  m}(r) Y_{\rm \ell m}
({\hat r})$ it is straightforward to
compute the corresponding $f_{\rm NL}$ contribution, and  
eventually the non-Gaussian temperature fluctuations
via equations (\ref{alm_eq}) and (\ref{alm_fnl_eq}).

Our simulations are provided in the HEALPix scheme (G{\'o}rski et
al. 1999) at a resolution of $N_{\rm side}=512$, giving a total of 3145728
pixels separated on average by $\theta_{\rm pix}=6.87$ arcmin.
We adopt a standard LCDM cosmological model, 
with the WMAP 5-year best fit parameters (Komatsu et al. 2009). 
An example of these realizations is shown in Figure \ref{sim_patch_fig} for a
small patch of the sky ($\simeq 10^{\circ}\times10^{\circ}$). Regions below a temperature threshold 
$\nu=0.50$ or above $\nu=-0.50$ 
are set to zero, where $\nu=\delta T / \sigma$,
with $\sigma$ being the
rms of the map and $\delta T$ the temperature anisotropy.
The left panel highlights the Gaussian case, the right panel 
shows the corresponding non-Gaussian scenario with
$f_{\rm NL}=500$. A Gaussian smoothing with a full width at half-maximum (FWHM) of 30 arcmin is
applied to those regions before clipping the field at $\nu = \pm 0.50$. 
Clearly, by visual inspection it is hard to
distinguish between the two maps,  
although one can easily show that their underlying skewness is quite different.

                          
\subsection{Excursion sets formalism in $f_{\rm NL}$ models} \label{excursion_set_formalism}

Given a CMB map with a temperature assigned to each point, an
excursion set is the ensemble of all pixels with temperatures
greater than a fixed threshold. The complementary excursion
set for temperatures lesser than a given level is symmetrically defined;
in the Gaussian case, it is expected to give the same results as the corresponding hot
excursion set. 
If the threshold under consideration is high enough, the excursion
set is composed of many disjoint groups of pixels, each group
surrounding one of the local maxima or temperature peak (see Figure
\ref{sim_patch_fig}). 
The excursion set regions 
are easily and unambiguously identifiable in the CMB sky
rather than the distributions of peaks, and
at high thresholds the number of maxima and excursion sets coincides
asymptotically. 
We are interested here in understanding how 
the number density and the clustering
statistics of the excursion regions are modified in the presence of local, 
and relatively weak, non-Gaussianity.
This theoretical framework will guide the interpretation of our 
numerical results presented in Section \ref{NG maps analysis}, from
a large set of non-Gaussian simulations. 

In a full Gaussian sky and in the absence of pixel noise, 
the number density of regions above (below) a
temperature threshold $\nu$ is simply given by:
\begin{equation}
n_{\rm pix}^{\rm G}(\nu) = {N_{\rm pix, tot}  \over 4 \pi} \cdot
{{\rm erfc} (\nu/\sqrt{2}) \over 2},
\label{nd_gauss_eq} 
\end{equation}
\noindent where $N_{\rm pix, tot} =
12 N_{\rm side}^2$ is the total number of pixels in the map, at a
resolution specified by the parameter $N_{\rm side}$.
Equation (\ref{nd_gauss_eq}) follows
immediately from an integration 
above (below) a level $\nu$ of a one-dimensional Gaussian PDF. 

In presence of non-Gaussianity of the local $f_{\rm NL}$ type, the theoretical
formalism for the number density is
complicated by the inclusion of an extra term which quantifies the
role of $f_{\rm NL}$.
Following Matsubara (1994, 2003) and Hikage, Komatsu \& Matsubara
(2006), for weak non-Gaussianity a
perturbative approach by the multidimensional Edgeworth expansion around a Gaussian
distribution function suggests that the expression for the number density will
acquire an additional term:
\begin{equation}
n_{\rm pix}^{\rm NG}(\nu) =  n_{\rm pix}^{\rm G}(\nu) +  n_{\rm
  pix}^{\rm f_{\rm NL}}(\nu)  
\label{nd_fnl_eq}
\end{equation}
where
\begin{equation}
n_{\rm pix}^{\rm f_{\rm NL}}(\nu) = {N_{\rm pix, tot} \over 4 \pi}
  \Big \{ {\sigma S^{(0)} \over 6 \sqrt{2 \pi}} 
  (\nu^2 -1)  e^{-\nu^2/2}            \Big \}.   
\label{nd_fnl_detail_eq}
\end{equation}
The skewness parameter
$S^{(0)} \equiv
\langle \delta T^3
\rangle /\sigma^4$ needs to be evaluated numerically; 
it contains the reduced bispectrum specific to the
non-Gaussian model -- simplified for $f_{\rm NL}$ constant, as given by
Komatsu \& Spergel (2001).  
Note that $S^{(0)}$ is an important parameter because it represents
the leading order contribution to the non-Gaussianity.
In fact $\sigma S^{(0)} = f_{\rm NL} \cdot A(\theta_{\rm S})$,
with $A(\theta_{\rm S})$ a numerical coefficient which depends on the
adopted smoothing $\theta_{\rm S}$.

\begin{figure}
\begin{center}
\includegraphics[angle=0,width=0.45\textwidth]{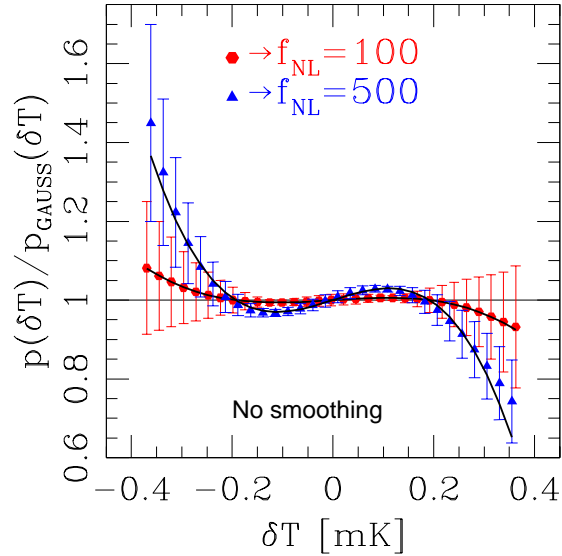}
\caption{CMB temperature distribution (mK units) in presence of weak local
  non-Gaussianity, when no smoothing is applied. Points in the figure are averages over 200
  non-Gaussian simulations with $f_{\rm NL}=100$ and $500$, errorbars
  are the corresponding 1$\sigma$ run-to-run estimates, and solid
  lines are from equation (\ref{one_d_pdf_fnl}) for the two different
  $f_{\rm NL}$ values. The average rms of $\delta T$ is 0.111 mK.} 
\label{temperature_smart_scale_fig}
\end{center}
\end{figure}

This implies that the underlying one-dimensional PDF in presence of
local non-Gaussianity is given by:
\begin{equation}
p(\mu) {\rm d} \mu \approx {1 \over \sqrt{2 \pi}}  e^{-\mu^2/2} \Big \{ 1 +
{\sigma S^{(0)} \over 6
} \mu (\mu^2 -3)
\Big \} {\rm d} \mu,
\label{one_d_pdf_fnl}
\end{equation}
where 
the first part on the right hand side of the equation is the usual Gaussian
contribution, the second is the non-Gaussian term, and  
$\mu=\delta T/
\sigma$ is now used to indicate the threshold level.
A plot of this distribution in units of the corresponding
Gaussian PDF is provided in Figure
\ref{temperature_smart_scale_fig}, for $f_{\rm NL}=100$ and $500$, when no smoothing is
applied. Points in the figure are averages over 200 realizations,
errorbars are the 1$\sigma$ run-to-run estimates from
the simulations, and
solid lines are from equation (\ref{one_d_pdf_fnl}). 
Note that, although the non-Gaussian term in (\ref{one_d_pdf_fnl}) is complicated
by the inclusion of $S^{(0)}$, $S^{(0)}$ itself is independent of the
threshold level; this will be important for the next considerations. 

With the one-dimensional PDF at hand, a number of well-known properties
in the context of Gaussian random fields,
such as the mean size and frequency of occurrence of the excursion sets above a given level
(Coles \& Barrow 1987; Kogut et al. 1995), can be easily generalized to $f_{\rm NL}$
models. We will present this analysis in a following paper, while here we
focus primarily on the pixel clustering statistics.

The correlation of the excursion sets above a threshold $\nu$
is given by (Kaiser 1984):
\begin{equation}
1+\xi_{\nu}(\theta) = P_2/P_1^2
\label{corr_smart}
\end{equation}
where
\begin{equation}
P_1 = \int_{\nu}^{\infty} p(\mu) {\rm d}\mu
\label{P1}
\end{equation}
and
\begin{equation}
P_2 = \int_{\nu}^{\infty} {\rm d}\mu_1 \int_{\nu}^{\infty}
{\rm d}\mu_2~p(\mu_1,\mu_2,w) 
\label{P2}
\end{equation}
with $p(\mu_1,\mu_2,w)$ being the two-dimensional PDF
and $w \equiv w  (\theta) = \langle \mu_1 \mu_2 \rangle$ the correlation.   

There have been attempts in the literature to generalize equation (\ref{corr_smart}) 
to non-Gaussian cases. For example, Berry (1973), Jones (1996) and
Barreiro et al. (1998) write $p(\mu_1,\mu_2,w)$ as:
\begin{equation} 
p(\mu_1,\mu_2,w)= p(\mu_1) \delta_{\rm D}(\mu_1-\mu_2) w + p(\mu_1) p(\mu_2) (1-w)
\label{2d_pdf_smart}
\end{equation}
so that (\ref{corr_smart}) is simply given by:
\begin{equation}
1+\xi_{\nu}(\theta) = w /P_1 + (1-w).
\label{cf_wrong_eq}
\end{equation}
Expression (\ref{cf_wrong_eq}) 
implies that one can fully
characterize the clustering statistics above (below) threshold
using only the knowledge of the one-dimensional PDF
(\ref{one_d_pdf_fnl}) and the correlation.
Unfortunately, this toy model cannot be applied in our context;
equation (\ref{cf_wrong_eq}) is valid when $w$ is small,
which is not true in our case. 

Instead, since we are interested in weak non-Gaussianity, we expect a
bivariate Edgeworth expansion to provide a reasonably good
description at low thresholds:
{\setlength\arraycolsep{2pt}
\begin{small}
\begin{eqnarray}
\lefteqn{p(\mu_1,\mu_2,w) {\rm d} \mu_1 {\rm d} \mu_2 \approx  {1 \over 2
    \pi \sqrt{1-w^2}} \exp \Big \{ -{\mu_1^2+\mu_2^2 - 2 \mu_1 \mu_2 w \over 2
  (1-w^2)} \Big \} \nonumber {}} \\
& & {} \times \Big [1 + \sigma S^{(0)} \Big ( {H_{30} + H_{03}
    \over 6} \Big ) + \lambda \Big ({H_{21}+H_{12} \over 2} \Big ) \Big ] {\rm d} \mu_1 {\rm d} \mu_2
\label{2d_edge_eq}
\end{eqnarray}
\end{small}}
where $\lambda = \langle \mu_1^2 \mu_2 \rangle \equiv \langle \mu_1
\mu_2^2 \rangle$ 
and
\begin{eqnarray}
H_{30}(\mu_1,\mu_2,w) &=& H_{03} (\mu_2,\mu_1,w) \nonumber \\
&=& {(\mu_1-w  \mu_2)^3 \over (1-w^2)^3}-{3(\mu_1-w \mu_2)\over (1-w^2)^2}  
\end{eqnarray}
\begin{eqnarray}
H_{21}(\mu_1,\mu_2,w) &=& H_{12} (\mu_2,\mu_1,w)  \nonumber \\
   &=& {2w(\mu_1 -
  w \mu_2) - (\mu_2-w \mu_1) \over (1-w^2)^2}  \nonumber \\
 & & + {(\mu_2-w \mu_1)
  (\mu_1-w \mu_2)^2 \over (1-w^2)^3}. 
\end{eqnarray}
Equation (\ref{2d_edge_eq}) is the two-dimensional version
of the
distribution (\ref{one_d_pdf_fnl}) -- see
also Kotz, Balakrishnan \& Johnson (2000) and Lam \& Sheth (2009).
Note that $w$ and 
$\lambda$ must be evaluated numerically.
By inserting (\ref{one_d_pdf_fnl}) and (\ref{2d_edge_eq}) into
(\ref{corr_smart}), it is possible to characterize the clustering
strength of pixels above/below threshold for weak non-Gaussianity.  

When $f_{\rm NL}=0$ (i.e. in the Gaussian limit) equation
(\ref{2d_edge_eq}) reduces to the usual bivariate Gaussian distribution, 
since $\sigma S^{(0)} \equiv 0$ and $\lambda \equiv 0$.
Therefore (\ref{corr_smart}) reduces to the well-known formula:
{\setlength\arraycolsep{2pt}
\begin{eqnarray}
\lefteqn{1+ \xi_{\nu}(\theta) \rightarrow 1+ \xi_{\nu}^{\rm G}(\theta)
\nonumber {}} \\
& & {} \equiv {\sqrt{2/\pi} \over {\rm erfc}^2(\nu/\sqrt{2})}
\int_{\nu}^{\infty} {\rm d} \mu ~e^{-\mu^2/2} {\rm erfc} \Big [ {\nu-w
 \mu \over \sqrt{2(1-w^2)}} \Big ]
\label{corr_smart_gauss}
\end{eqnarray}}
where 
\begin{equation}
w \equiv w(\theta) = \langle \delta T_1
\delta T_2 \rangle / \sigma^2 \rightarrow
C(\theta)/C(0)
\end{equation}
with
\begin{equation}
C(\theta) = \sum_{\rm \ell} 
{(2\ell+1) \over 4 \pi} C_{\rm \ell} W_{\rm \ell}^{smooth}
P_{\rm \ell}^{0}(\cos \theta).
\label{cs} 
\end{equation}
\noindent $C_{\rm \ell}$ is the input power spectrum, and $W_{\rm \ell}^{smooth}$
is the window function which includes all the additional smoothing.  

                          
\subsection{Relation to other topological estimators} \label{history}

The excursion set statistics belongs to a more general class of geometrical estimators, which
retain information on the spatial distribution of the non-Gaussian signal. 
In this respect, it is related to many other commonly used
topological estimators.
For example, since the distribution of peaks with CMB temperatures above/below
a given threshold is a subset of the pixel distribution, there is a direct  
correspondence between the excursion sets and the peak statistics.
In presence of weak non-Gaussianity, 
it is relatively straightforward
to repeat the steps illustrated in the previous section
for the peak, rather than the pixel ensemble. In fact,
once the one- and two
dimensional non-Gaussian PDFs are known (equations \ref{one_d_pdf_fnl} and \ref{2d_edge_eq}),
one only needs to impose an extra condition in order to select local
maxima, but much of the logic remains the same.
Hence, analytic expressions for the number density and for the
clustering strength above/below threshold can be
obtained for the peak statistics
as well.
We present a more detailed investigation of the
peak clustering statistics, extended to non-Gaussian models, in a
forthcoming publication; for an exhaustive treatment of the Gaussian
case see instead Bond \& Efstathiou (1987).

Similarly, other topological or geometrical estimators which utilize
information concerning the morphology of the density structure    
are also directly related to the excursion set statistics.
This is for example the case of the 
Minkowski functionals (Schmalzing \& Gorski 1998; Winitzki \& Kosowsky 1998; 
Banday, Zaroubi \& Gorski 2000; Hikage et al. 2006, 2008b; Matsubara
2010); the number density defined
in Section \ref{excursion_set_formalism} is effectively the first Minkowski functional
(i.e. fraction of total area above the threshold), besides
some normalization factors.
The genus itself (another Minkowski functional) and its derived statistics (Coles 1988;
Gott et al. 1990; Smoot et al. 1994; Colley \& Gott 2003; Park 2004;
Gott et al. 2007) are also directly related to the excursion
sets formalism. This is because
the genus, being the number of isolated hot spots minus the
number of isolated cold spots, can be obtained from the
contours for a given threshold temperature and can be parametrized
by the area fraction above the threshold -- which is given by equation
(\ref{nd_fnl_eq}) for the
pixel ensemble, in the weak non-Gaussian limit. 



\section{Constraining non-Gaussianity with the excursion set statistics}  \label{NG maps analysis}

\begin{figure*}
\begin{center}
\includegraphics[angle=0,width=0.29\textwidth]{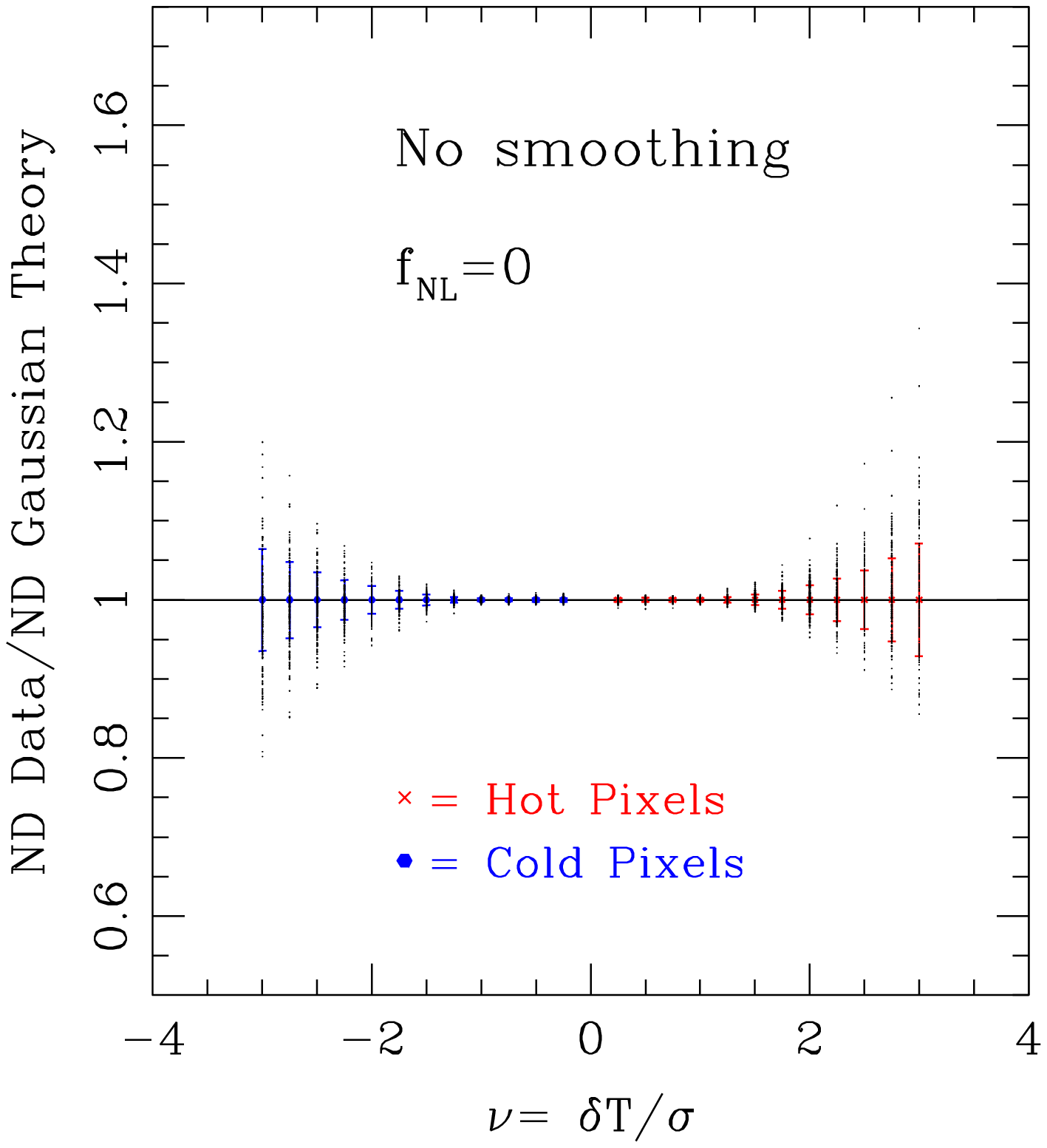}
\includegraphics[angle=0,width=0.29\textwidth]{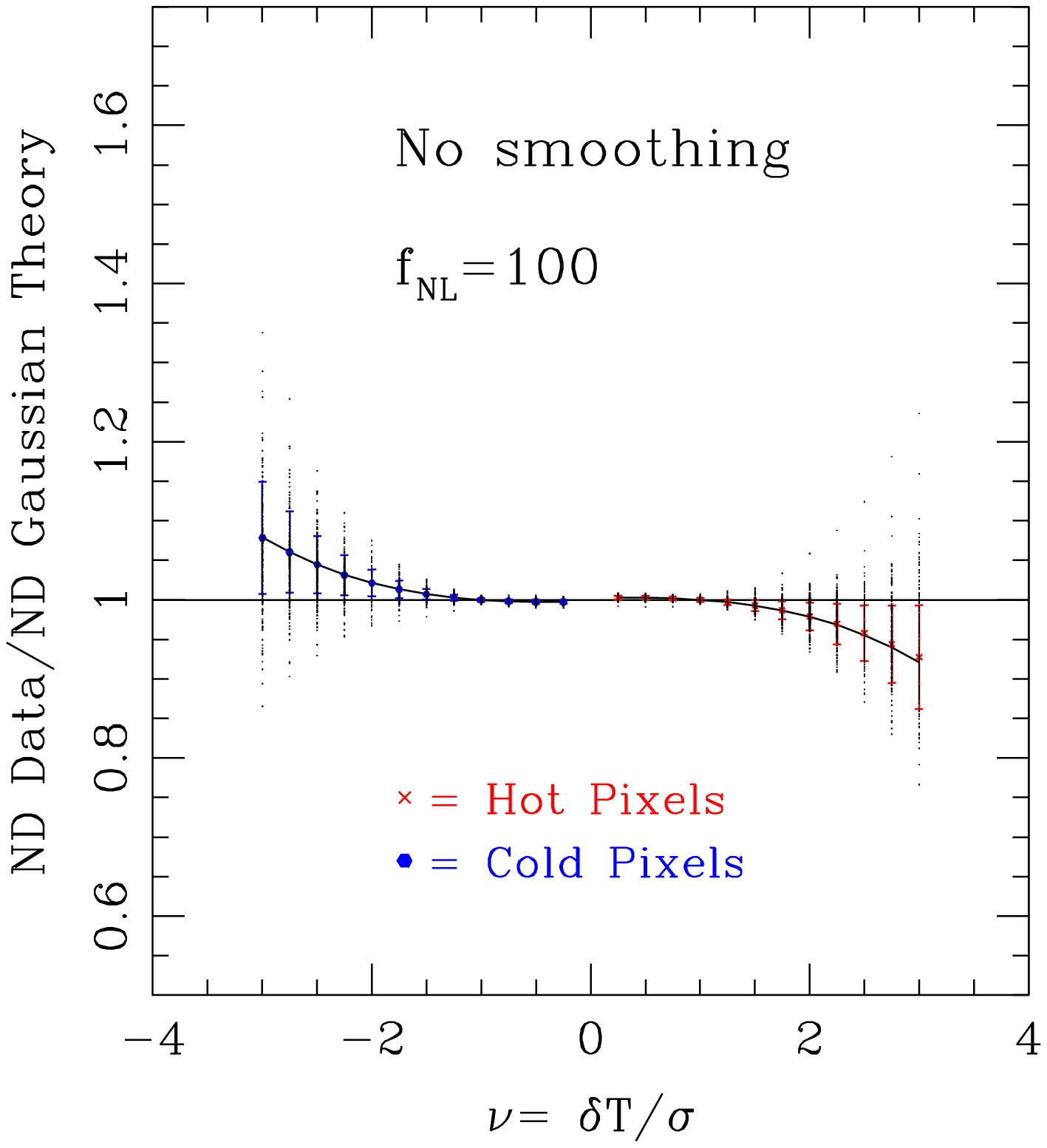}
\includegraphics[angle=0,width=0.29\textwidth]{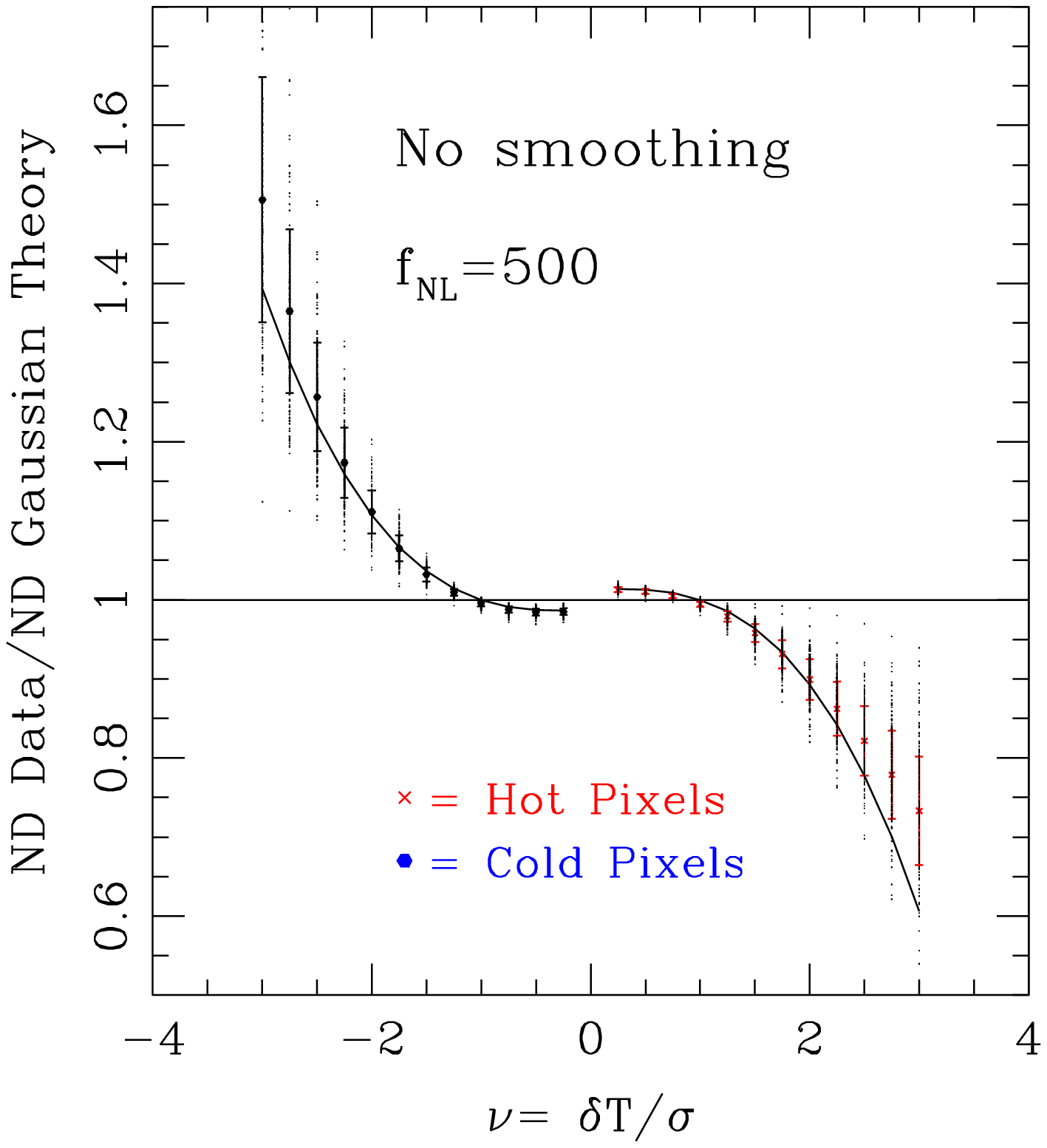}
\includegraphics[angle=0,width=0.29\textwidth]{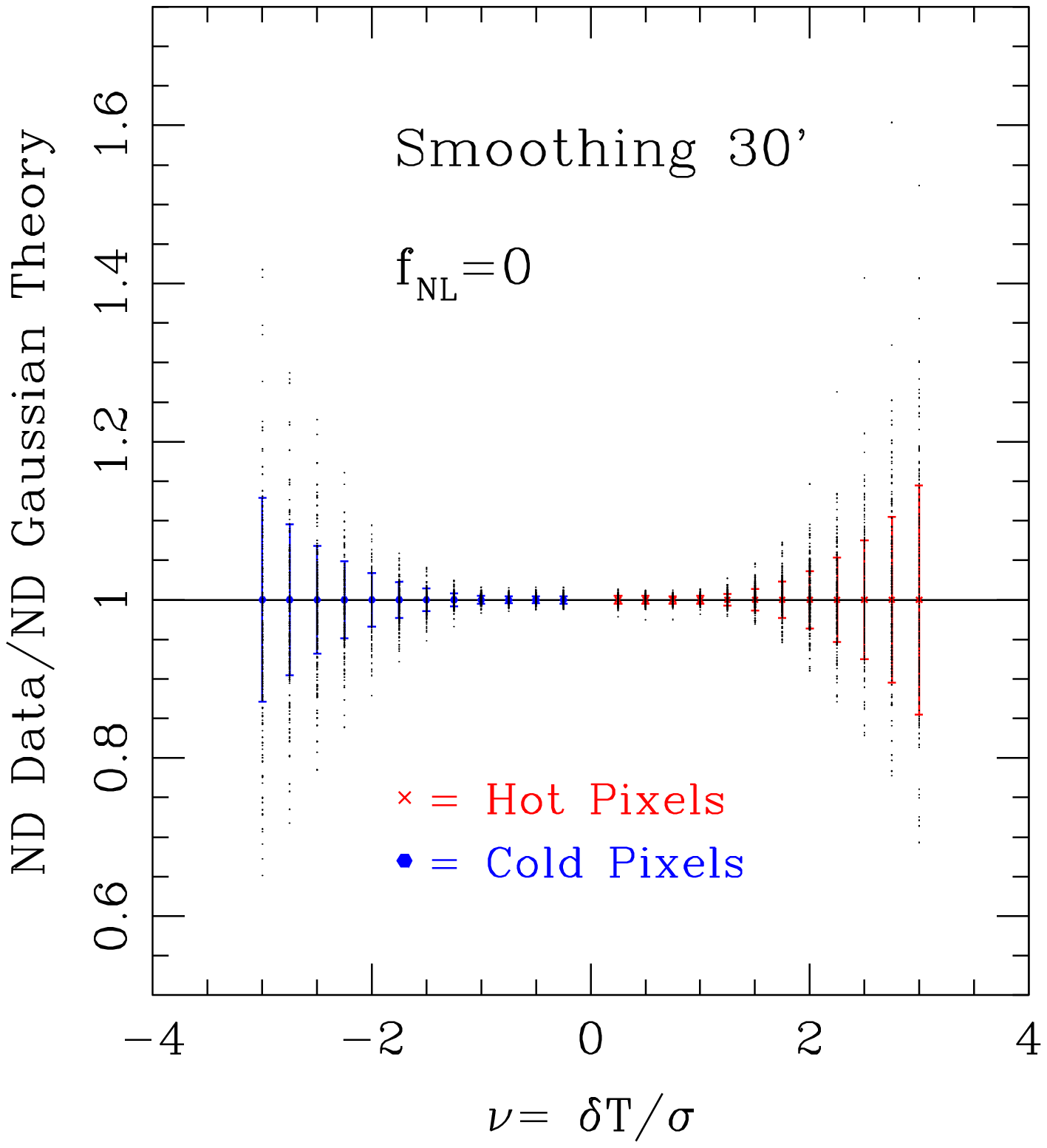}
\includegraphics[angle=0,width=0.29\textwidth]{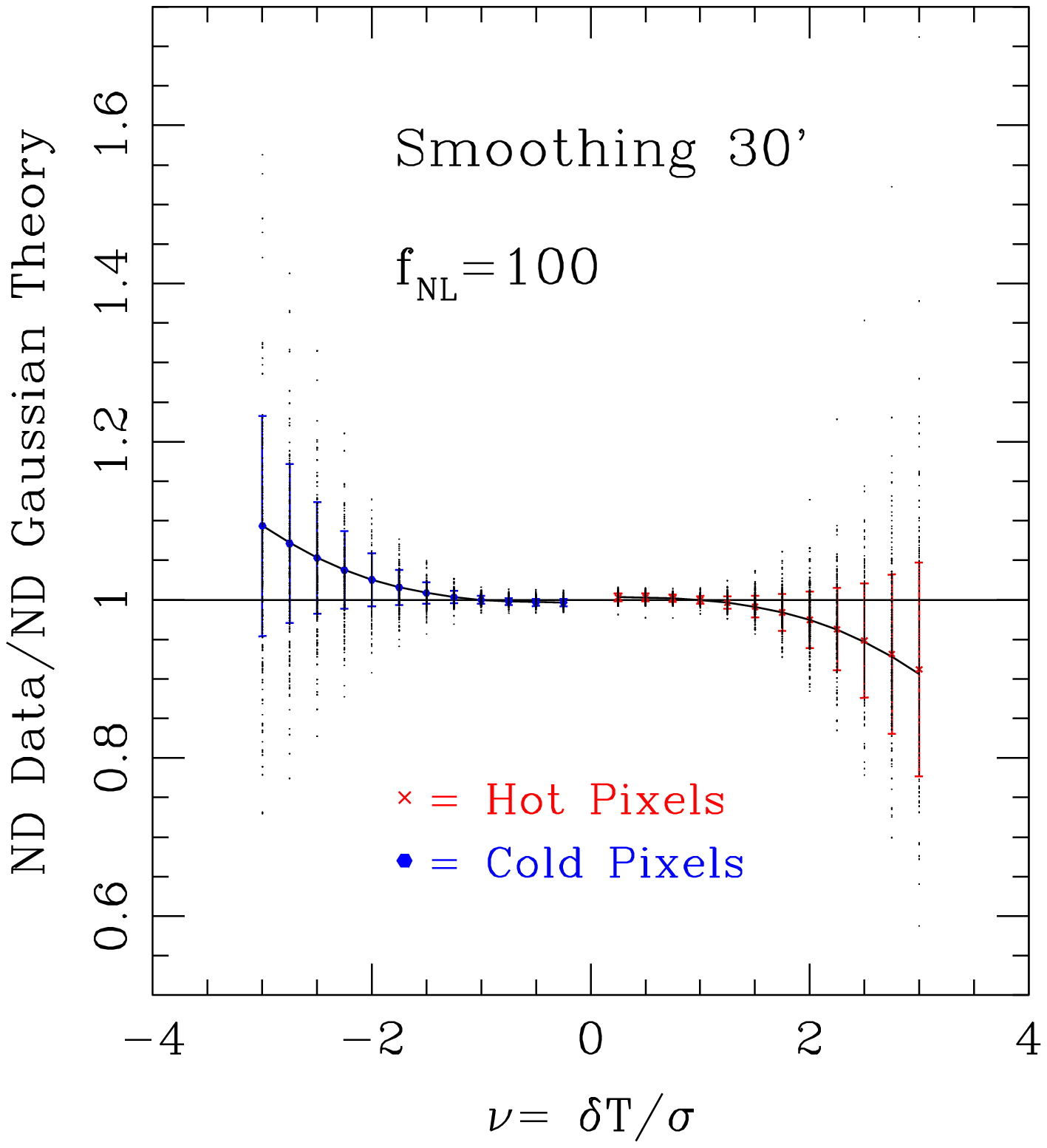}
\includegraphics[angle=0,width=0.29\textwidth]{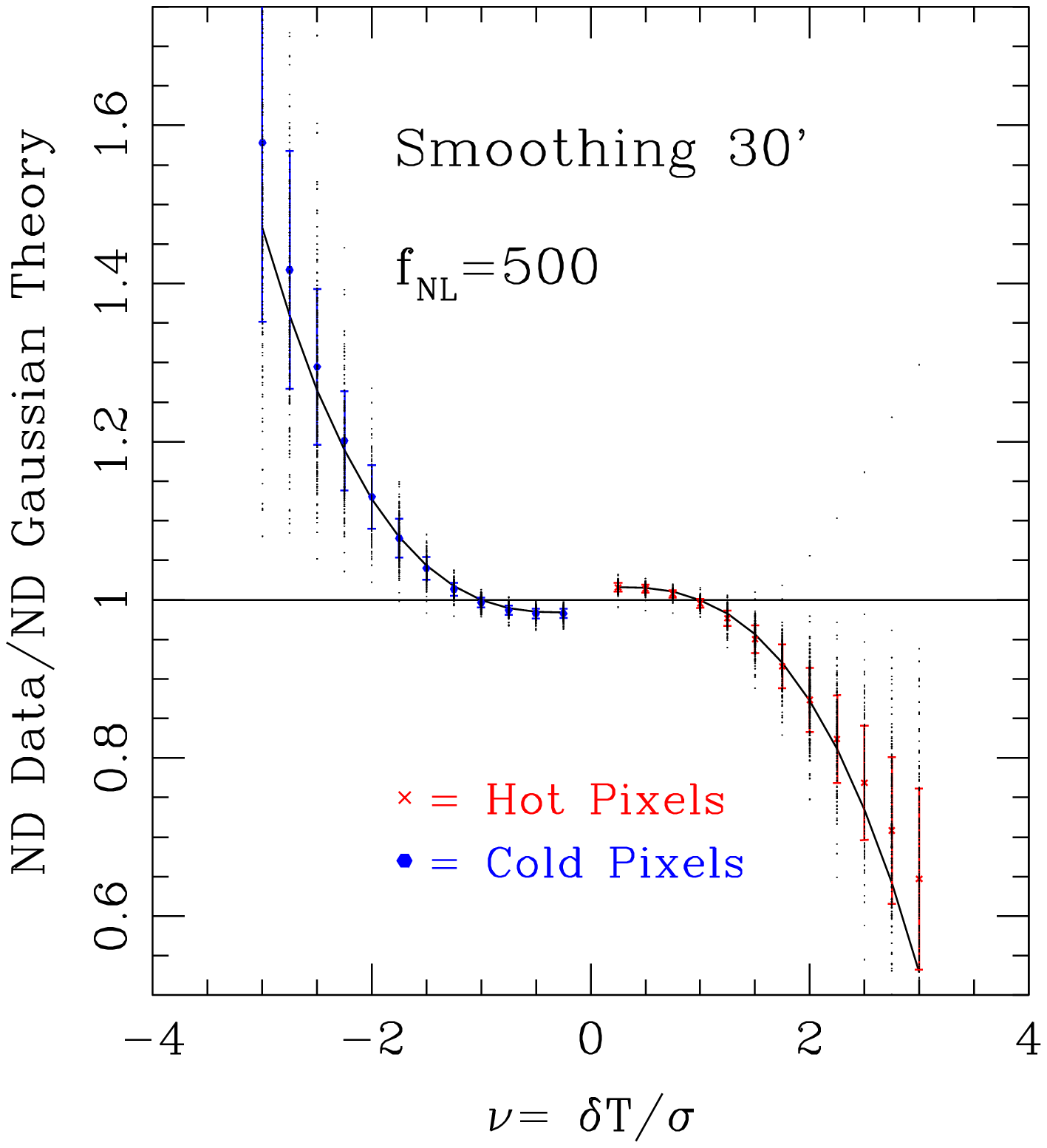}
\includegraphics[angle=0,width=0.29\textwidth]{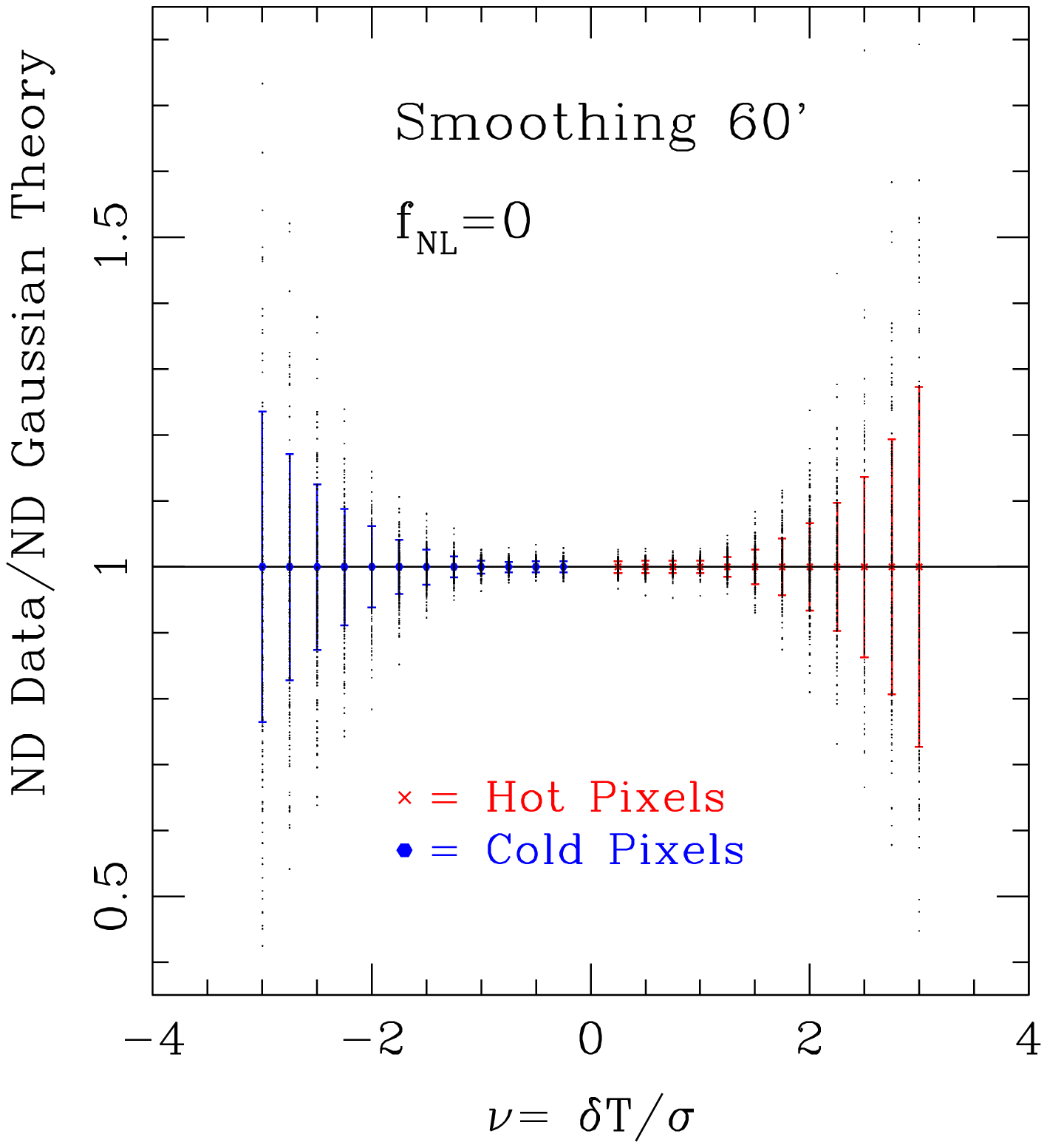}
\includegraphics[angle=0,width=0.29\textwidth]{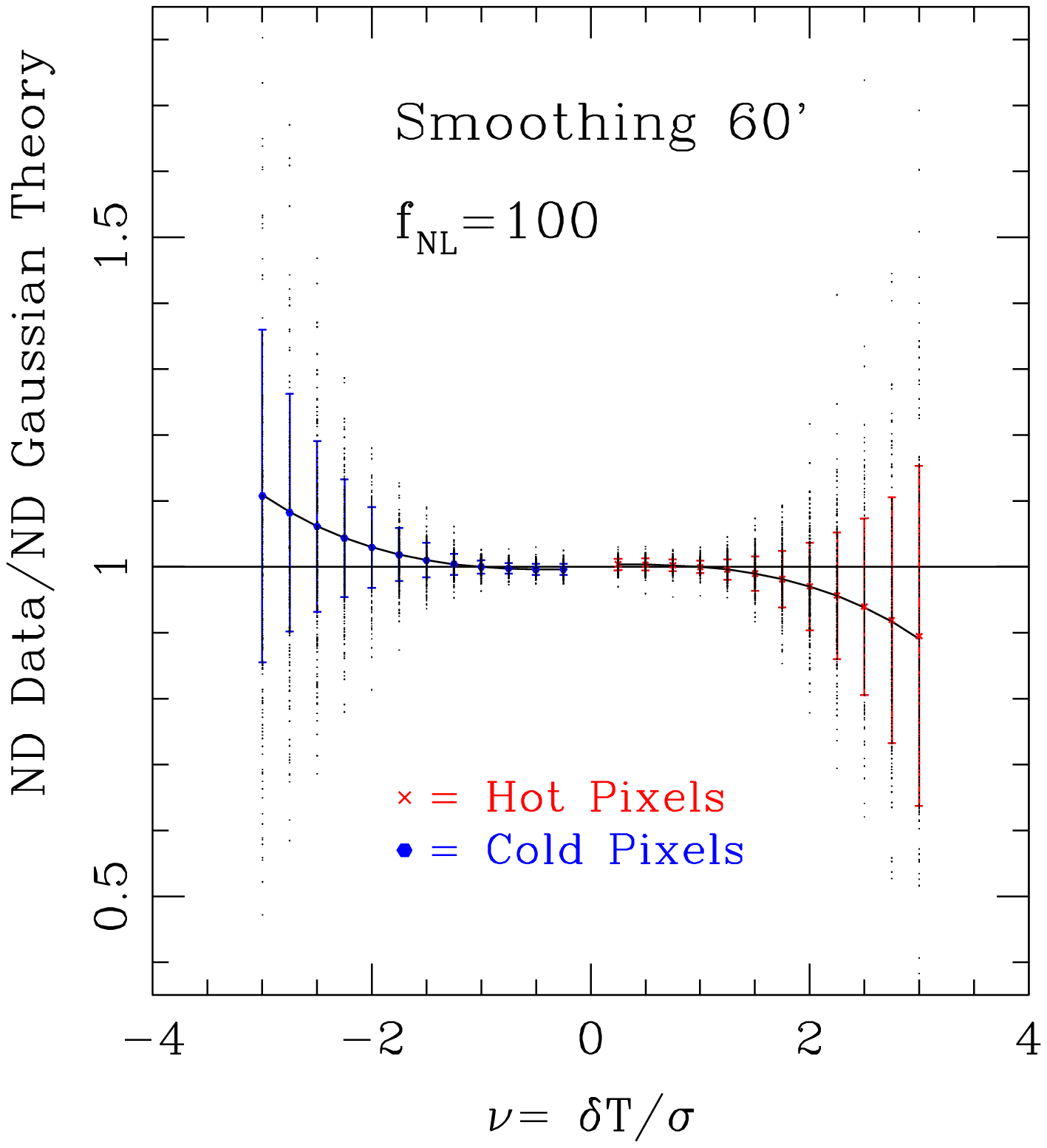}
\includegraphics[angle=0,width=0.29\textwidth]{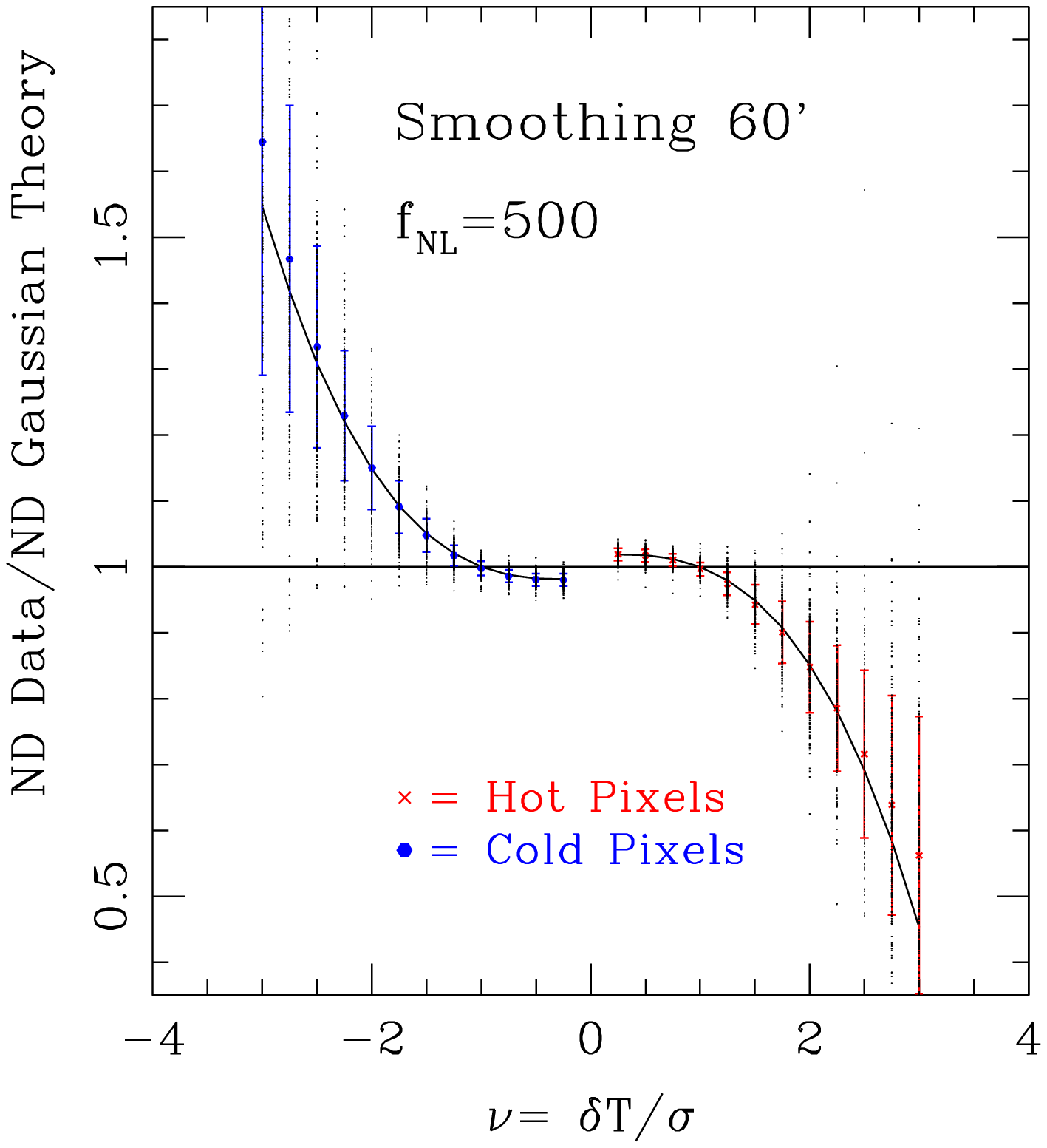}
\caption{Number density of pixels in a Gaussian case (left panels) or
  in $f_{\rm NL}$ models (in the middle panels $f_{\rm NL}=100$; in
  the right ones $f_{\rm NL}=500$). A Gaussian smoothing with
  FWHM=30' and 60' is applied in the central and lower panels, respectively,
  at $N_{\rm side}=512$. Errorbars are the
  1$\sigma$ run-to-run estimates from 200 maps. Solid curves
  are theory predictions from equations (\ref{nd_gauss_eq}), (\ref{nd_fnl_eq}) and (\ref{nd_fnl_detail_eq}).}
\label{nd_smart_scale_fig}
\end{center}
\end{figure*}

In this section we present numerical results for the number density
and for the clustering strength of pixels
above/below threshold, calculated from non-Gaussian simulations.
We also perform a thorough statistical analysis to evaluate the
sensitivity of the two observables to the level of non-Gaussianity
and to the smoothing resolution.
Our calculations are averaged over 200
full-sky CMB realizations, and the associated errorbars are the 1$\sigma$ run-to-run
estimates. The effects of noise and of other experimental artifacts,
as well as confusion effects due to secondary non-Gaussianities,
are not addressed in this analysis; rather, in this first
work our main goal is to characterize the intrinsic non-Gaussian CMB
signal within the excursion set statistics. However, in the Appendices \ref{noise_analytic}, \ref{errors_analytic}
and \ref{spurious_ng}, we briefly explain 
how to include these complications in our theoretical framework. 


\subsection{Number density of the excursion set pixels} \label{nd_subsection}

\begin{figure}
\begin{center}
\includegraphics[angle=0,width=0.50\textwidth]{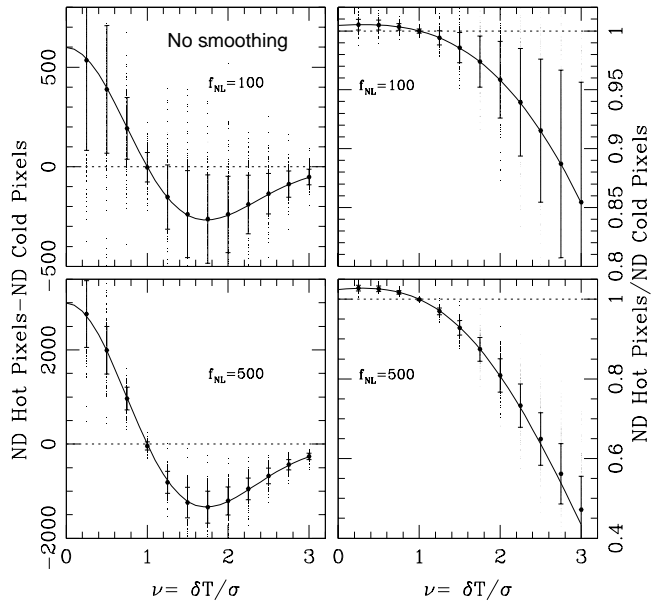}
\caption{Difference (left) and ratio (right) between hot
  and cold excursion set regions, at corresponding temperature
  thresholds, when no smoothing is applied. In the top panels $f_{\rm
  NL}=100$, in the bottom ones $f_{\rm NL}=500$. Solid lines are
  theoretical expectations derived from equations (\ref{nd_fnl_eq}) and (\ref{nd_fnl_detail_eq}).  
  At $\nu=1$, a transition area in the number density is clearly visible.} 
\label{nd_difference_ratio_fig}
\end{center}
\end{figure}
\begin{figure*}
\begin{center}
\includegraphics[angle=0,width=0.29\textwidth]{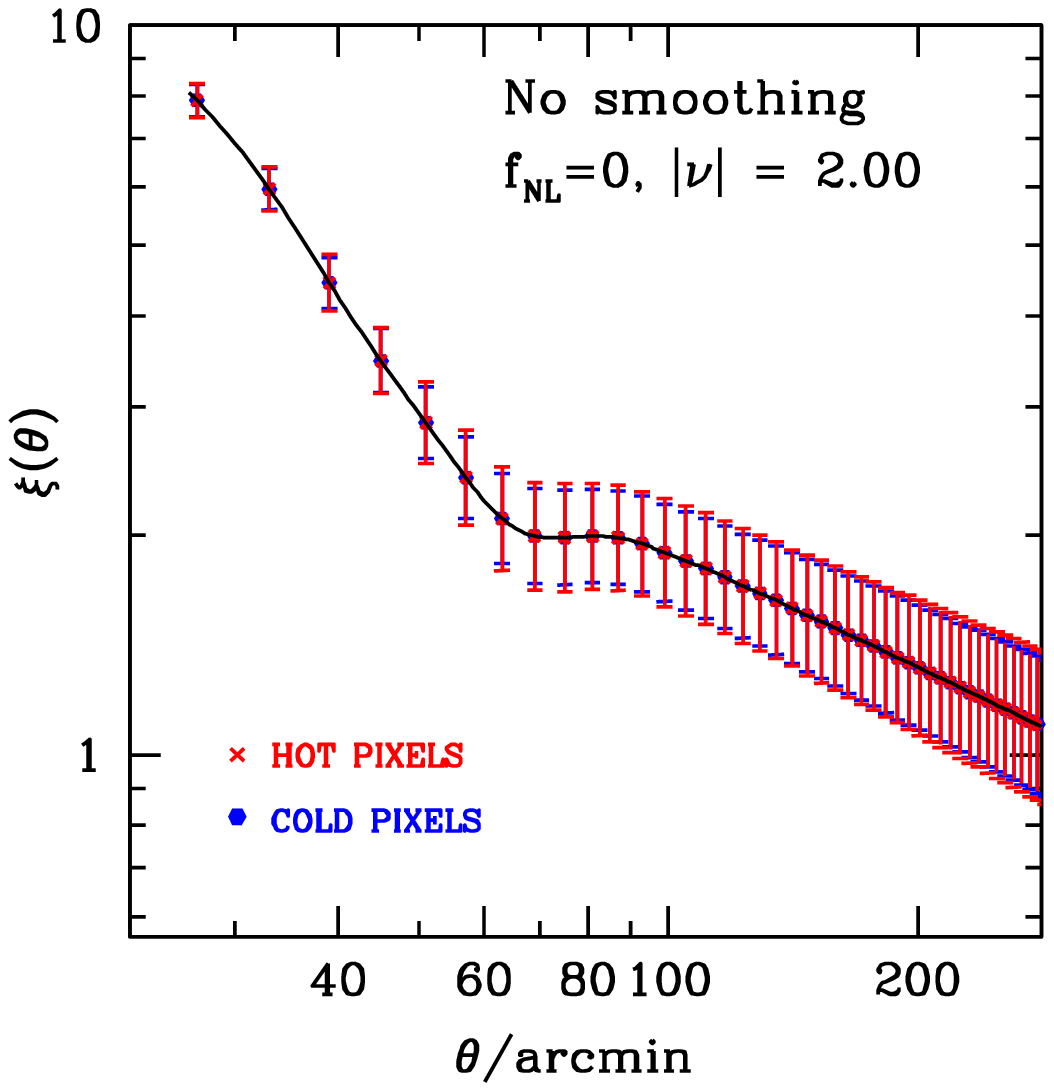}
\includegraphics[angle=0,width=0.29\textwidth]{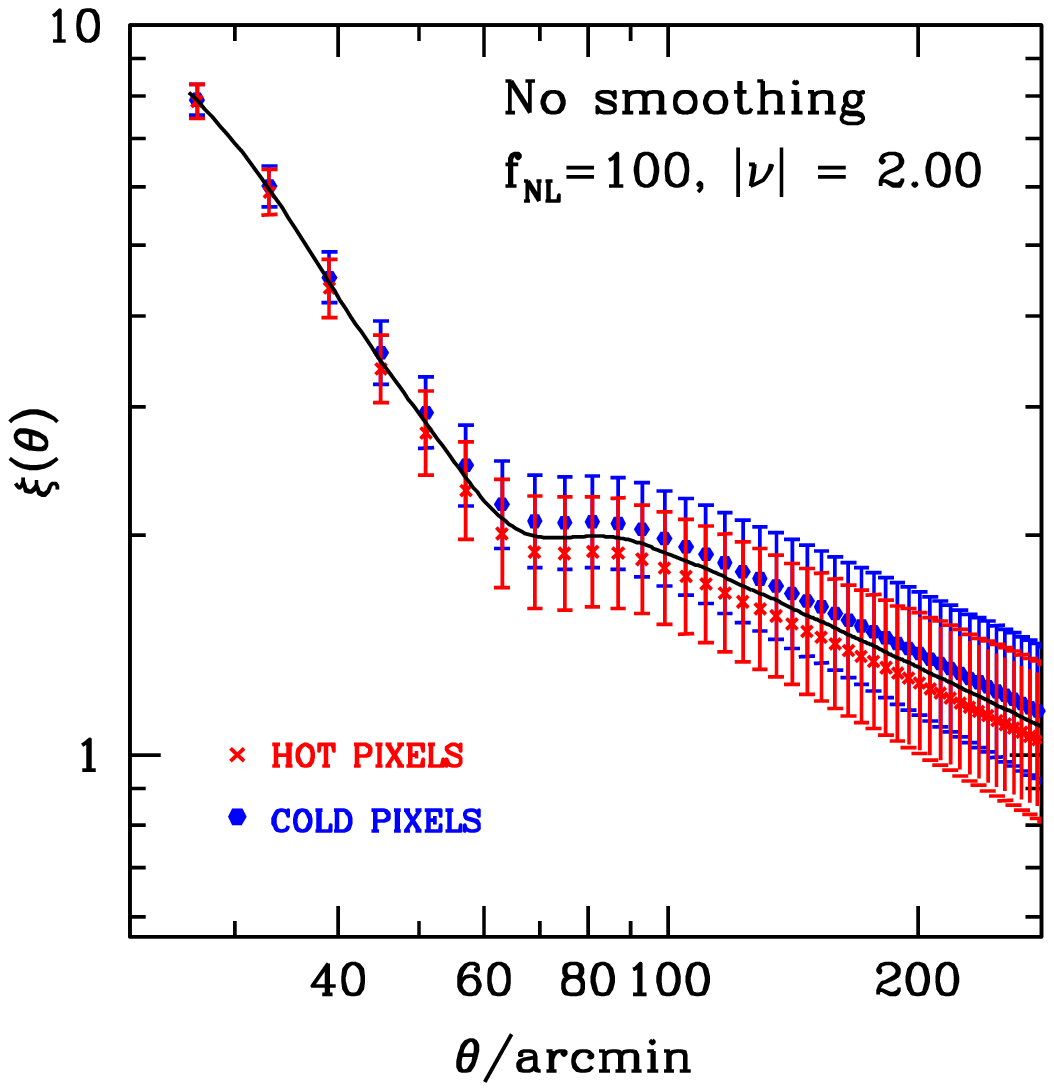}
\includegraphics[angle=0,width=0.29\textwidth]{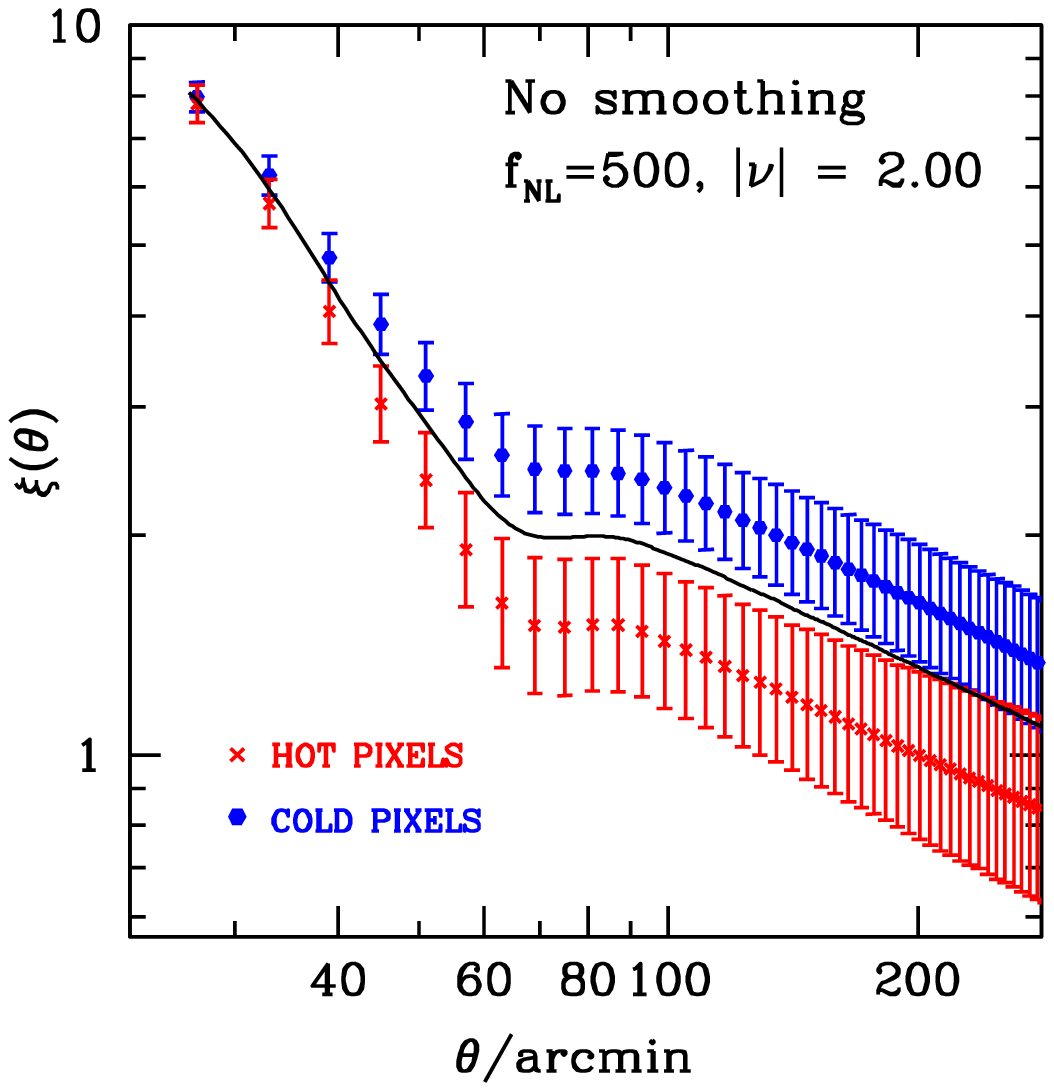}
\includegraphics[angle=0,width=0.29\textwidth]{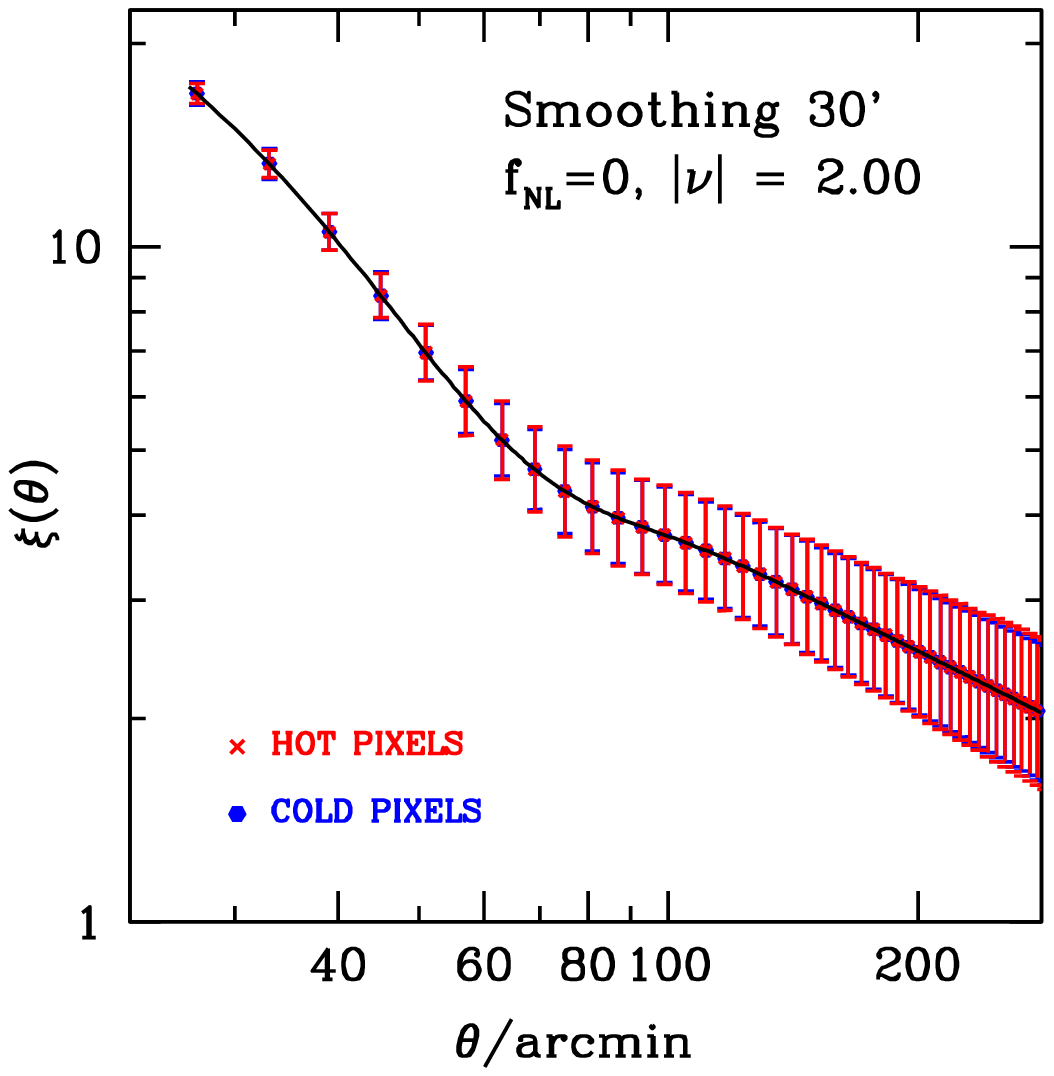}
\includegraphics[angle=0,width=0.29\textwidth]{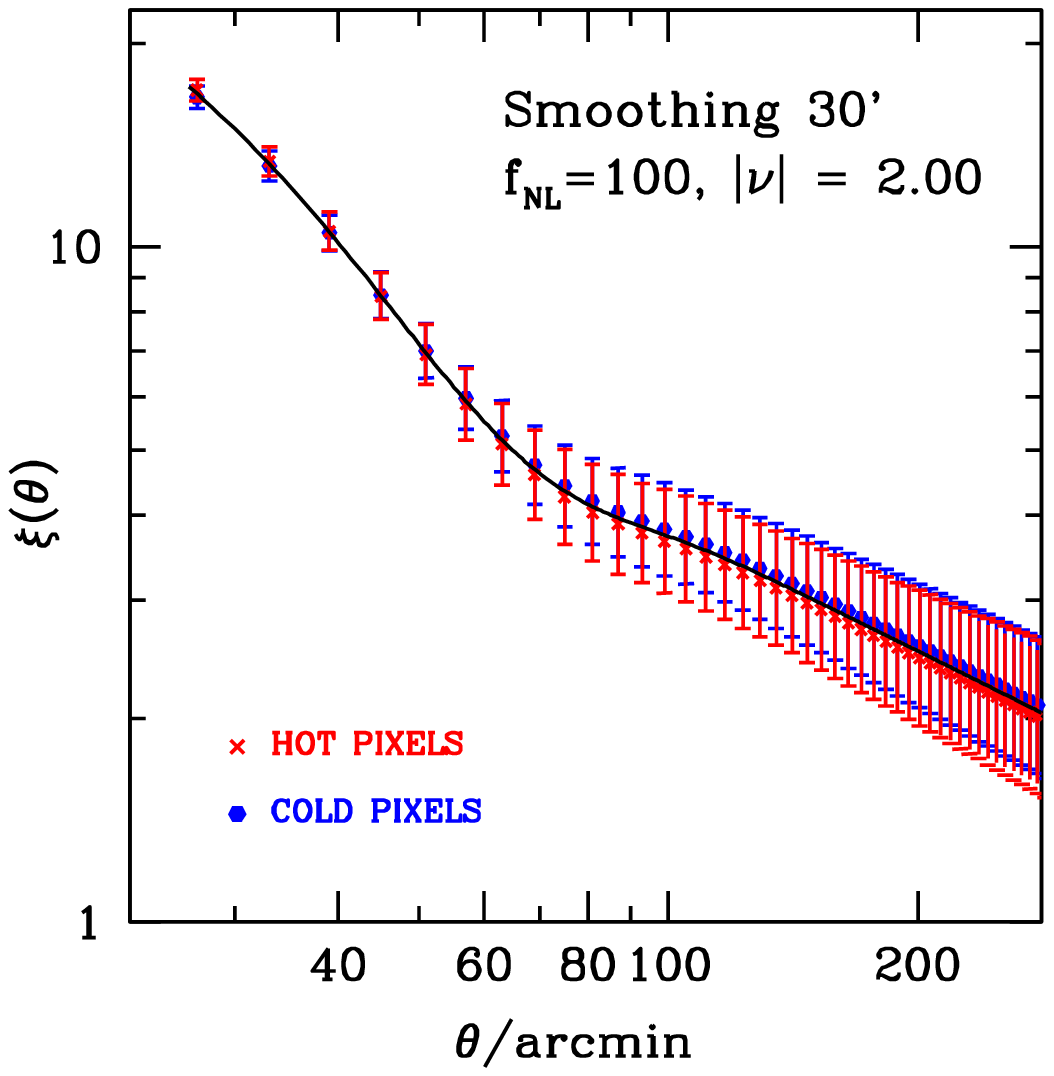}
\includegraphics[angle=0,width=0.29\textwidth]{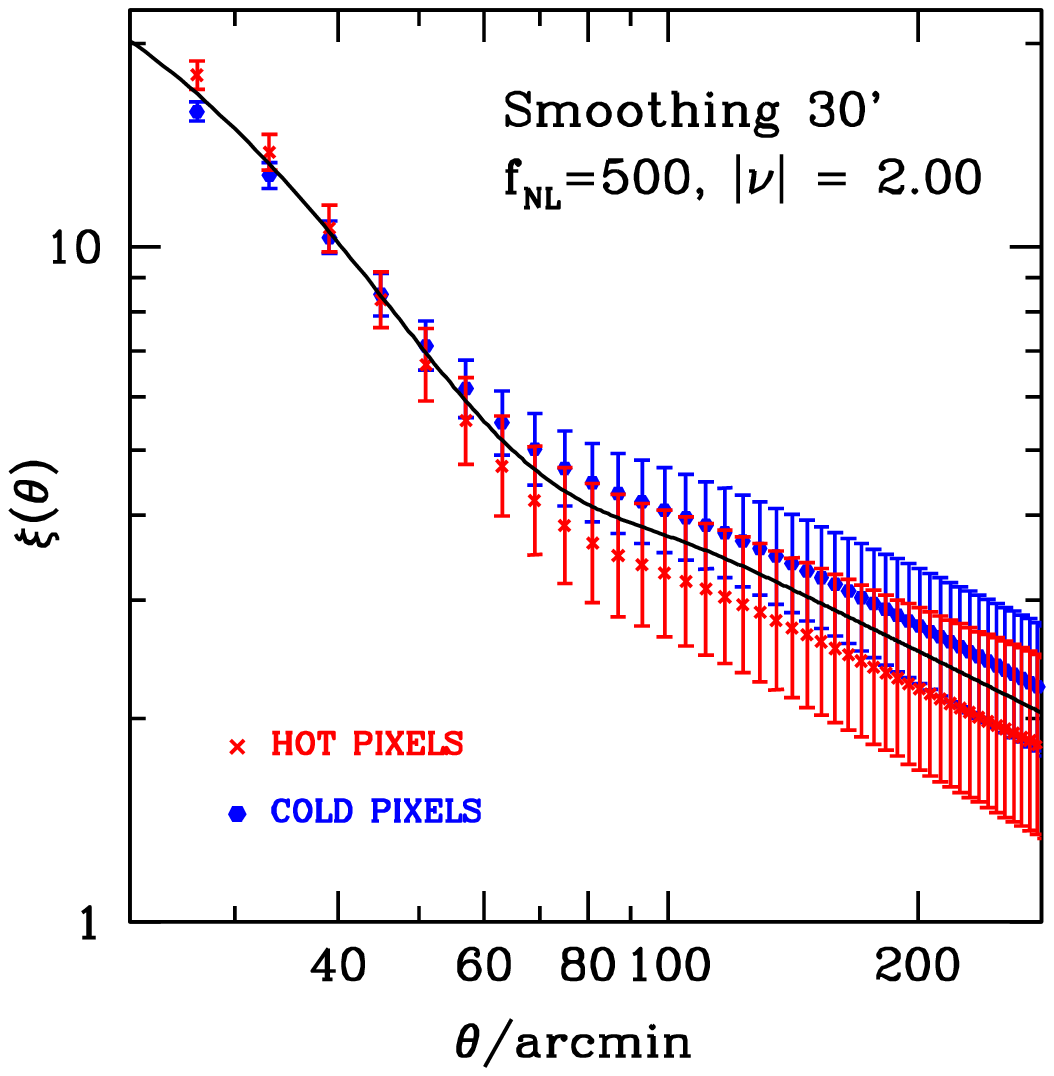}
\includegraphics[angle=0,width=0.29\textwidth]{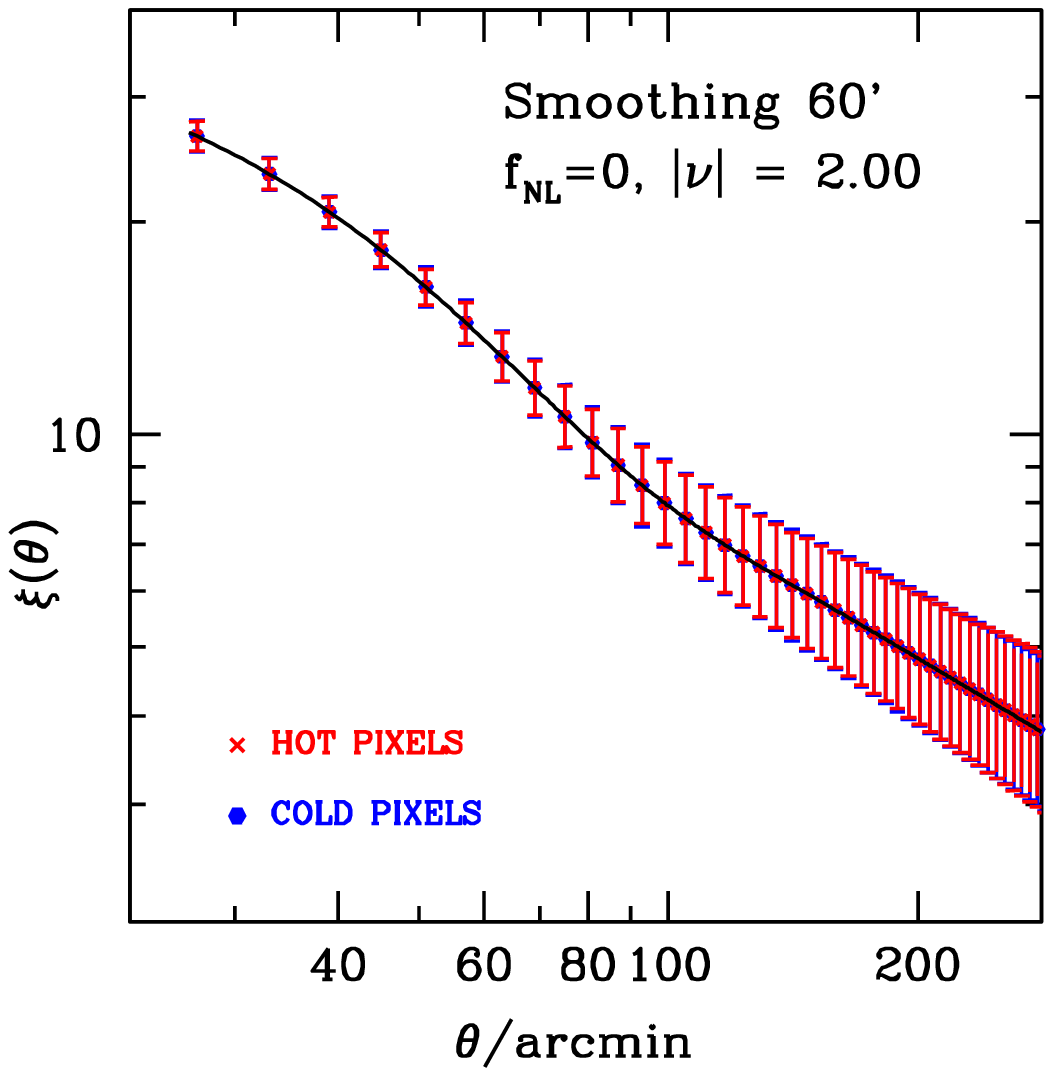}
\includegraphics[angle=0,width=0.29\textwidth]{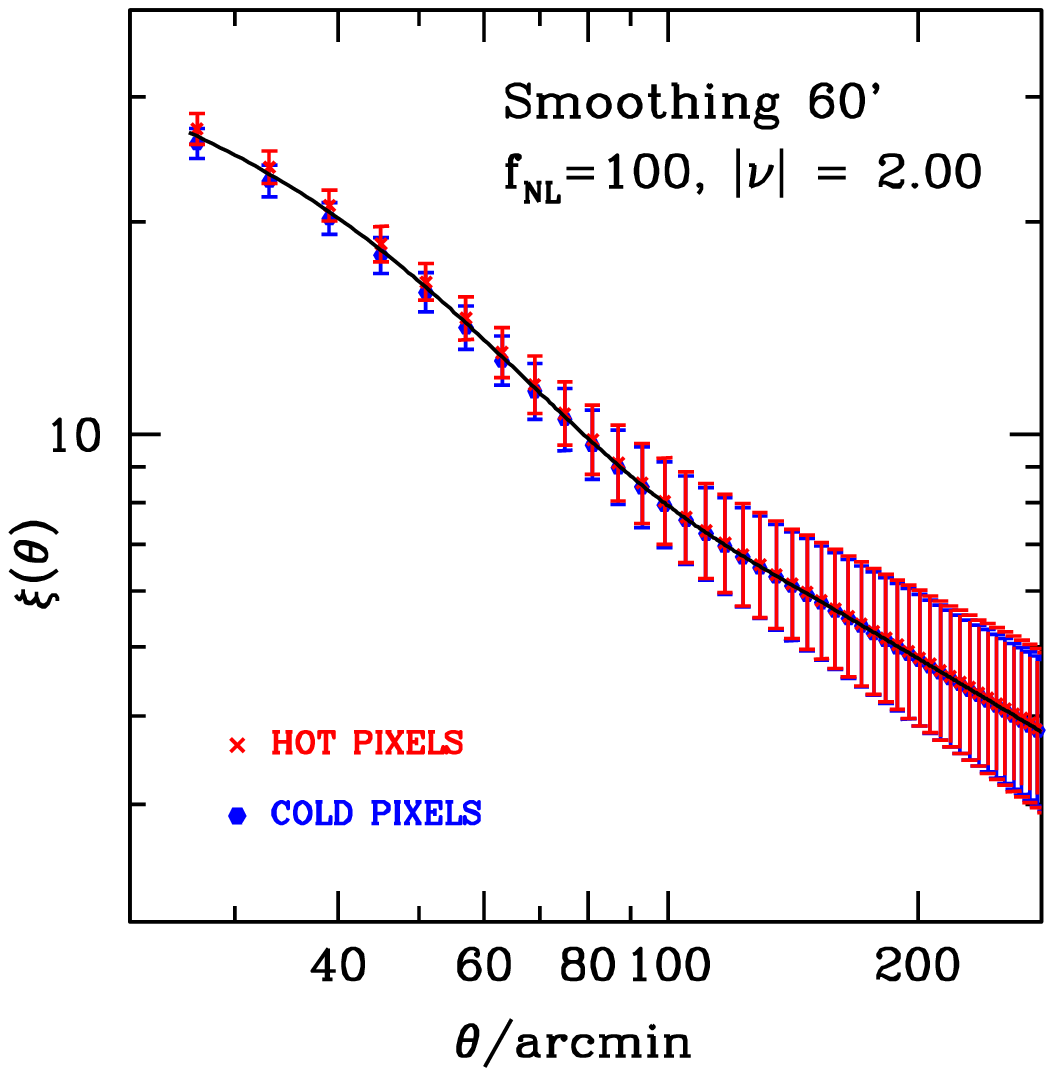}
\includegraphics[angle=0,width=0.29\textwidth]{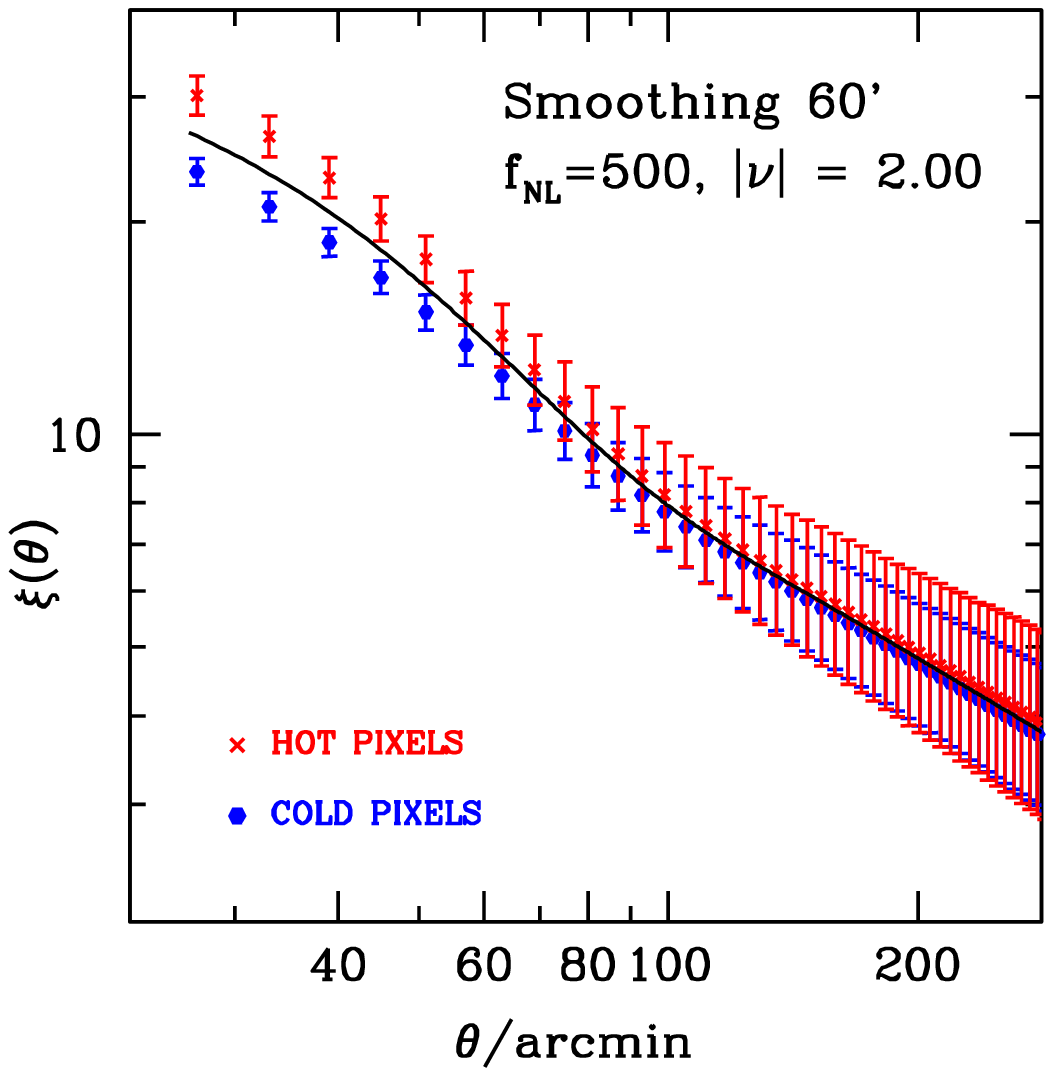}
\caption{Clustering strength of pixels above/below threshold when
  $|\nu|=2.00$, a regime particularly sensitive to a non-Gaussian
  signal of the local  $f_{\rm NL}$ type.  Left panels represent the Gaussian case, in the middle
  panels $f_{\rm NL}=100$, in the right panels $f_{\rm NL}=500$.
  A Gaussian smoothing with FWHM of 30 and 60 arcmin is applied in the
  middle and bottom panels, respectively. Solid lines are analytic
  predictions in the Gaussian limit, from equation (\ref{corr_smart_gauss}).}
\label{clustering_fig}
\end{center}
\end{figure*}

Figure \ref{nd_smart_scale_fig} shows the variation with $f_{\rm NL}$ of the pixel number density
above (below) threshold, normalized by the
expectation from a Gaussian theory (equation \ref{nd_gauss_eq}). No smoothing is
applied in all the top panels, while a Gaussian smoothing with FWHM
of 30 and 60 arcmin is
applied in the intermediate and bottom panels, respectively, at a resolution $N_{\rm side}=512$.
Solid curves are analytic
predictions for $f_{\rm NL}$ type non-Gaussianity from equations
(\ref{nd_gauss_eq}-\ref{nd_fnl_detail_eq}); they are in very good agreement
with our numerical results.
Note that due to a relatively small
number of maps considered, in practice at higher thresholds the Gaussian mean undergoes a small
shift because of statistical fluctuations. We have accounted for this
effect in our calculations, and shifted all the corresponding non-Gaussian means by the same
amount, as done in Chingangbam \& Park (2009). This is the reason why
in the figure and in the following ones
(\ref{nd_difference_ratio_fig}-\ref{clustering_sigma_units_fig}) 
all the Gaussian expectations lie \textit{exactly} on a line.
Clearly, this procedure does not affect the relative distance
between Gaussian and non-Gaussian means, the quantity we want to
characterize here; hence, our results are independent of this small rescaling. 

A number of interesting features can
be inferred from Figure \ref{nd_smart_scale_fig}. First, the existence of two
regions where the non-Gaussian contribution appears to be more significant,
namely at relatively low thresholds ($\nu=0.25,
0.50$) or around $\nu=2.00$. Second, the fact (never pointed out so far
in the literature) that \textit{there are} optimal
thresholds which maximize the local non-Gaussianity, as well as others which
do not allow for a distinction between the Gaussian and the
non-Gaussian case.
This is expected from equations
(\ref{nd_fnl_eq}) and (\ref{nd_fnl_detail_eq});
in particular, when $\nu=1$ then
$n_{\rm pix}^{\rm f_{NL}} \equiv 0$ and 
$n_{\rm pix}^{\rm NG}  \equiv n_{\rm pix}^{\rm G}$. Therefore, in this
statistical framework 
levels around $\nu=1$ are not sensitive to departures
from Gaussianity of the $f_{\rm NL}$ local type.
Third, at higher thresholds a positive
$f_{\rm NL}$ causes
an enhancement of the number density of the cold pixels and reduces
that of the hot ones, while the opposite trend
happens when $\nu<1$. This effect is more evident for larger
$f_{\rm NL}$ values. We will use these findings to devise a new
optimal statistical test  
in the next subsection.  
An additional Gaussian smoothing 
increases the errorbars in the number density calculations, and 
slightly reduces the effect just described. 

Figure \ref{nd_difference_ratio_fig} displays the difference (left
panels) and the ratio (right panels)
between the number density of hot and cold
pixels, at corresponding temperature thresholds, when no smoothing is applied.
Top panels highlight the case of $f_{\rm NL}=100$,  bottom panels are for
$f_{\rm NL}=500$. Solid curves
show the analytic predictions, which are easily derived from equations (\ref{nd_fnl_eq}) and (\ref{nd_fnl_detail_eq}).
Again, we find a very good agreement between numerical results and
analytical expectations.
At $\nu=1$, a `transition area' in 
the number density is clearly visible, 
particularly when we consider the difference between hot and cold excursion set
regions.

\subsection{Statistical test derived from the number density} \label{nd_stat_subsection}


\begin{figure*}
\begin{center}
\includegraphics[angle=0,width=0.31\textwidth]{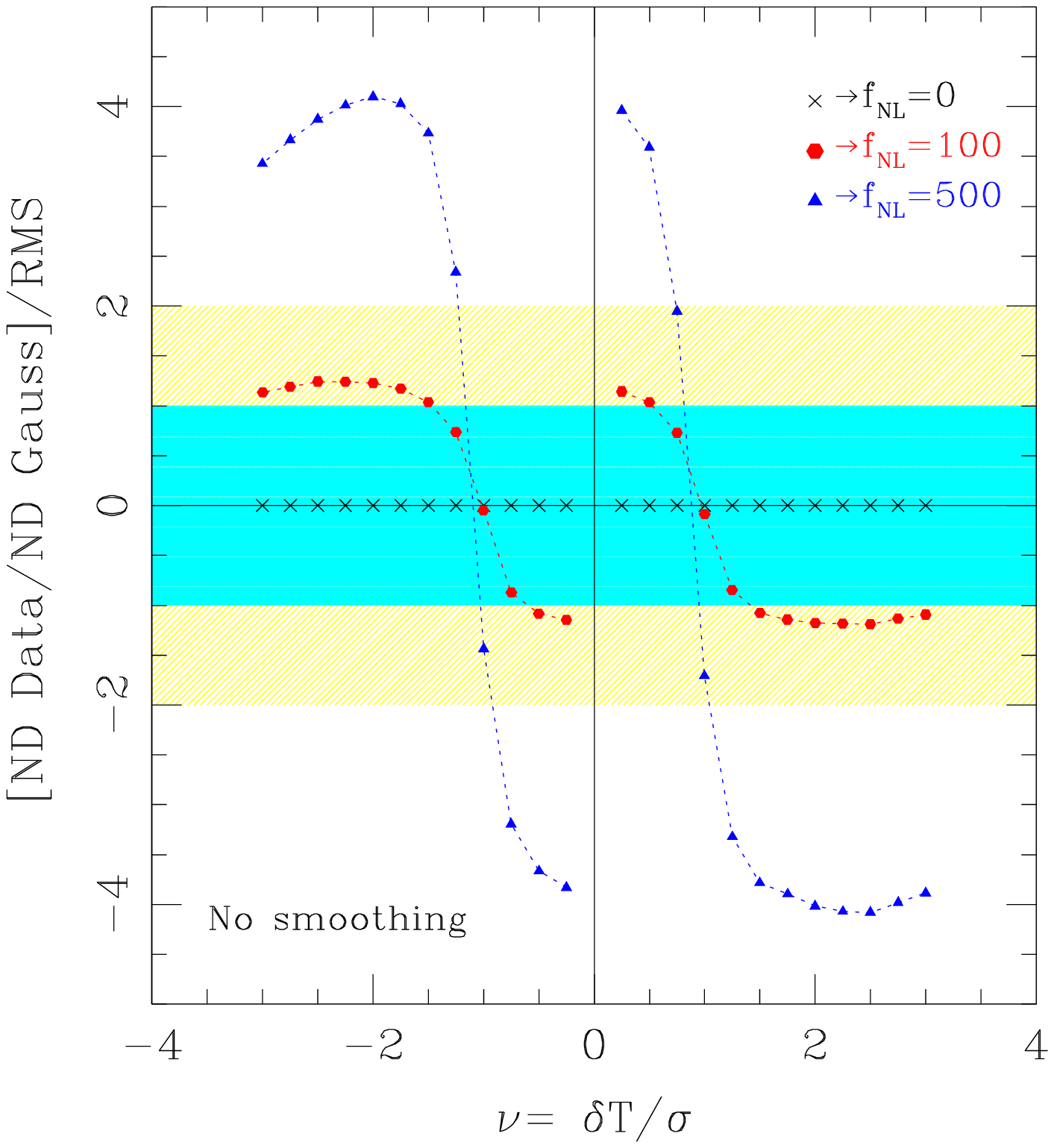}
\includegraphics[angle=0,width=0.32\textwidth]{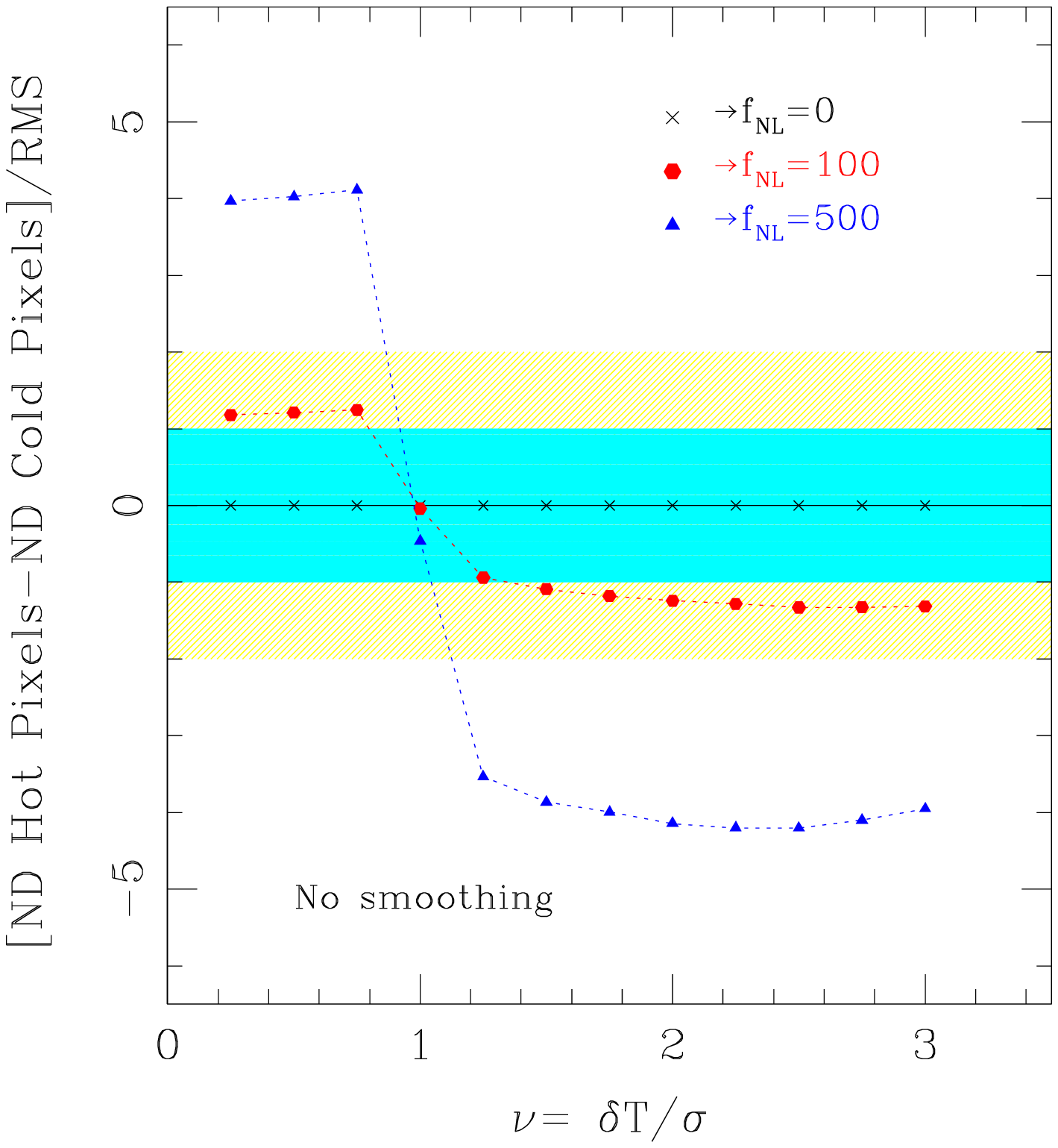}
\includegraphics[angle=0,width=0.32\textwidth]{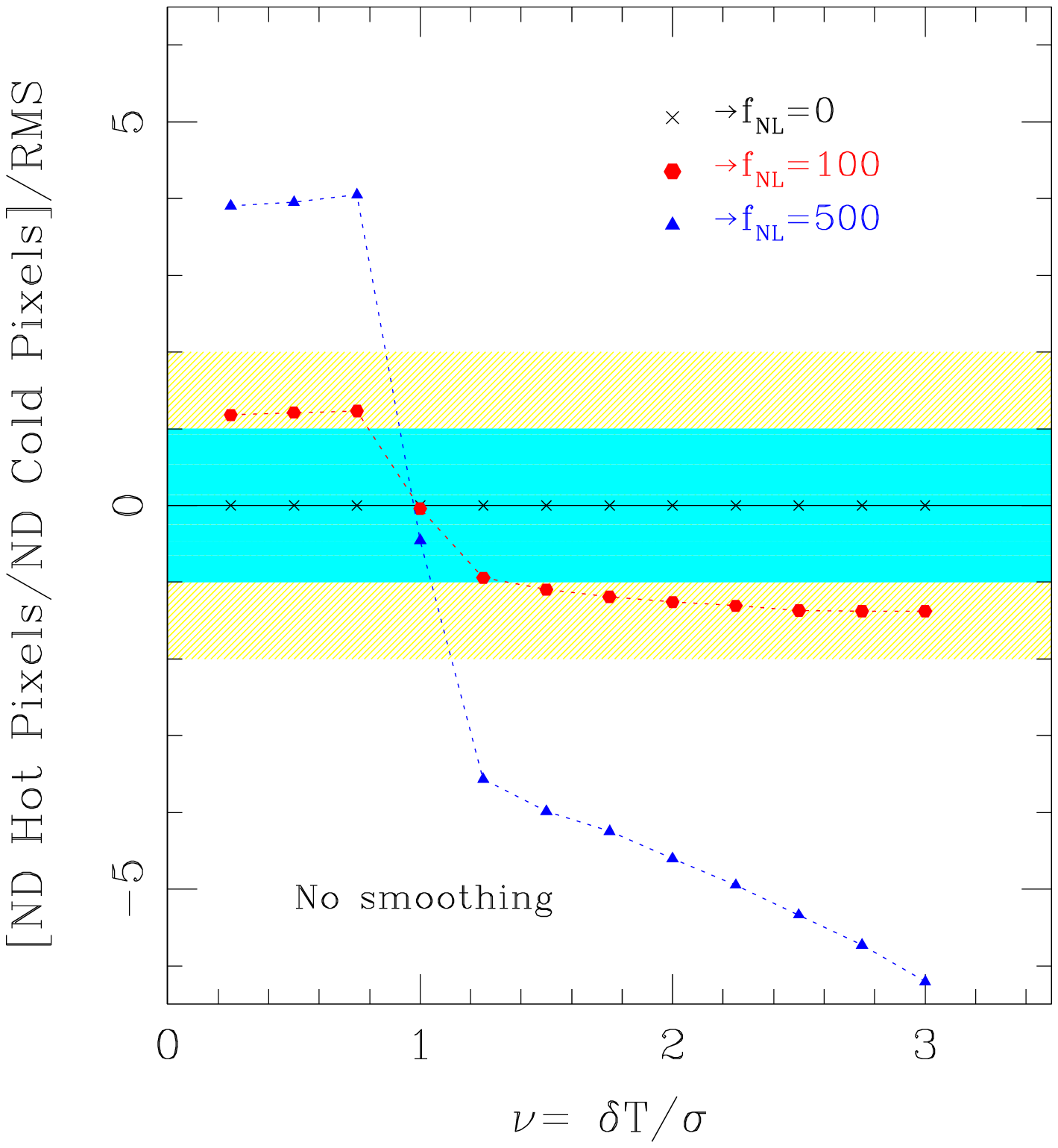}
\caption{Reinterpretation of Figure \ref{nd_smart_scale_fig} (left panel) and Figure \ref{nd_difference_ratio_fig} (middle and right panels) 
in terms of the run-to-run associated errors, in order to quantify the
sensitivity of the number density to local non-Gaussianity. 
Different values of $f_{\rm NL}$ are displayed, as indicated in the
plots, when no smoothing is applied. 
Departures from Gaussianity are maximized around $\nu=0.25,0.50$ or around
$\nu=2.00, 2.25$ while
areas close to $\nu=1.00$ are insensitive to non-Gaussianity of the local type.}
\label{nd_sigma_distances_fig}
\end{center}
\end{figure*}

\begin{figure*}
\begin{center}
\includegraphics[angle=0,width=0.85\textwidth]{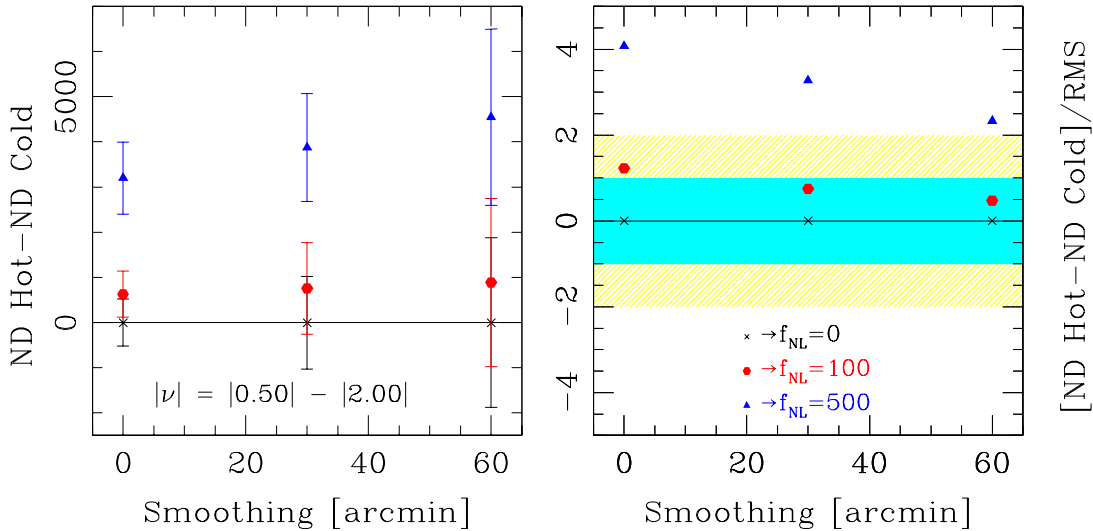}
\caption{Composite quantity for the pixel number density as a
  function of the smoothing scale,
  defined by equation (\ref{nd_composite_eq}) in the main text. The left panel is in real units, with
  errorbars estimated from 200 realizations; the right panel shows
  similar quantities but in RMS units as in Figure \ref{nd_sigma_distances_fig}. 
  The two areas where the non-Gaussian sensitivity
  is maximized (i.e. $|\nu|=0.50$ and $|\nu|=2.00$) are combined, in
  order to boost the departure from Gaussianity.}
\label{nd_combined_fig}
\end{center}
\end{figure*}

The conclusions drawn from
Figures \ref{nd_smart_scale_fig} and \ref{nd_difference_ratio_fig}
can be expressed in a more quantitative form as follows.
If we assume $n_{\rm pix}^{\rm NG}$ to be the possible non-Gaussian
discriminator, we can
plot the number density measurements 
in terms of their errorbars.
In other words, we can normalize all the
points in Figures \ref{nd_smart_scale_fig} and
\ref{nd_difference_ratio_fig} by their run-to-run associated errors, and
quantify their `distance' from the expected Gaussian predictions. 
This is quite convenient, as it allows one to realize
which thresholds are particularly sensitive to a local non-Gaussian
signal, and to determine
the exact values of $\nu$ which maximize departures from Gaussianity.

In Figure \ref{nd_sigma_distances_fig} we
show the case when no smoothing is applied and
reinterpret in this context the number density per Gaussian units (left panel),
the difference (middle panel) and the ratio (right panel) between the abundance of
hot and cold pixels. 
Shaded areas represent the
1 and 2$\sigma$ errors, while different symbols are
used for different values of $f_{\rm NL}$, as specified in the plots.
When $\nu=0.25,0.50$ or $\nu=2.00, 2.25$ departures
from Gaussianity are maximized: they exceed the 1$\sigma$ level for
$f_{\rm NL}=100$. Instead, the transition area
around $\nu=1$ is insensitive to a non-Gaussian signal of the $f_{\rm NL}$
type. At higher thresholds, departures are
more significant with increasing $f_{\rm NL}$; however,
severe pixel-noise and poor statistics (too few excursion sets)
prevent them from being reliable non-Gaussian indicators.

Although the sensitivity of the first skewness parameter $S^{(0)}$ to $f_{\rm
 NL}$, and so of the number density itself,
is much worse than that of the angular bispectrum (Komatsu \& Spergel
2001), the previous findings suggest that we could construct a 
derived quantity which amplifies the $f_{\rm NL}$ contribution. 
This is achieved by combining
two thresholds, where departures from Gaussianity are most
significant. Namely,  
\begin{equation}
n_{\rm hc}^{\rm NG} = n_{\rm hc}^{+} - n_{\rm hc}^{-} 
\label{nd_composite_eq}
\end{equation}
where
\begin{equation}
n_{\rm hc}^{+} = n_{\rm pix}^{\rm NG}(\nu=0.50)-n_{\rm pix}^{\rm NG}(\nu=2.00)
\end{equation}
\begin{equation}
n_{\rm hc}^{-} = n_{\rm pix}^{\rm NG}(\nu=-0.50)-n_{\rm pix}^{\rm NG}(\nu=-2.00).
\end{equation}
Figure \ref{nd_combined_fig} shows measurements of this quantity from
the simulations, as a function of the
smoothing scale adopted. 
The left panel is real units, the right panel is in errorbar units
as in Figure \ref{nd_sigma_distances_fig}.
By combining the two optimal levels
the sensitivity slightly improves, but still remains at the 1$\sigma$
level for $f_{\rm NL}=100$. This is
because the associated errorbars at different thresholds are correlated.

\subsection{Clustering strength of the excursion set pixels} \label{clustering_subsection}


Next, we consider the clustering strength of the excursion set
regions. We analyze in detail the case of $\nu=2.00$, where 
according to our previous findings
the sensitivity to a local $f_{\rm NL}$ type non-Gaussianity is maximized.

Figure \ref{clustering_fig} shows measurements
of the hot and cold pixel correlations above/below
threshold from the simulations. 
Left panels highlight the 
Gaussian case, intermediate panels are for 
$f_{\rm NL}=100$, and in  the right panels  
$f_{\rm NL}=500$.
A smoothing with FWHM of 30 and 60 arcmin is applied
in the central and bottom panels, respectively.
Solid lines are analytic predictions in the Gaussian
limit, i.e. equation (\ref{corr_smart_gauss}). 

The various unweighted correlation functions are calculated as
explained in Rossi et al. (2009). 
In particular, the number of random pairs is computed by distributing random points
on a unit sphere, and then by pixelizing them
in the HEALPix scheme at the same resolution of the maps.
The random realization used contains at least 10 times more
points than the simulated samples.
All the errorbars are estimated
directly from 200 realizations. Note that the 
scatter between different runs can be quite significant, especially 
at relatively large angular scales.

From Figure \ref{clustering_fig} it is evident that,  
for a positive and large $f_{\rm NL}$ value,
the clustering of the cold pixels is enhanced with respect to that of
the hot ones. This peculiar feature is a distinct signature of
non-Gaussianity of the local $f_{\rm NL}$ type. It is most prominent at angular scales of 
about $\theta=75'$, the Doppler transition
due to the sharp turn-down in the power spectrum at $l \simeq 1500$ (because of
the thickness of the last scattering surface). 
The asymmetry in the excursion set clustering is not surprising: it is
expected from the corresponding number
density behavior (Figures \ref{nd_smart_scale_fig} and
\ref{nd_difference_ratio_fig}), 
and from the shape of the non-Gaussian potential (equation \ref{fnl_expansion_eq}). 
In fact, at thresholds $\nu>1$ the cold pixel abundance is amplified with
$f_{\rm NL}$, while the number density of the hot pixels is reduced. 
This causes the difference in the clustering. Also, the
quadratic term in (\ref{fnl_expansion_eq}) is insensitive to a change
in sign, hence the asymmetry between hot and cold regions.  
However, when $f_{\rm NL}=100$ this feature is small, 
as shown in the middle panels of Figure \ref{clustering_fig}.
Therefore, at $N_{\rm side}=512$ the Gaussian correlation function of the excursion sets
is not easily distinguishable
from a non-Gaussian one, if the model of
non-Gaussianity is of the local $f_{\rm NL}$ type. Turning the
argument around, the excursion set clustering
statistic does not provide accurate constraints on 
Gaussianity itself (at least with this particular non-Gaussian model
in mind), contrary to what was previously thought
(i.e. Kogut et al. 1995; Barreiro et al. 1998; Heavens \& Sheth 1999; Heavens \& Gupta 2001).
Working at higher resolution would be 
more advantageous, as at $N_{\rm side}=512$ the predicted errorbars
from our simulations are quite large -- 
although they are rather pessimistic estimates, because they are based on 200 runs only.
Moreover, smoothing the maps has a more dramatic effect on the clustering
of the excursion set regions, rather than on their abundance: 
the larger-scale power is suppressed, while the
small-scale strength is enhanced. 
This results in an overall suppression of the
clustering feature previously described, particularly when $f_{\rm NL}=100$ (see the bottom panels in
Figure \ref{clustering_fig}).
Note that as we increase the smoothing, the clustering of hot
and cold pixels makes a transition between FWHM 30 and 60 arcmin, and
the clustering behavior is reversed.  
At larger smoothing values, we expect the hot pixels to cluster more
at larger $\theta$.   
This indicates that large values of FWHM will still be useful for
comparison of this quantity with real data.
We are addressing this issue in a forthcoming paper, where we deal
with the detectability of these non-Gaussian features from a real dataset.

While in Figure \ref{clustering_fig} solid lines are analytic predictions in the Gaussian
limit from equation (\ref{corr_smart_gauss}),
in Appendix \ref{edgeworth_proxy} we show an example of how well the Edgeworth approximation
(i.e. equation \ref{2d_edge_eq}) works for the clustering, when $f_{\rm NL}=100$.   
In equation (\ref{2d_edge_eq}), the second and third terms inside the square
bracket become important when $f_{\rm NL}$ is non-zero. While
$S^{(0)}$ is straightforward to compute, the term $\lambda$ is much
more complicated since it involves the computation of the full-sky three-point
function, and it is beyond the scope of this paper. However, we find
that ignoring the
$\lambda$ term still gives relatively good agreement with the results
from the simulations.  

\begin{figure}
\begin{center}
\includegraphics[angle=0,width=0.50\textwidth]{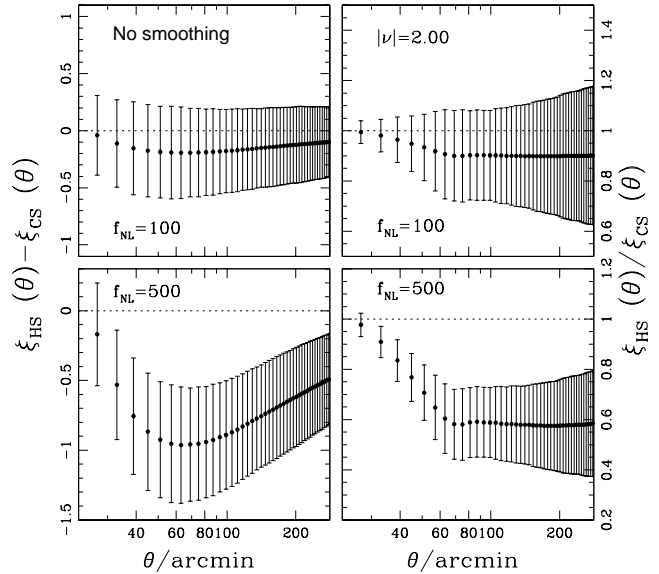}
\caption{Clustering difference (right panels) and ratio (left panels)
  between hot and cold excursion set regions at $|\nu|=2.00$, when no smoothing is
  applied. In the top panels $f_{\rm NL}=100$, in the bottom ones
  $f_{\rm NL}=500$.  See the main text for more details.} 
\label{clustering_difference_ratio_fig}
\end{center}
\end{figure}

In Figure \ref{clustering_difference_ratio_fig} we show the clustering
difference (left panels) and ratio (right panels) between hot
and cold excursion set regions, for two significant values of $f_{\rm NL}$. 
No smoothing is applied.
This is done in parallel with the number
density case (Figure \ref{nd_difference_ratio_fig}).
When $f_{\rm NL}=100$, the estimated sizes of our errorbars suggest again
that a clustering analysis is not ideal 
to detect departures from Gaussianity, although the behavior at
$f_{\rm NL}=500$ is quite peculiar.
In the next subsection we propose an optimized statistical test based
on the clustering strength, which aims at maximizing the difference between the
Gaussian and the non-Gaussian cases.

\subsection{Statistical test derived from the clustering strength} \label{clustering_stat_subsection}


\begin{figure*}
\begin{center}
\includegraphics[angle=0,width=0.31\textwidth]{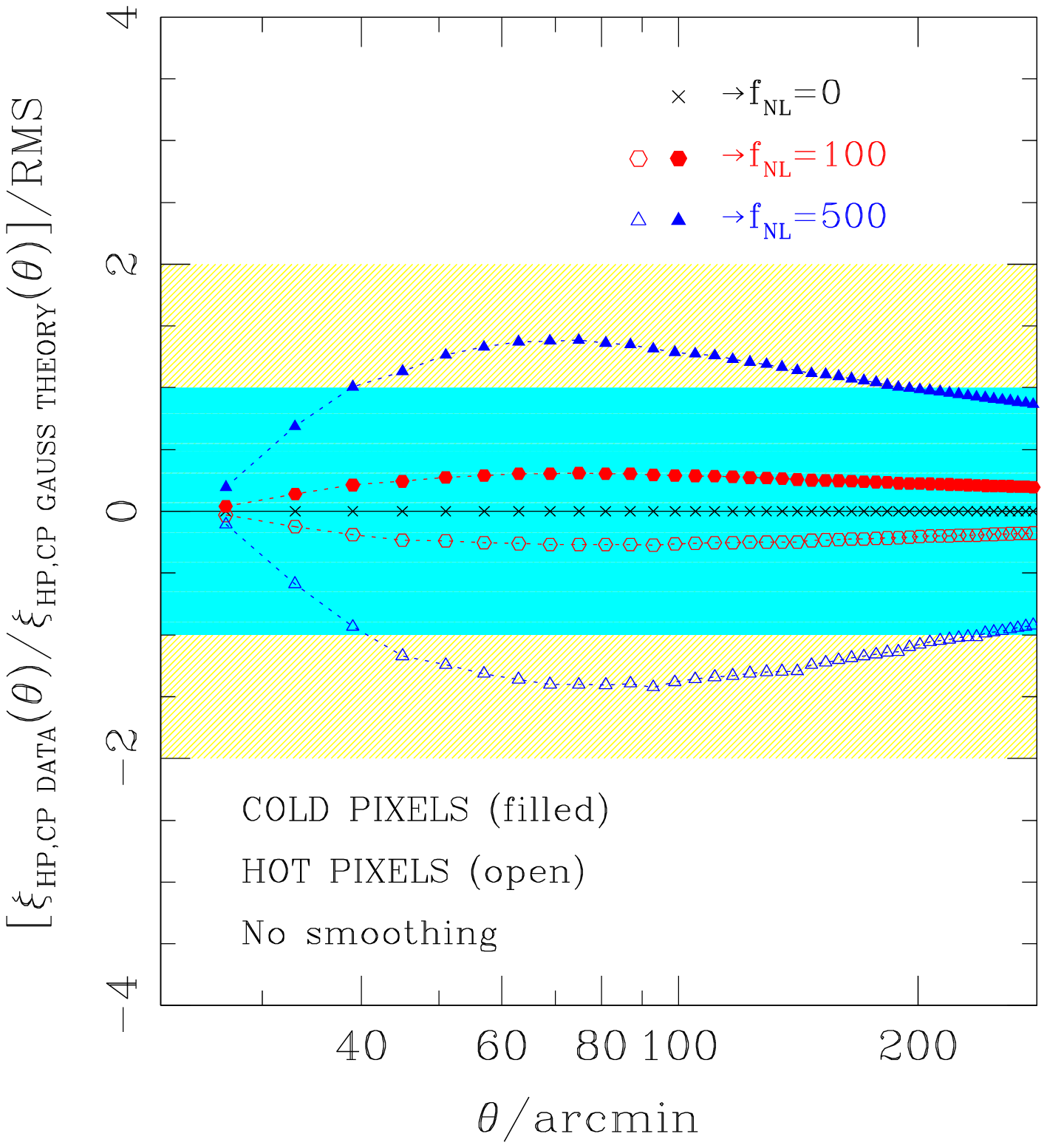}
\includegraphics[angle=0,width=0.31\textwidth]{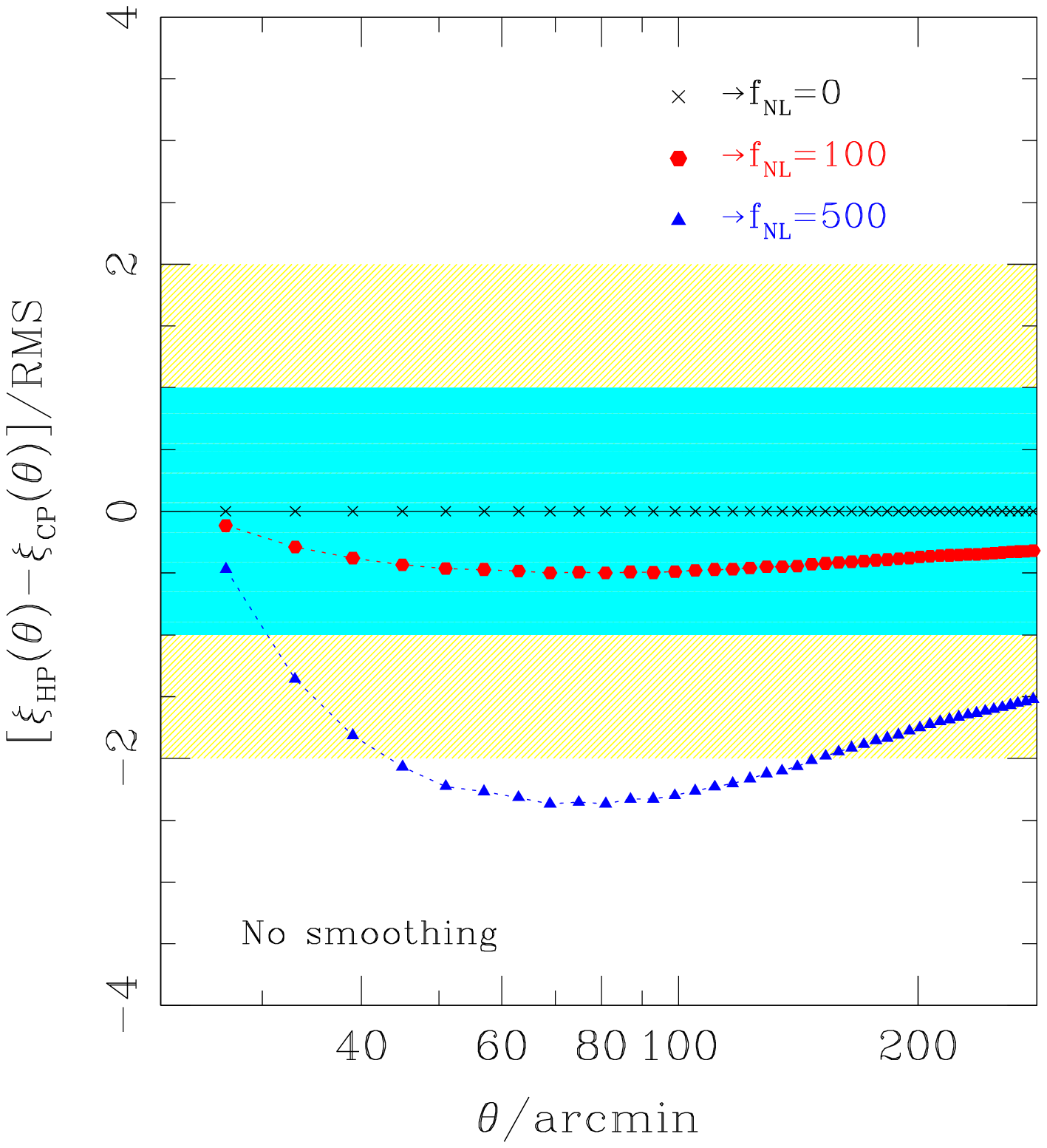}
\includegraphics[angle=0,width=0.31\textwidth]{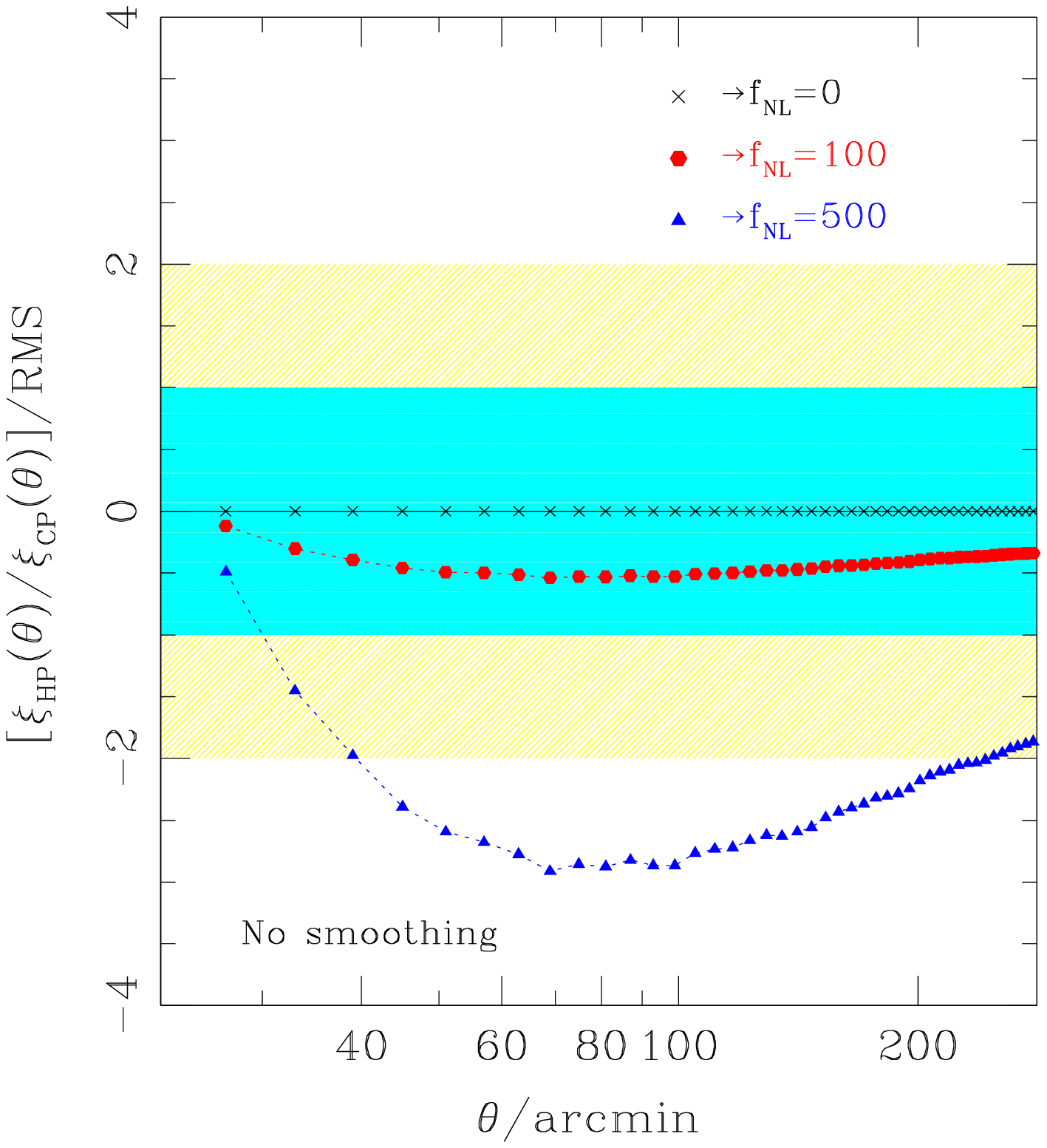}
\caption{Reinterpretation of Figure \ref{clustering_fig} (left panel) and Figure \ref{clustering_difference_ratio_fig} (middle and right panels) 
in terms of the run-to-run associated errors, in order to quantify the
sensitivity of the excursion set clustering strength to local non-Gaussianity. 
Different values of $f_{\rm NL}$ are displayed, as indicated in the
plots, as a function of the angular
separation $\theta$. No smoothing is applied. 
Only when $f_{\rm NL}$ is of the order of 500 there is a
noticeable difference in the clustering, which exceeds the 
2$\sigma$ level around the Doppler scale ($\theta \simeq 75'$) when $|\nu|=2.00$.} 
\label{clustering_sigma_units_fig}
\end{center}
\end{figure*}

The conclusions drawn from
Figures \ref{clustering_fig} and \ref{clustering_difference_ratio_fig}
can be expressed in a more quantitative form, as
was done for the number density in Section \ref{nd_stat_subsection}.
Assuming now $\xi_{\rm \nu}(\theta)$ to be the non-Gaussian
discriminator, Figure \ref{clustering_sigma_units_fig} shows
in errorbar units the correlation strength of the excursion set regions per
Gaussian expectations (left panel),
the difference (middle panel) and the ratio (right panel) between the clustering of
hot and cold pixels when $|\nu|=2.00$. No smoothing is applied.
Shaded areas represent the
1 and 2$\sigma$ zones, while different symbols are
used for different values of $f_{\rm NL}$, as specified in the plots. 
When $f_{\rm NL}=100$, departures from
Gaussianity lie always below the 1$\sigma$ level, unlike for the abundance case 
(see Figure \ref{nd_sigma_distances_fig}
for a direct comparison). 
The situation does not improve significantly
if we consider the clustering difference or ratio
between hot and cold patches.
Only when $f_{\rm NL}=500$ there is a
noticeable effect, which exceeds 
2$\sigma$ around the Doppler scale, at $\theta \simeq 75$ arcmin. 

The scatter in the clustering strength is mainly due to cosmic variance, which causes large fluctuations
among different full-sky realizations ($m$- and $\ell$-modes). As a
result, errorbars are large. To minimize its effect,
we propose a statistical test, which 
involves the clustering information alone. The procedure can be
summarized as follows.

\begin{enumerate}
\item Consider a non-Gaussian CMB temperature map  
    and extract its power spectrum.
\item Use equation (\ref{corr_smart_gauss}) to compute
  the corresponding analytic expectation for
 the pixel correlations above/below a threshold $\nu$, as if the map were thought
 to be Gaussian. 
 Denote this quantity as $\xi_{\rm \nu, Gauss}^{\rm NG}$; it is
the same for hot and cold excursion set regions, in the
Gaussian statistics. 
\item Compute the hot and cold correlation
   functions $\xi_{\rm \nu, h}^{\rm NG}$ and $\xi_{\rm \nu, c}^{\rm
   NG}$ directly from the map, at the same
  threshold level.
\item Construct the quantities:
\begin{equation}
\xi_{\rm \nu, h}^{\rm NG} = (\xi_{\rm \nu, h}^{\rm NG} -  \xi_{\rm
  \nu, Gauss}^{\rm NG})/\xi_{\rm
  \nu, Gauss}^{\rm NG} 
\label{cf_test_smart_hp_eq}
\end{equation}  
\begin{equation}
\xi_{\rm \nu, c}^{\rm NG} = (\xi_{\rm \nu, c}^{\rm NG} -  \xi_{\rm
  \nu, Gauss}^{\rm NG})/\xi_{\rm
  \nu, Gauss}^{\rm NG} 
\label{cf_test_smart_cp_eq}
\end{equation} 
and plot them as a function of the angular separation $\theta$.
\item Repeat this procedure for the entire set of non-Gaussian maps,
     with different $f_{\rm NL}$ values.
\end{enumerate}

\begin{figure*}
\begin{center}
\includegraphics[angle=0,width=0.49\textwidth]{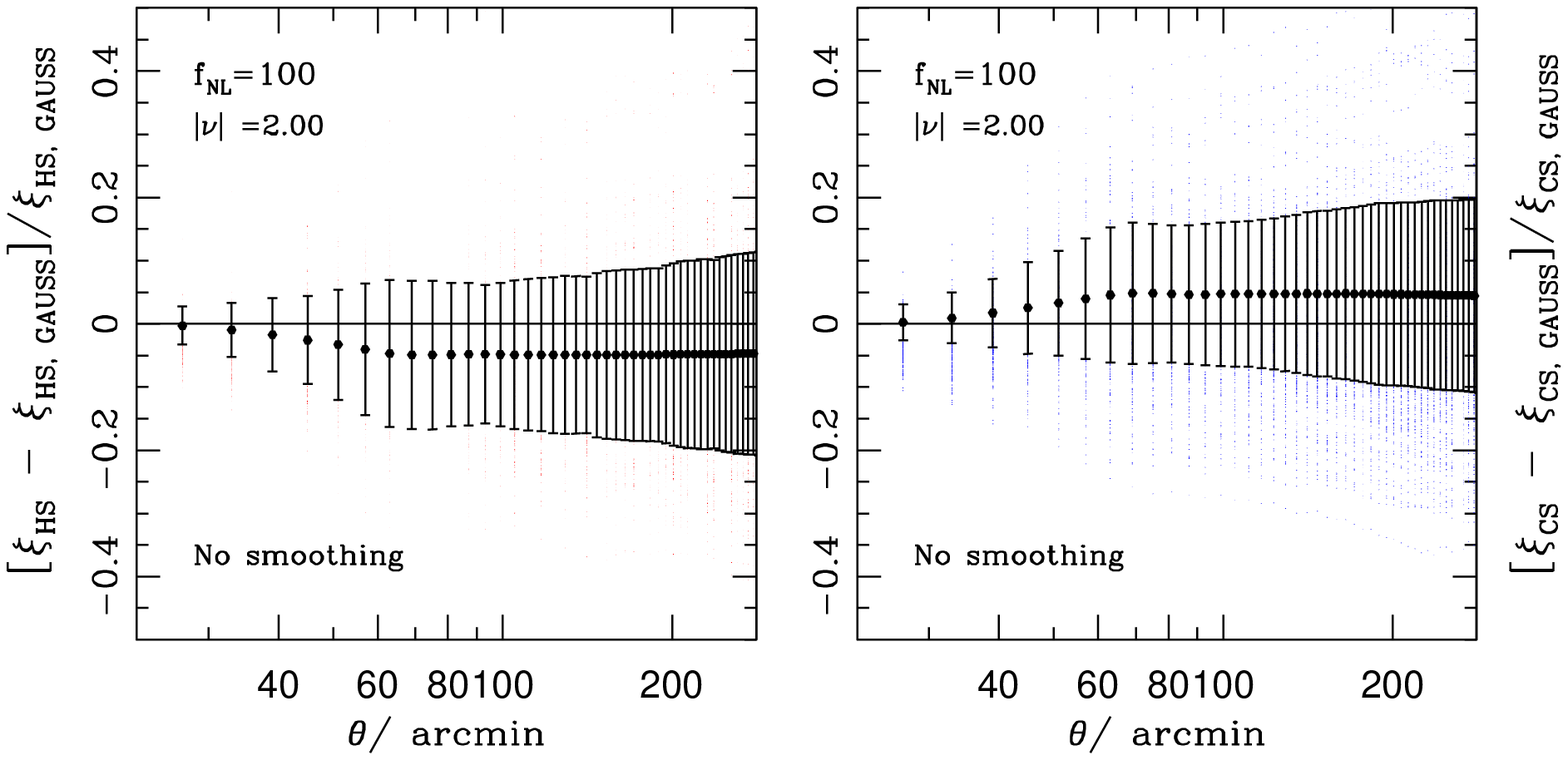}
\includegraphics[angle=0,width=0.49\textwidth]{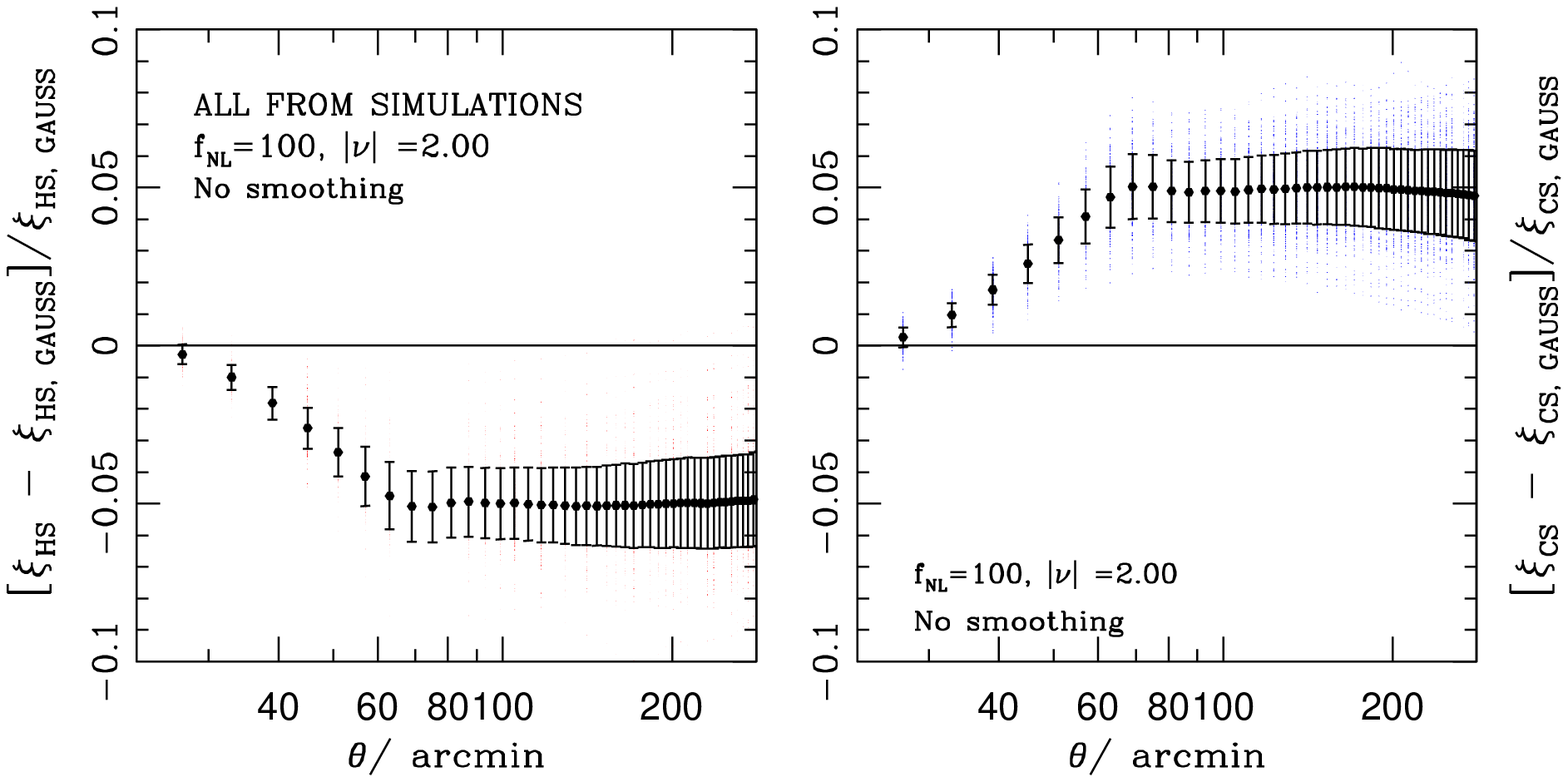}
\includegraphics[angle=0,width=0.49\textwidth]{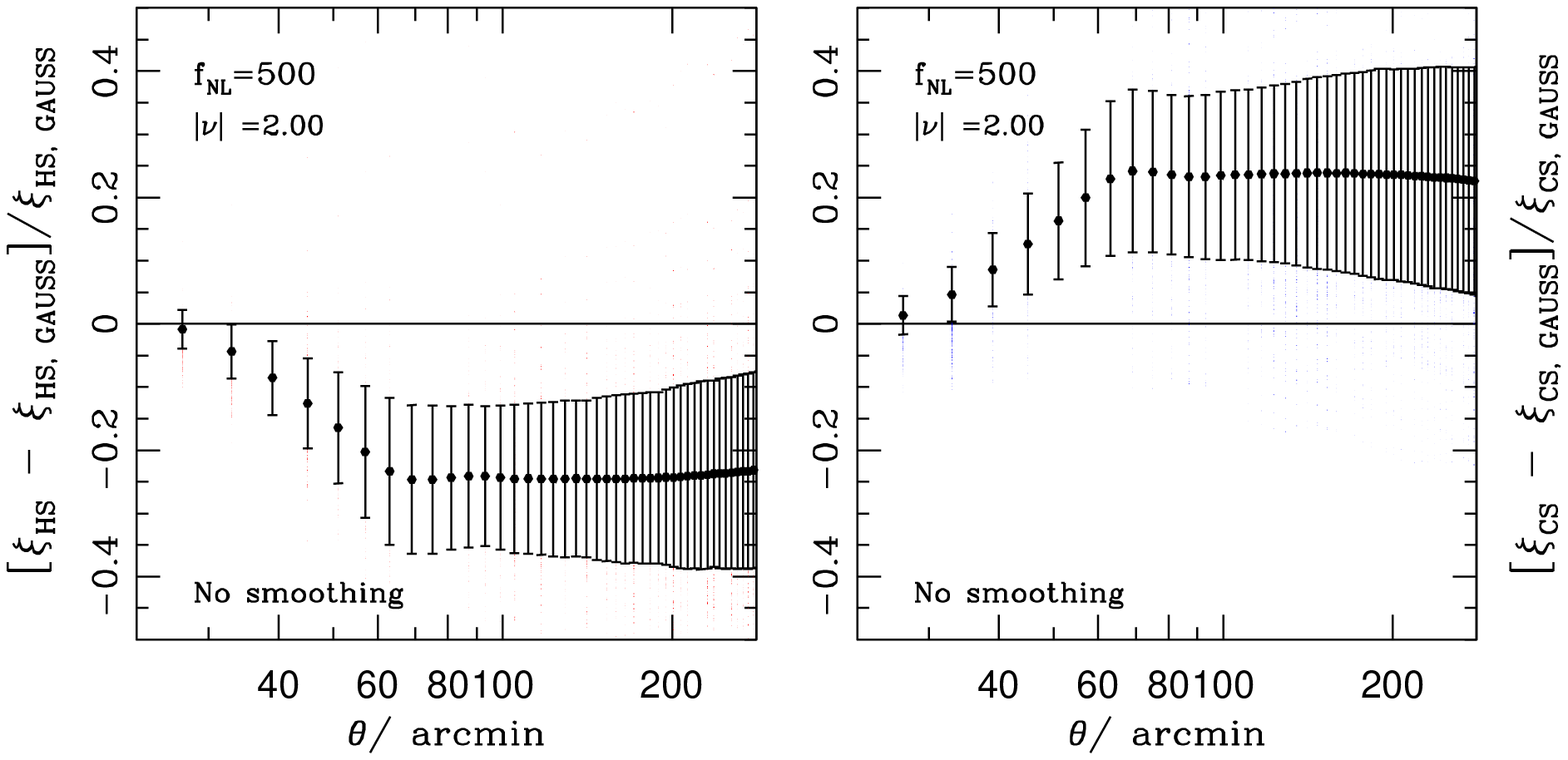}
\includegraphics[angle=0,width=0.49\textwidth]{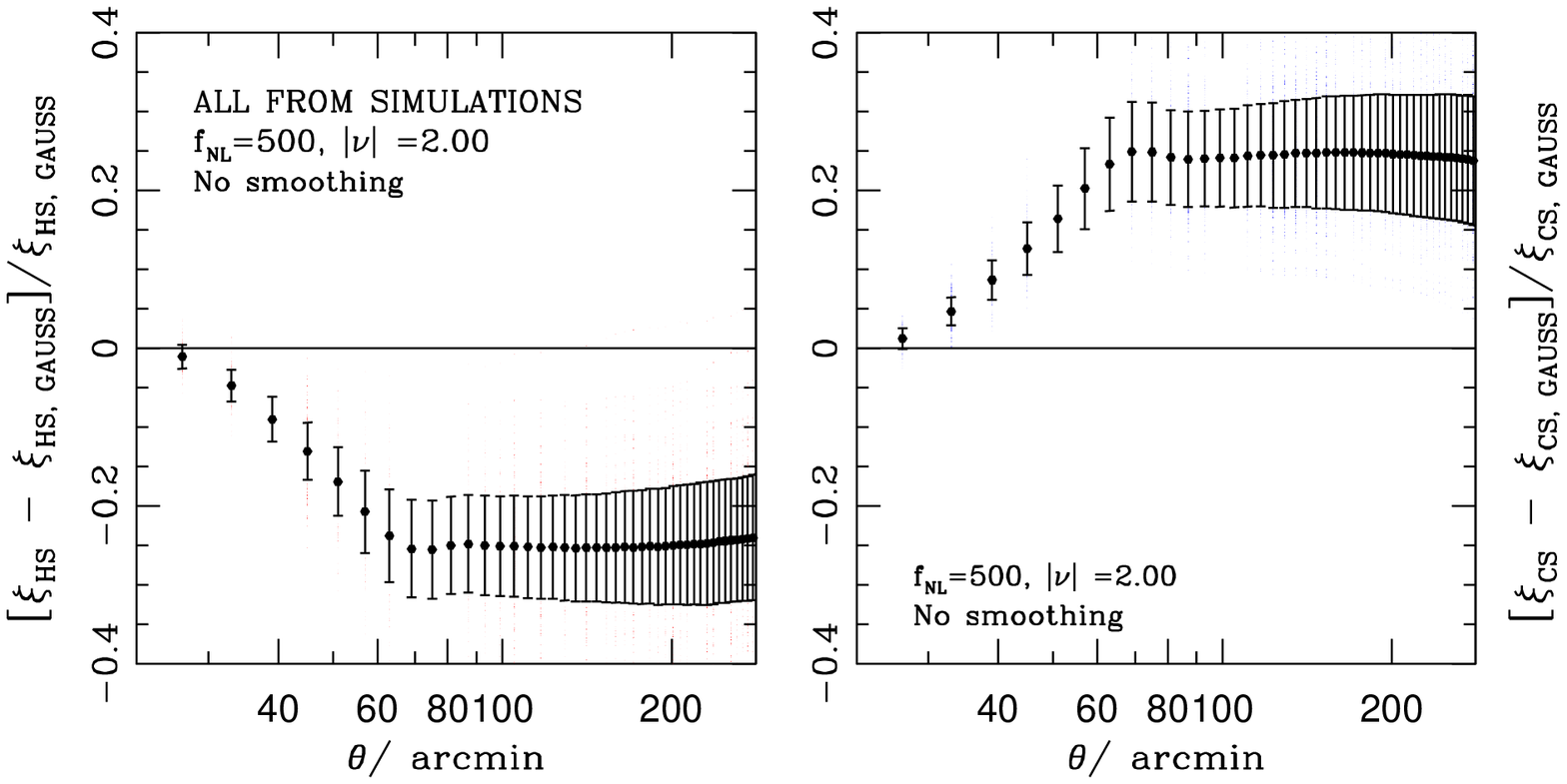}
\caption{Statistical test which 
involves the clustering information alone, as explained in the main
text. [Left] Measurements of the quantities (\ref{cf_test_smart_hp_eq}) and (\ref{cf_test_smart_cp_eq}) 
from the non-Gaussian simulations, with the corresponding Gaussian
theoretical prediction  $\xi_{\rm \nu, Gauss}^{\rm NG}$ estimated from the same maps.
[Right] Same as the left panels, but now $\xi_{\rm \nu, Gauss}^{\rm NG}$
is computed from the corresponding Gaussian maps with the same random seeds
of the non-Gaussian ones. In this way, the cosmic variance effect is completely cancelled.}
\label{clustering_smart_smarter_test_fig}
\end{center}
\end{figure*}

Results of these measurements are shown 
in the left part of Figure \ref{clustering_smart_smarter_test_fig};
$f_{\rm NL}=100$ in the top panels, $f_{\rm NL}=500$ in the bottom
ones. By subtracting the power spectrum contribution in the
numerator of (\ref{cf_test_smart_hp_eq}) and
(\ref{cf_test_smart_cp_eq}), the cosmic variance is partially reduced
because fluctuations in the corresponding $\ell$-modes are cancelled.
Unfortunately, the scatter in the
$m$-modes is still quite significant, so that at $f_{\rm NL}=100$ it
is not possible to distinguish a weakly non-Gaussian signal from a
Gaussian one using the clustering information alone.

The idealized situation is presented in the right part of
the same figure. Here, we replace $\xi_{\rm \nu, Gauss}^{\rm NG}$ by 
its direct measurement from the corresponding Gaussian
map with the same random seed. 
In other words, we do not use equation (\ref{cf_test_smart_hp_eq}) and
the non-Gaussian power spectrum to determine $\xi_{\rm \nu, Gauss}^{\rm NG}$.
Instead, we produce a Gaussian map with the same random seed
of the non-Gaussian one, and use it to compute the
theoretical expectation. This is repeated for the entire set of
non-Gaussian realizations.
The procedure is essentially the CMB analogous of what has been proposed by Seljak (2009)
for the LSS. In fact, in this way the cosmic variance effect is
completely cancelled;  
even at $f_{\rm NL}=100$, a clustering analysis would
then provide $\sim 5\%$ difference at the Doppler peak with respect to the
Gaussian case.

The situation described here is clearly ideal, because only
the first procedure (left panels in Figure
\ref{clustering_smart_smarter_test_fig}) 
can be performed from a real dataset.
However, it provides some important insights:
cosmic variance is the real limit and obstacle within this statistical
framework. If we could somehow control the 
fluctuations in the $m$-modes, then the excursion set analysis would
provide a powerful tool to detect non-Gaussianity.
The problem is that the CMB alone does not allow one to compare `tracers' at
different epochs (as for the LSS case), and so to eliminate completely the effect
of cosmic variance.



\section{Conclusions}  \label{NG_conclusions}


We have extended and applied the statistics of the excursion sets
to models with primordial non-Gaussianity of the $f_{\rm NL}$ local type.
While in presence of Gaussian initial conditions many statistics based
on geometrical and topological properties of the CMB temperature have
been developed and well-studied, to date
fewer analyses have been focused on geometrical properties of the CMB
radiation in the presence of primordial
non-Gaussianity. In particular, our work is the first
extension of the excursion set formalism to local $f_{\rm NL}$ type
non-Gaussianity.

From a large set of simulated full-sky non-Gaussian maps,
we computed the number density and the spatial clustering of 
CMB patches above/below a temperature threshold (Section \ref{NG maps
analysis}). We found that a positive value of 
$f_{\rm NL}$
enhances the number density of the cold CMB excursion sets (Figures
\ref{nd_smart_scale_fig},
\ref{nd_difference_ratio_fig}) along with their
clustering strength (Figures \ref{clustering_fig},
\ref{clustering_difference_ratio_fig}) and reduces that of the hot ones. 

We performed a thorough statistical analysis to evaluate the
sensitivity of the two observables to the level of non-Gaussianity
and to the smoothing resolution.
We also provided the analytical formalism to
interpret our results (Section \ref{NG theory}). 
Expressions for the one- and two-dimensional PDFs (Equations \ref{one_d_pdf_fnl}
and \ref{2d_edge_eq}) 
were obtained from a perturbative approach by the multidimensional Edgeworth expansion around a Gaussian
distribution function, and 
used to characterize the abundance and clustering statistics as a function of
$f_{\rm NL}$ (Equations \ref{nd_fnl_eq}, \ref{corr_smart}, \ref{corr_smart_gauss}). 
We showed that there are optimal
thresholds which maximize the local $f_{\rm NL}$ non-Gaussianity ($\nu
= 0.25, 0.50$ and $\nu=2.00, 2.25$), as well as others ($\nu=1.00$) which
do not allow for a distinction between the Gaussian and the
non-Gaussian signals (Figures \ref{nd_sigma_distances_fig} and \ref{clustering_sigma_units_fig}).
We devised a new statistical test based of the number density (Section \ref{nd_stat_subsection}),  
which combines
two thresholds where departures from Gaussianity are most
significant (Figure \ref{nd_combined_fig} and Equation \ref{nd_composite_eq}).
We also proposed a new procedure aimed at minimizing the effect of
cosmic variance (Section \ref{clustering_stat_subsection}),
which involves the clustering information alone (Figure
\ref{clustering_smart_smarter_test_fig}, Equations \ref{cf_test_smart_hp_eq} and \ref{cf_test_smart_cp_eq}).

Although we focused here on $f_{\rm NL}$ models of the local
type, the statistical tools developed are more
general and can be applied to describe any other type of non-Gaussianity.
A typical example is represented by the curvaton model, for which the cubic term
indicated as $g_{\rm NL}$ can be large, while $f_{\rm NL}$ can be negligible. Our technique can be applied to this case
 as well, and it is the subject of a forthcoming publication.

This work was primarily motivated by our previous finding (Rossi et al. 2009),
namely a remarkable difference in the
clustering of hot and cold pixels at relatively small angular scales
from the WMAP 5-yr data. We analyzed the possibility
that this discrepancy may arise from primordial
non-Gaussianity of the local $f_{\rm NL}$ type (Section
\ref{clustering_subsection}), and concluded that only a large value of
$f_{\rm NL}$ would provide such a difference (Figure
\ref{clustering_fig}). 
Cosmic variance plays a
crucial role within this statistical framework, so that 
the Gaussian correlation function of the excursion sets
is not easily distinguishable
from the non-Gaussian one, contrary
to what was previously thought. 
In fact, while a distinct signature in the clustering of hot and cold pixels 
clearly emerges for a large $f_{\rm NL}$ non-Gaussianity, particularly at angular
scales of about 75 arcmin (around the Doppler peak), as expected this feature is reduced  
when $f_{\rm NL}=100$. The clustering behavior is also strongly affected by the smoothing angle.
These findings suggest that Gaussianity itself cannot be
accurately constrained from the excursion set clustering statistics.
In fact, if in principle the use of pixel-pixel correlation functions as a test of Gaussianity is
very powerful, because there are no free parameters once the
underlying power spectrum has been measured, this may not be the case if
the non-Gaussian model is of the $f_{\rm NL}$ local type, and $f_{\rm NL}$ is small.

Our study was focused on a few selected values of
thresholds and two different statistics, so that the predicted
constraints on $f_{\rm NL}$ are wider than what one would get by
combining several threshold levels and different smoothing angles.
In this respect, our predicted constraints from the excursion
sets are compatible with those of Smidt et al. (2010), obtained from
the trispectrum. 

Since cosmic variance is the main obstacle in the analysis,
we are considering derived statistics which could potentially  beat
its effect and maximize the non-Gaussian contribution. 
It is also important to adopt different and complementary 
statistical approaches, and  
not just a single view, because
there is no such statistics which describes fully and
uniquely the non-Gaussian nature of a sample. 
To this end, a lot of effort
has recently gone into developing optimal estimators, and in this sense our statistical technique belongs to a class of topological estimators 
which may be considered ``sub-optimal'' for measuring non-Gaussianity.
However,  in
reality all the geometrical methods
complement and ``diagnose'' results obtained with bispectrum or
trispectrum estimators. Moreover, geometrical techniques are  
often model-independent, easy to implement, with low
computational cost, and
they can retain information on the spatial distribution of the
non-Gaussian signal. Also, they provide useful analytic insights and
physical intuition. 
For example, the derivation and implementation of the analytical
formula for the CMB Minkowski functionals in the limit of weak
non-Gaussianity (Hikage, Komatsu \&
Matsubara 2006; Matsubara 2010) has allowed to obtain limits on
various models, for which the optimal estimators are difficult to
implement; at the moment, a limit on the primordial non-Gaussianity in
the isocurvature perturbation is available \textit{only} from the Minkowski
functionals (Hikage et al. 2009). 
Note also that 
the concept of ``optimal'' is often misleading, as it requires a posteriori
knowledge of the type of non-Gaussianity which is, at least in principle, unknown.
The main question, instead, is whether or not it is possible to improve 
limits on $f_{\rm NL}$ using the CMB data only.

Including realistic effects in our simulations, such as inhomogeneous
noise, point source contamination or foregrounds, so that we can compare
our predictions with current observations, 
is subject of ongoing work (we provide some discussion in Appendices \ref{noise_analytic} and
\ref{spurious_ng}).  We present results of these
investigations in a companion
paper, where we are also consider more terms in the expansion (\ref{fnl_expansion_eq}).
Application of the formalism presented in Section \ref{excursion_set_formalism} 
to peak rather than pixel statistics is a straightforward exercise,
and is also the subject of another forthcoming publication.

The Planck satellite with its increased sensitivity and resolution is
expected to improve the measurements of most cosmological parameters
by several factors compared to WMAP, and in synergies with future
galaxy surveys (Colombo, Pierpaoli \& Pritchard 2009).
In fact, Planck gains a factor of 2.5 in angular resolution and up to 10 in
instantaneous sensitivity with respect to WMAP, and it is nearly
photon noise limited in the CMB channels (100-200 GHz). 
Repeating this analysis at the Planck resolution
may then provide more
stringent limits on $f_{\rm NL}$ from the excursion set statistics,
and is also the subject of work in progress.



\section*{Acknowledgments}

We dedicate this paper to the memory of KIAS president Prof. Hyo Chul Myung, who passed
away on February 11, 2010.
We thank an anonymous referee for helpful comments and suggestions.
We acknowledge the support of the Korea Science and Engineering
Foundation (KOSEF) through the Astrophysical Research Center for the
Structure and Evolution of the Cosmos (ARCSEC).
We acknowledge the use of the \textit{L}egacy \textit{A}rchive for 
\textit{M}icrowave \textit{B}ackground \textit{D}ata \textit{A}nalysis 
(\textit{LAMBDA}), support for which is provided by the National 
Aeronautics and Space Administration (NASA) Office of Space Science. 
Some of the results in this paper have been derived using the 
\textit{HEALPix} package (G\'orski et al. 1999). 
The computation in the paper was done on the QUEST cluster at KIAS.




\appendix



\section{Analytic non-Gaussian predictions for the excursion set pixel clustering}  \label{edgeworth_proxy}


\begin{figure}
\begin{center}
\includegraphics[angle=0,width=0.40\textwidth]{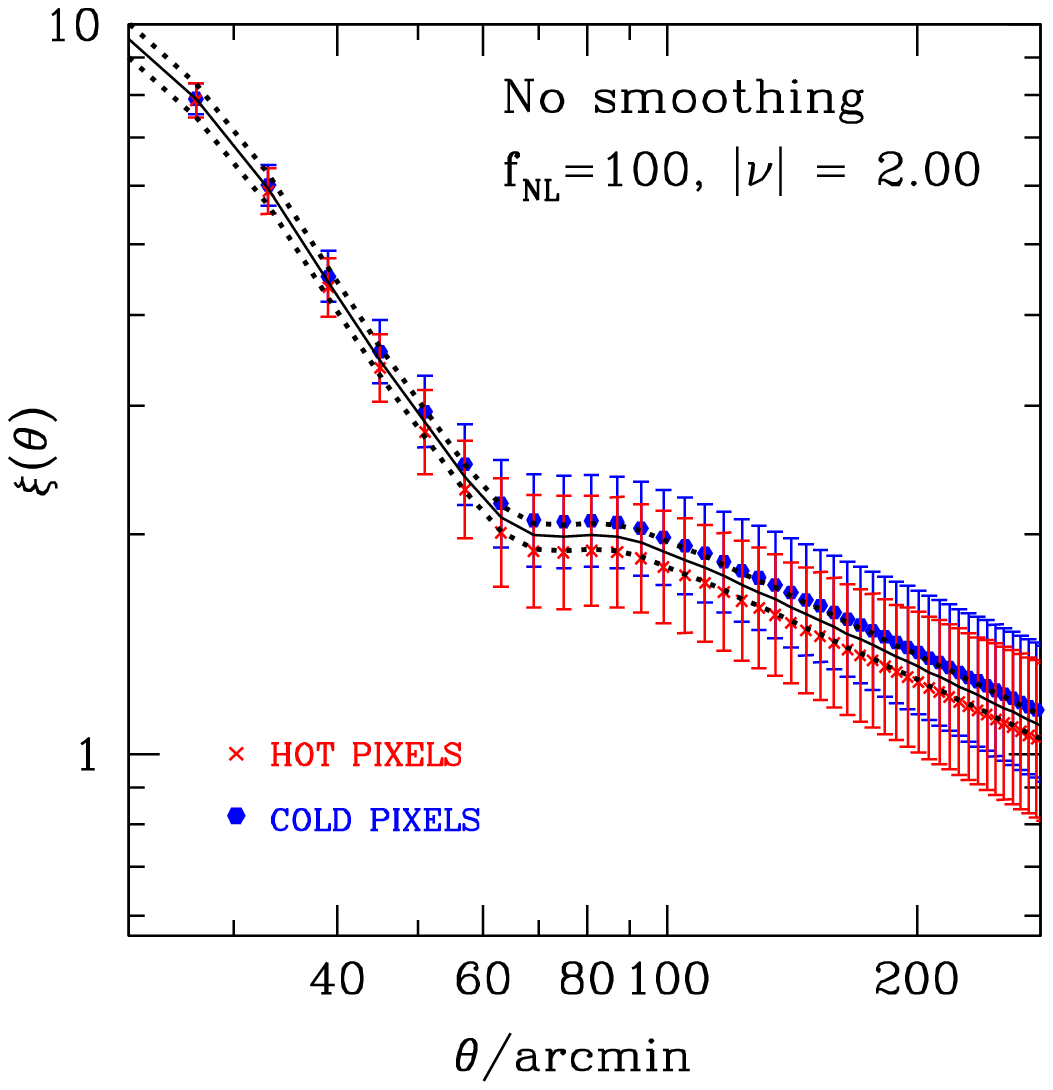}
\caption{Example of how well the Edgeworth approximation
works in describing the CMB excursion set clustering when $f_{\rm NL}=100$.
Points are hot and cold pixel correlations above/below $|\nu|=2.00$
measured from the simulations, when no smoothing
is applied.
Solid line is the Gaussian analytic prediction
(equation \ref{corr_smart_gauss}). Dotted lines are 
the non-Gaussian analytic expectations obtained by using equation
(\ref{2d_edge_eq}) and by ignoring
the $\lambda$ term.
The agreement between numerical results and 
theory predictions is reasonably good only at large angular scales $\theta$.}
\label{2d_edge_theory_fig}
\end{center}
\end{figure}

The multidimensional Edgeworth expansion is a convenient way of
approximating a PDF in terms of its cumulants. It is a true asymptotic
expansion, so that the error is well-controlled; it can be used to
describe weak non-Gaussianity.

Figure \ref{2d_edge_theory_fig} 
shows an example of how well the Edgeworth approximation
works for the CMB excursion set clustering.
Points in the figure are 
hot and cold pixel correlations above/below
a threshold $|\nu|=2.00$ from the simulations, when $f_{\rm NL}=100$ and no smoothing
is applied.
Solid line is the Gaussian analytic prediction from
equation (\ref{corr_smart_gauss}); dotted lines are 
the non-Gaussian analytic expectations obtained by using equation
(\ref{2d_edge_eq}) in (\ref{P2}) and (\ref{corr_smart}), and by ignoring
the $\lambda$ term.
The agreement between numerical results and 
theory predictions is still good at large angular scales $\theta$.
However, for small values of $\theta$ we expect $\lambda \rightarrow
\sigma S^{(0)}$,
hence this term becomes important in (\ref{2d_edge_eq}); 
this is why in Figure \ref{2d_edge_theory_fig}
the analytic prediction fits poorly in that regime.  
On the opposite, at higher threshold levels and when $\sigma S^{(0)}$
becomes large the Edgeworth expansion cannot be used.



\section{Inhomogeneous noise and partial sky coverage}  \label{noise_analytic}

The number density and the correlation strength of pixels above/below a temperature threshold
$\nu$ can be generically expressed by
\begin{equation}
 n_{\rm pix}(\nu) = { N_{\rm pix, tot} \over 4 \pi} \cdot P_{1}, 
\label{nd_gen}
\end{equation}
\begin{equation}
  1 + \xi_{\rm \nu}(\theta) = P_2/P_1^2,   
\label{corr_gen}
\end{equation}
\noindent where $N_{\rm pix, tot}$ is the total number of pixels, and
$P_1$ and $P_2$ are defined in equations (\ref{P1}) and
(\ref{P2}). By inserting the corresponding one- and
two-dimensional PDFs in those equations, and by using their
output in (\ref{nd_gen}) and (\ref{corr_gen}), one can readily characterize the pixel number density and the
clustering statistics above/below threshold in the fully Gaussian
case or the weak non-Gaussian limit (equations \ref{nd_gauss_eq}--\ref{corr_smart_gauss}).

The analytic expressions derived in the main text apply to
full-sky intrinsic CMB signal; the effect of noise is not included.
However, with the formalism introduced by Rossi et
al. (2009) we can also describe analytically the excursion set clustering in the weak
non-Gaussian limit, in presence of inhomogeneous noise. 
Maintaining the same notation, we
indicate the observed value in a pixel by $D = T - \langle T \rangle
\equiv \delta T = s + n$, which is the sum of the true signal $s$ plus noise
$n$, both of which have mean zero. We
consider a model in which the signal is homogeneous and may have 
spatial correlations 
whereas the noise, independent of 
the signal, may be inhomogeneous and have spatial 
correlations. We 
denote $p(D)$ the observed one-point distribution of $D$, 
$p(s)$ the distribution of $s$ with rms $\sigma_{\rm S}$, 
$p(\sigma_{\rm n})$ the distribution of the rms value of the noise 
in a pixel, 
and $p(n|\sigma_{\rm n})$ the distribution of the noise when the rms 
value of the noise is $\sigma_{\rm n}$. 
The one-point observed distribution is 
\begin{eqnarray}
p(D) &=& \int {\rm d}s~p(s) \int {\rm d}n~p(n) \delta_{\rm D} (s+n=D) \nonumber \\
     &=& \int {\rm d}s~p(s) \int {\rm d}n \int {\rm d}\sigma_{\rm n}~p(n|\sigma_{\rm
     n})~p(\sigma_{\rm n}) \nonumber \\
     & & \times~\delta_{\rm D}(s+n=D) \nonumber \\
     &=& \int {\rm d}\sigma_{\rm n}~p(\sigma_{\rm n}) \int
     {\rm d}s~p(s)~p(D-s|\sigma_{\rm n}) \nonumber \\
   &=& \int  {\rm d}\sigma_{\rm n}~p(\sigma_{\rm n})~p(D|\sigma_{\rm n}),
\label{pD}
\end{eqnarray}
\noindent where $\delta_{\rm D}$ is the Dirac delta.
The fraction of pixels above some temperature threshold $D_{\rm t}$ is
\begin{eqnarray}
 f(D_{\rm t}) &=& \int_{D_{\rm t}}^\infty {\rm d}D\, p(D)
        = \int {\rm d}\sigma_{\rm n}\, p(\sigma_{\rm n})\,
            \int_{D_{\rm t}}^\infty {\rm d}D\, p(D|\sigma_{\rm n}) \nonumber \\
        &=& \int {\rm d}\sigma_{\rm n}\, p(\sigma_{\rm n})\, f(D_{\rm
        t}|\sigma_{\rm n}).
\label{fDt}
\end{eqnarray}
Similarly, for two pixels separated by the angular distance $\theta$,
or having correlation $w \equiv w(\theta)$,
the two-point observed distribution is specified by:
{\setlength\arraycolsep{0.5pt}
\begin{eqnarray}
\lefteqn{p(D_1,D_2,w) = \int {\rm d}s_1 \int
  {\rm d}s_2~p(s_1,s_2,w)
  \int {\rm d}n_1 \int {\rm d}n_2 {}} \nonumber \\ 
& & {} ~~~ \times p(n_1,n_2)~\delta_{\rm D} (s_1
  +n_1=D_1)~\delta_{\rm D} (s_2+n_2=D_2) \nonumber \\
& & {} = \int {\rm d}s_1 \int {\rm d}s_2~p(s_1,s_2,w) \int {\rm d}n_1
  \int {\rm d}n_2 \int {\rm d}\sigma_1 \nonumber \\
& & {} ~~~ \times \int {\rm d}\sigma_2
  ~p(n_1,n_2|\sigma_1,\sigma_2)~p(\sigma_1,\sigma_2,w)~\delta_{\rm D} (s_1+n_1=D_1) \nonumber \\
& & {} ~~~ \times~\delta_{\rm D} (s_2+n_2=D_2) \nonumber \\
& & {}  = \int {\rm d}\sigma_1 \int {\rm d}\sigma_2\,
  p(\sigma_1,\sigma_2,w)
           \int {\rm d}s_1 \int {\rm d}s_2\, p(s_1,s_2,w)  \nonumber  \\
& & {} ~~~ \times p(D_1-s_1|\sigma_1)\,p(D_2-s_2|\sigma_2) \nonumber  \\
& & {} = \int {\rm d}\sigma_1 \int {\rm d}\sigma_2\,
                         p(\sigma_1,\sigma_2,w)\,
                         p(D_1,D_2,w|\sigma_1,\sigma_2)
\label{pD1D2}
\end{eqnarray}}
\noindent where 
\begin{eqnarray}
 p(D_1,D_2,w|\sigma_1,\sigma_2) &=& \int {\rm d}s_1 \int {\rm d}s_2\,
 p(s_1,s_2,w) \nonumber \\
 &\times& p(D_1-s_1|\sigma_1)\,p(D_2-s_2|\sigma_2). 
\label{pD1D2|sig1sig2}
\end{eqnarray}

Since $\mu=\delta T/ \sigma \equiv s/\sigma_{\rm S}$, where $\mu$ is the variable
used in the main text to indicate the threshold level, then $p(s) {\rm d}s \equiv p(\mu) {\rm d} \mu$ and $p(s_1,
s_2, w) {\rm d}s_1 {\rm d} s_2 \equiv p(\mu_1, \mu_2, w) {\rm d} \mu_1 {\rm d} \mu_2$.
Therefore we can use the PDFs (\ref{one_d_pdf_fnl}) and (\ref{2d_edge_eq}) to
characterize (\ref{pD}), (\ref{fDt}) and (\ref{pD1D2}) 
in the weak non-Gaussian limit, when inhomogeneous
noise is present.
Once (\ref{pD}), (\ref{fDt}) and (\ref{pD1D2}) are known, then 
the pixel number density and the clustering above/below threshold can be
inferred from (\ref{nd_gen}) and
(\ref{corr_gen}), where now
\begin{equation}
P_1 = \int_{D_{\rm t}}^{\infty} p(D) {\rm d}D \equiv f(D_{\rm t})
\label{P1_noise}
\end{equation}
and
\begin{equation}
P_2 = \int_{D_{\rm t}}^{\infty} {\rm d}D_1 \int_{D_{\rm t}}^{\infty}
{\rm d}D_2~p(D_1,D_2,w). 
\label{P2_noise}
\end{equation}

The corresponding Gaussian limiting case has been presented in detail in Rossi et al. (2009). 
In particular, if 
$p(s_1,s_2,w)$ is bivariate Gaussian with
$\langle s_1^2\rangle = \langle s_2^2\rangle = \sigma_{\rm S}^2$,
$\langle s_1s_2\rangle = C_{\rm S}(\theta)$ as defined in equation (\ref{cs}),
and the noise $p(n|\sigma_{\rm n})$ is Gaussian with variable rms
$\sigma_{\rm n}$, then
{\setlength\arraycolsep{0pt}
\begin{eqnarray}
\lefteqn{p(D_1,D_2,w|\sigma_1,\sigma_2) = \frac{1}{2 \pi \sqrt{||C||} }
 e^{-\frac{1}{2}D^{\rm T} \cdot C^{-1} \cdot D} {}} \\
& & {} =  {1\over 2 \pi \sigma_{\rm D}^2 \sqrt{\alpha_1 \alpha_2 - w^2} }\
    {\rm exp}\left\{-{\alpha_2 D_1^2 + \alpha_1 D_2^2 - 2w\, D_1D_2\over
               2\sigma_{\rm D}^2\,(\alpha_1 \alpha_2 -
 w^2)}\right\} \nonumber
\label{pD1D2_gauss}
\end{eqnarray}}
\noindent with $\alpha_1 = (\sigma_{\rm S}^2 + \sigma_1^2)/\sigma_{\rm D}^2$,
$\alpha_2  = (\sigma_{\rm S}^2 + \sigma_2^2)/\sigma_{\rm D}^2$, and 
\begin{equation}
w = {C_{\rm S}(\theta) + 
C_{\rm N}(\theta) \over \sigma_{\rm D}^2},
\end{equation}
\begin{equation}
C_{\rm N}(\theta) = \sum_{\rm \ell} {(2\ell+1)\over 4\pi}\,
             C_{\rm \ell}^{\rm N}\,
                      W^{\rm smooth}_{\rm \ell}\
             P_{\rm \ell}^0(\cos \theta),
\label{ctheta_noise}
\end{equation}
\noindent where $C$ is the covariance matrix of the temperature field,
$\sigma_{\rm D}^2$ the variance of D, 
$C_{\rm \ell}^{\rm N}$ the power spectrum of the noise map, 
and $W^{\rm smooth}_{\rm \ell}$ the
additional smoothing due to finite pixel size, 
optional Gaussian beam smoothing and mask influence.
Note in fact that in presence of incomplete sky coverage one needs to add an extra
window function in (\ref{cs}) and in (\ref{ctheta_noise}), 
according to the geometry of the survey, to account for
extra-correlations introduced by the mask.
If the noise is spatially uncorrelated, then clearly $C_{\rm
  N}(\theta) = 0$ and therefore
$w \equiv C_{\rm S}(\theta)/\sigma_{\rm D}^2$.
In the approximation where $\sigma_1=\sigma_2$, rms noise varies
spatially on scales much larger than those of interest, then
$\alpha_1=\alpha_2$.  The ``standard'' approximation, rms noise
independent of position, has $\alpha_1=\alpha_2=1$.

Properly characterizing all these experimental complications, when
primordial non-Gaussianity is assumed, 
is the next step in our analysis; it will be presented in a forthcoming publication. 
In particular, in order to compute (\ref{pD}), (\ref{fDt}) and (\ref{pD1D2}) 
in the weak non-Gaussian limit, a detailed knowledge of the noise distributions is required; namely,
$p(\sigma_{\rm n})$, $p(n|\sigma_{\rm n})$ and their corresponding two-dimensional
expressions must be specified. 
For example, those PDFs can be directly measured from a real dataset
and/or given by the specifics of the experiment (i.e,
WMAP, Planck, etc.), as was the case in Rossi et al. (2009).

However, it is
straightforward to predict what happens
in the presence of Gaussian white noise (independent of position, with
rms $\sigma_{\rm N}$). 
In fact, in this case  the effective rms of the CMB map increases; it
is given by $\sigma \equiv \sigma_{\rm D} =
\sqrt{\sigma_{\rm S}^2 + \sigma_{\rm N}^2 }$. Hence, there will be a
slight shift in the threshold level $\nu = D/\sigma_{\rm D}$, but all the equations derived in
the main text are still applicable -- provided that one replaces
$\sigma_{\rm S}$ with the effective rms of the map, $\sigma_{\rm D}$.
Handling inhomogeneous noise is more complicated and will be presented
separately, as the overall effect on the pixel number density and
clustering critically depends on the detailed characteristics of the noise.



\section{Analytic error estimates}  \label{errors_analytic}

\begin{figure}
\begin{center}
\includegraphics[angle=0,width=0.40\textwidth]{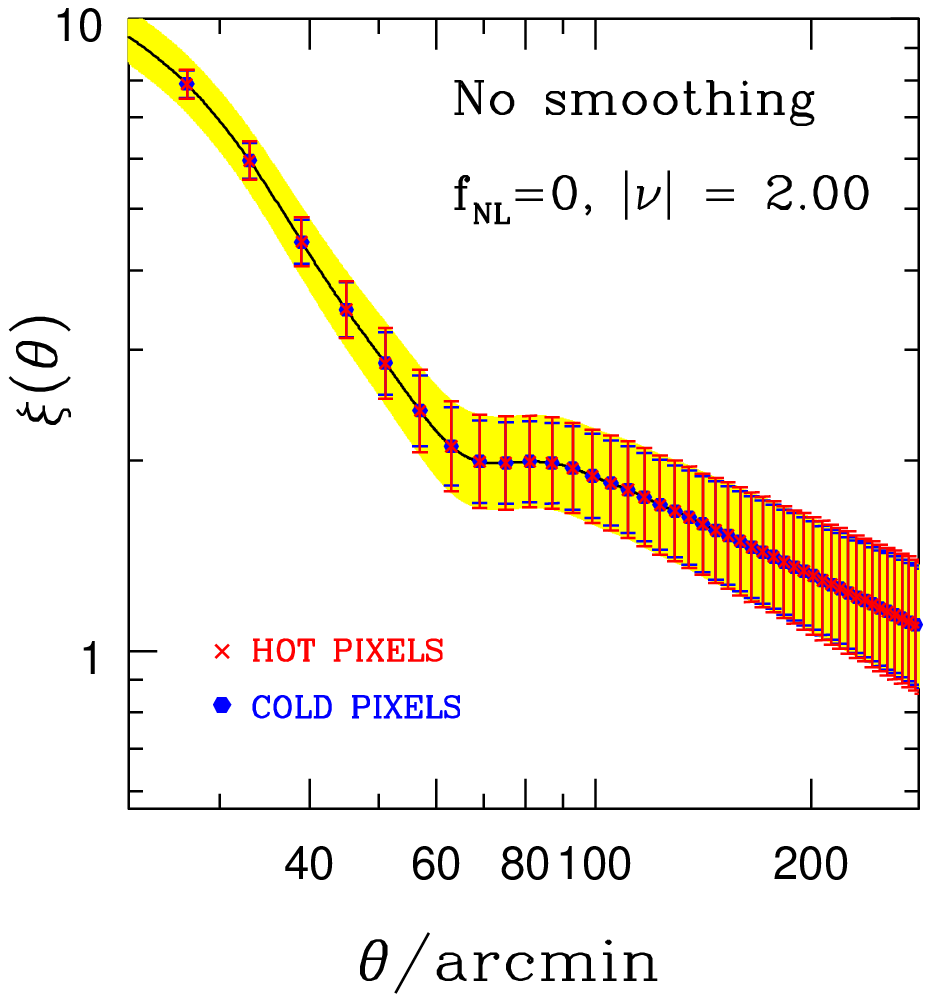}
\caption{Analytic estimate of the errors for the pixel
     correlation function above/below threshold (shaded area) derived from
     equation (\ref{delta_xi_theta}), in the Gaussian limit.
     Numerical errorbars are also shown; they are in good
     agreement with the theoretical predictions. The threshold level is $|\nu|=2.00$.}
\label{errors_analytic_vs_sim_fig}
\end{center}
\end{figure}

For a Gaussian random field,
the uncertainties in the pixel number density and in the correlation function above/below threshold 
can be evaluated analytically from
the optimal variance limit, which contains cosmic variance, instrumental
noise, and finite bin size effects. Details can be found in Rossi et al.
(2009). In essence, 
the ultimate accuracy with which the CMB power spectrum can
be determined at each $\ell$ is given by (Knox 1995):
\begin{equation}
 \Delta C_{\rm \ell} = \sqrt{{2 \over (2\,\ell+1)f_{\rm sky}}}\, \Big [
    C_{\rm \ell} + { 4\,
    \pi\, \sigma_{\rm N}^2 \over N_{\rm pix, tot} \, W_{\rm \ell}^{\rm instr}} \Big ]\,
\label{opt_var}
\end{equation}
\noindent where $W_{\rm \ell}^{\rm instr}$ is the instrumental window function and $f_{\rm sky}$ the fraction of
the sky covered by the experiment. 
The uncertainty in the angular correlation function for
narrow bins in $\theta$ is then:
\begin{eqnarray}
  \Delta C (\theta) &=&
  \Big\{ \sum_{\rm \ell} \Big | {\partial C(\theta) \over \partial
  C_{\rm \ell}}  \Big |^2 \, \Delta C_{\rm \ell}^2 \Big\}^{1/2} \nonumber \\
  &=& \Big\{ \sum_{\rm \ell} {(2 \ell+1)
  \over 8 \pi^2 f_{\rm sky}} |P_{\rm \ell}^0(\cos \theta)|^2(W^{\rm
  instr}_{\rm \ell}
  \cdot W^{\rm smooth}_{\rm \ell})^2 \times \nonumber \\
  & & \times \Big[{2 \pi C_{\rm \ell}^{\star} \over \ell(\ell+1)} +
  {\Omega_{\rm pix} \sigma_{\rm N}^2 \over W^{\rm instr}}
  \Big]^2\Big\}^{1/2}
\label{delta_C_theta}
\end{eqnarray}
\noindent where $C_{\rm \ell}^{\star} = \ell(\ell+1) C_{\rm \ell}/2
\pi$ and $\Omega_{\rm pix} = \theta_{\rm pix}^2$ is the pixel area.
If the bin size is
not infinitesimal, one needs to make a small correction -- which is
negligible for the scales we are interested in (see Rossi et al. 2009 for more details).
The uncertainties in the correlation function above/below threshold are finally
derived from:
\begin{equation} 
\Delta \xi_{\nu} (\theta) = \Big | {\partial \xi_{\nu} (\theta) \over \partial
   C(\theta)}  \Big | \, \Delta C(\theta)   
\label{delta_xi_theta}.
\end{equation}
An example of how well this analytic relation works in the Gaussian
limit is shown in Figure \ref{errors_analytic_vs_sim_fig}
by the shaded area. The result is compared with 
numerical estimates (errorbars), when the full-sky intrinsic CMB
signal is considered, in absence of pixel noise.  
As evident from the figure, the agreement between
theoretical expectations from equation (\ref{delta_xi_theta}) and numerical
predictions is good. 

In presence of primordial local non-Gaussianity, the situation is more
complicated. In principle, one can still follow the previous steps and derive
similar analytic expressions. However, when $f_{\rm NL} \ne 0$ the
power spectrum $C_{\rm \ell}$
is different from the Gaussian case, and the full-sky two-point angular correlation
function $C(\theta)$ cannot be expressed as simply as in (\ref{cs});
one needs to account for 
extra correlations introduced by the primordial non-Gaussianity. If noise
is also included, the situation is even more complicated -- as presented in Appendix
\ref{noise_analytic}. 
If we instead neglect all the contribution
from bispectrum to modify the power spectrum and the angular correlation function as for large $f_{\rm NL}$, then 
(\ref{delta_xi_theta}) will have an additional term proportional to $f_{\rm NL}$; the extra non-Gaussian part can be inferred from 
the two-dimensional PDF (\ref{2d_edge_eq}), i.e.  a term proportional to
\begin{equation}
\propto S^{(0)} \Big ( {H_{30} + H_{03}
    \over 6} \Big ) + O(f_{\rm NL}).
\end{equation}
However, we feel that estimating the errorbars by using  a large set of simulations is more accurate, for any 
arbitrary value of non-Gaussianity and in particular when $f_{\rm NL}$ is small. 
This is why
we do not attempt to derive analytic uncertainties in the weak non-Gaussian limit;
rather, our approach is to estimate errors using a large set of 
numerical simulations, with some guidance provided by the  
analytic expressions (\ref{opt_var}-\ref{delta_xi_theta}) valid in the Gaussian regime. 



\section{Spurious non-Gaussianities}  \label{spurious_ng}

Even for standard ``optimal'' estimators like the bispectrum or trispectrum, the problem of
non-Gaussianities arising from non-primordial sources is very challenging.  
There are many different contaminants which can be confused as
primordial non-Gaussian signals. Those include (1) \textit{instrumental
effects}, such as beam asymmetries, inhomogeneous noise, masks or incomplete
sky coverage, (2) \textit{astrophysical contaminants}, such as point sources,
foregrounds, presence of voids or anomalous cold spots, (3) \textit{secondary anisotropies}, such as  
the Integrated Sachs-Wolfe effect (ISW) and lensing, and so forth.
In certain cases the contamination is negligible, in other cases it may be
severe but one can account for it. There are also situations in which the
spurious non-Gaussianity is hard to account for, or its contribution
remains still unclear. 

Regarding instrumental effects, the inhomogeneity of the noise is the
most critical problem. However, in Appendix \ref{noise_analytic} we
showed how to extend our formalism in presence of inhomogeneous noise. 
Given a good knowledge of the experimental beams, window functions and
noise absolute calibration, it is possible to separate its
constribution from a primordial non-Gaussianity. 
Another possible contaminant is introduced by partial sky
coverage: for instance, edge effects due to pixels which lie very close to the mask could
potentially induce undesirable non-Gaussianities, but their effect
can be carefully modelled.

As far as astrophysical contaminants, the presence of low-density
regions in the southern Galactic cap (cold spots), and the contribution by 
unknown point sources or anomalous foreground emissions, are all possible
sources of confusion. Their accurate measurement is therefore crucial. 
In particular, uncertainties
in the foreground template model used for the foreground subtractions
may introduce anomalies at the percentage level, since Galactic
foregrounds are non-Gaussian and anisotropic. 
While an optimal estimator
has a clear framework to asses the amount of contribution
from secondary sources, in our case we may account for point source contaminations in two ways: theoretically, and by using simulations. 
For example, from a theoretical point of view, the following calculation shows how to quantify for the
contamination in the correlation function. Denote with the subscripts $P$ a point source, and with $T$
the pixel temperature; use $N$ for the respective number of pairs. Assume
no correlations between point sources (i.e. $\xi_{\rm PP} = 0$) and
neglect possible cross-correlations (i.e. $\xi_{\rm TP} = 0$).
The overall observed unweighted correlation function is then:
\begin{eqnarray}   
1 + \xi_{\rm obs} &=& {\sum_{\rm i} w_{\rm i} (1+\xi_{\rm i}) \over \sum_{\rm i} w_{\rm i}} \nonumber \\
&=& {N_{\rm TT} (1+\xi_{\rm TT})+ N_{\rm PP} (1+\xi_{\rm PP})+ N_{\rm TP} (1+\xi_{\rm TP})
  \over N_{\rm TT} + N_{\rm PP} + N_{\rm TP}} \nonumber \\
&\simeq& {N_{\rm TT} (1+\xi_{\rm TT})+ N_{\rm PP} + N_{\rm TP}
   \over N_{\rm TT} + N_{\rm PP} + N_{\rm TP}} \nonumber \\
&=& {N_{\rm TT} (1+\xi_{\rm TT})
   \over N_{\rm TT} + N_{\rm PP} + N_{\rm TP}} +  
   {N_{\rm PP}+N_{\rm TP} \over N_{\rm TT} + N_{\rm PP} + N_{\rm TP}} \nonumber \\
&\simeq& {N_{\rm TT} (1+\xi_{\rm TT})
   \over N_{\rm TT} + N_{\rm PP} + N_{\rm TP}} 
\label{xiobs}
\end{eqnarray}
\noindent since $N_{\rm TT} \gg N_{\rm PP}+ N_{\rm TP}$. Therefore one can write: 
\begin{equation}   
1 + \xi_{\rm obs} \simeq \gamma (1+\xi_{\rm TT})
\label{xiobs_approx}
\end{equation}
\noindent where
\begin{equation}   
\gamma = {N_{\rm TT} \over N_{\rm TT}+N_{\rm PP}+N_{\rm TP}}
\label{gamma}
\end{equation}
\noindent and $N_{\rm TT} = n_{\rm T}(n_{\rm T}-1)/2$, $N_{\rm PP} = n_{\rm P}(n_{\rm P}-1)/2$,
$N_{\rm TP} = n_{\rm T} n_{\rm P}$, with $n_{\rm T}$ the number of
effective temperature pixels and $n_{\rm P}$ the number of spurious
undetected point sources (equivalent to bad pixels). 
With a simulation approach, we can also quantify very accurately the 
contamination induced by point sources. This is achieved by adding 
point sources to the mock maps, and by repeating the same analysis as for the uncontaminated case.
A comparison between the two situations allows one to quantify the degree of contamination.

Spurious non-Gaussianities could also arise from secondary
anisotropies, such as gravitational lensing, cosmic reionization, Sunyaev-Zel'dovich, 
Sachs-Wolfe or Ostriker-Vishniac effects (see Komatsu 2010 for a
recent review). Phase transitions in the early Universe may
also introduce a new source of non-Gaussianity, difficult to
disentangle from a primordial non-Gaussian signal.
In particular, the most serious contamination of the local $f_{\rm
  NL}$ model is represented by the coupling between the ISW and the
weak gravitational lensing: in fact, the coupling between small and large
scales creates a local form bispectrum of non-primordial
origin. Recently, Hanson et al. (2009) have shown that the lensing-ISW
coupling can cause a bias in the $f_{\rm NL}$ parameter on the order of $\Delta
f_{\rm NL} \simeq 10$.   
However, in general the primordial non-Gaussian signal can be
separated from non-Gaussian secondary anisotropies on scales relevant
for WMAP and Planck.

All these effects
are of course very important, before one can claim a pure detection of a
pzrimordial non-Gaussianity. We are planning to address all these issues in
detail, when we apply our techniques to a real dataset. 


\label{lastpage}
\end{document}